\documentclass[%
 amsmath,amssymb,
 aps,
reprint
]{revtex4-2}

\usepackage{graphicx}
\usepackage{dcolumn}
\usepackage{bm}
\usepackage[colorlinks,citecolor=blue]{hyperref}

\usepackage{outlines}
\usepackage{braket}
\usepackage{float}
\usepackage{comment}
\usepackage{makecell}
\usepackage{mathtools}
\usepackage[dvipsnames]{xcolor}
\usepackage{tabularx}
\usepackage{url}
\usepackage[normalem]{ulem} 
\usepackage{siunitx}
\usepackage{array}
\usepackage{dsfont}
\usepackage{amsthm}
\usepackage{xpatch}
\usepackage{xcolor}
\usepackage{titletoc}   
\usepackage{booktabs,tabularx}

\usepackage[utf8]{inputenc}
\usepackage[T1]{fontenc}
\usepackage{lmodern}

\xpatchbibmacro{doi+eprint+url}{%
  \printfield{doi}%
}{%
  \iffieldundef{doi}{}{\href{https://doi.org/\StrSubstitute{\thefield{doi}}{ }{}}{\nolinkurl{\thefield{doi}}}}%
}{}{}

\begin{document}

\title{Ultracold Mechanical Quantum Sensor for Tests of New Physics}

\author{Andraž Omahen$^{1,2}$}
\thanks{These authors contributed equally to this work. \newline \href{mailto:aomahen@phys.ethz.ch}{aomahen@phys.ethz.ch}}
\author{Simon Storz$^{1,2}$}
\thanks{These authors contributed equally to this work. \newline \href{mailto:aomahen@phys.ethz.ch}{aomahen@phys.ethz.ch}}
\author{Marius Bild$^{1,2}$}%
\author{Dario Scheiwiller$^{1,2}$}%
\author{Matteo Fadel$^{1,2}$}
\author{Yiwen Chu$^{1,2}$}%
\thanks{\href{mailto:yiwen.chu@phys.ethz.ch}{yiwen.chu@phys.ethz.ch}}
\affiliation{$^{1}$Department of Physics, ETH Z\"{u}rich, 8093 Z\"{u}rich, Switzerland}
\affiliation{$^{2}$Quantum Center, ETH Z\"{u}rich, 8093 Z\"{u}rich, Switzerland}


\begin{abstract} 

Initialization of mechanical modes in the quantum ground state is crucial for their use in quantum information and quantum sensing protocols. In quantum processors, impurity of the modes' initial state affects the infidelity of subsequent quantum algorithms. In quantum sensors, excitations out of the ground state contribute to the noise of the detector, and their prevalence puts a bound on rare events that deposit energy into the mechanical modes. In this work, we measure the excited-state populations of GHz-frequency modes in a high-overtone bulk acoustic wave resonator (HBAR). We find that the population of the first excited state can be as low as $P_p$~=~(1.2$\pm$5.5)$\times10^{-5}$, corresponding to an effective temperature of \SI{25.2}{\milli\kelvin}, which are upper bounds limited by imperfections in the measurement process. These results compare favorably to the lowest populations measured in superconducting circuits. Finally, we use the measured populations to constrain the amplitude of high-frequency gravitational waves, the kinetic mixing strength of ultra-light dark matter, and non-linear modifications of the Schr\"{o}dinger equation describing wavefunction collapse mechanisms. Our work establishes HBARs as a versatile resource for quantum state initialization and studies of fundamental physics.
\end{abstract}

\maketitle

 Circuit quantum acoustodynamics (cQAD) systems allow us to control the mechanical modes of massive objects at the single-quantum level \cite{Chu2017,Satzinger2018,Arrangoiz-Arriola2019}. Their rapid development in recent years has inspired many proposals for their use as building blocks of quantum processors \cite{Qiao2023, Wollack2022, vanThiel2025} or quantum sensors and detectors \cite{Linehan2024,Trickle2025}. The latter can be used to explore a wide range of important open questions in different fields of physics, from the existence of high-frequency gravitational waves \cite{Aggarwal2025} to how quantum effects manifest in macroscopic, massive systems \cite{Schrinski2023}.

$\hbar$BARs, which are cQAD devices based on HBARs \cite{Chu2018}, are especially well-suited for both quantum information processing and sensing applications. Compared to other GHz-frequency mechanical systems such as surface acoustic waves or phononic crystals, HBARs have several unique properties. First, with a mode volume of approximately \SI{e6}{\micro\meter^3} and an effective mass of a few \SI{}{\micro\gram}~\cite{Bild2023}, the mechanical resonator constitutes a macroscopic quantum system~\cite{Schrinski2023,Fadel2025}, which is crucial for coupling to signals such as gravitational waves and dark matter. Second, not only have HBAR modes themselves been shown to exhibit extremely high quality factors with minimal dephasing~\cite{luo2025}, the coherences of superconducting qubits are not significantly compromised by their coupling to HBARs~\cite{Garcia-Belles2025}. This has allowed for the preparation of complex quantum states of motion and high-fidelity quantum operations~\cite{Bild2023,vonLuepke2024,Marti2024}. Third, HBAR modes are encapsulated inside a 3-dimensional bulk crystal, which is beneficial for thermalization to the surrounding thermal bath and initialization in the quantum ground state at the mK temperatures. This last property of HBAR modes makes them particularly promising as low-noise quantum sensors, quantum resources with low state-initialization errors, and even ancilla modes for cooling other quantum systems such as superconducting qubits.

In this work, we measure the excited-state populations $P_p$ of HBAR modes and use them to place bounds on a number of fundamental physics signals. Our experiment uses a quantum protocol that transfers excitations in the mechanical modes to a superconducting qubit, whose first excited state population $P_q$ is then measured~\cite{Geerlings2013, Jin2015}. We further determine that our measurements are limited by slow drifts of the device properties, in particular the qubit decay and decoherence rate, which leads to an over-estimation of $P_q$. Nevertheless, under favorable measurement conditions, we observe $P_p$~=~(1.2$\pm$5.5)$\times10^{-5}$, which is the lowest excited-state population in any quantum system reported in the MHz or GHz range. Our results represent a proof-of-principle demonstration that $\hbar$BARs can be used as versatile quantum sensors in a variety of fundamental physics studies.

\begin{figure*}
  \begin{center} \includegraphics[width=1.8\columnwidth]{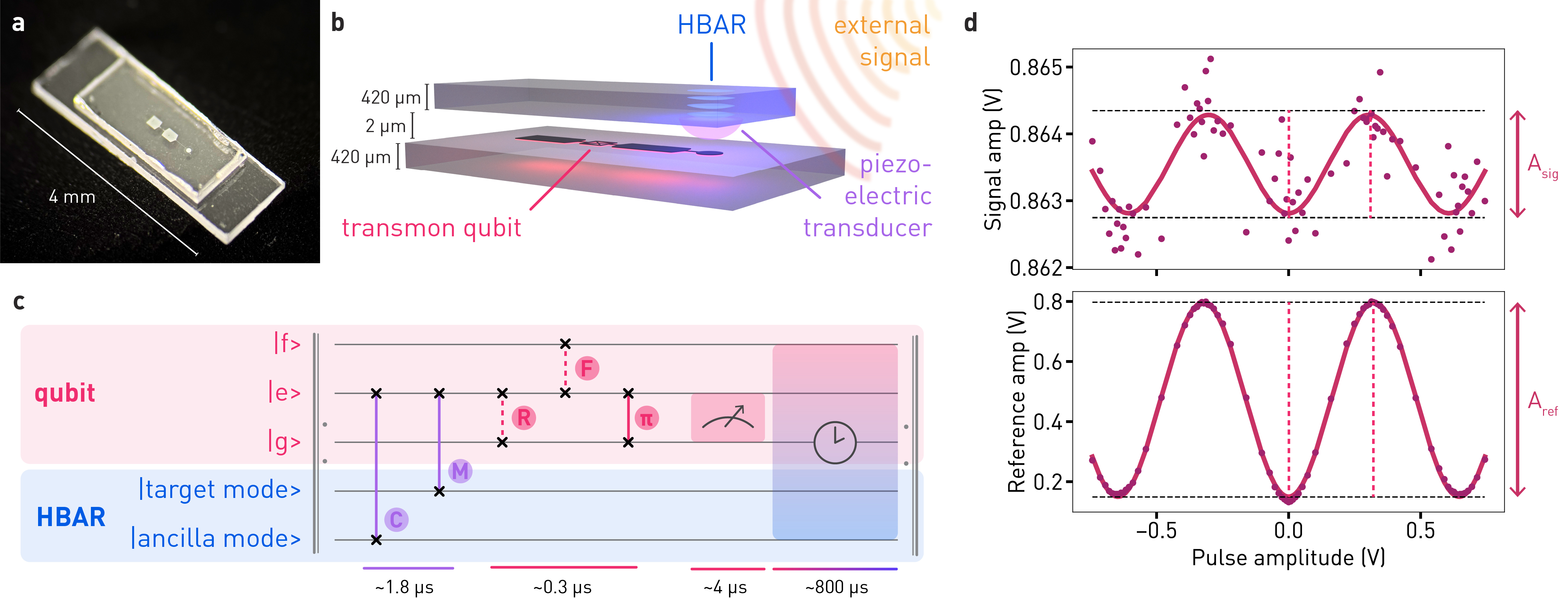}
  \caption{\textbf{Experimental System and Protocol.} \textbf{a}, Photograph of a high-overtone bulk acoustic resonator (HBAR; Sapphire (Al$_2$O$_3$)), bonded on top of a transmon qubit (Al on Sapphire). \textbf{b}, Schematic of the device. An external signal (e.g. a gravitational wave) excites phonons in the HBAR. A superconducting circuit is used to read out the excited-state population of the phonons using a piezo-electric transducer. \textbf{c}, Experimental protocol, as described in the text. Horizontal lines indicate different states of the qubit or modes of the HBAR, accordingly. Vertical lines correspond to quantum gates between such states. \textbf{d}, Example data for qubit e-f-subspace Rabi oscillations with (below) and without (above) a reference pulse.}
  \label{fig:protocol}
  \end{center}
\end{figure*}


\section*{Experimental System and Procedure}

We employ a system consisting of an HBAR coupled to a superconducting transmon qubit, as in former work \cite{Yang2024}. The device is comprised of two sapphire chips, flip-chip bonded together, as shown in Fig.~\ref{fig:protocol}(a). The upper chip hosts the HBAR (Fig.~\ref{fig:protocol}(b)), which supports localized phonon modes. The lower chip contains the superconducting transmon qubit, used to create, control, and read out quantum states of the acoustic modes. Coupling between the two systems is mediated by a piezoelectric AlN transducer deposited on the acoustic chip. The spherical dome shape of the transducer confines acoustic modes with a Gaussian beam profile, which have a waist of approximately $\mu$~=~\SI{30}{\micro\meter} and a length of L~=~\SI{435}{\micro\meter}, yielding a free spectral range of $\omega_{\text{FSR}}\approx$~\SI{12.5}{\mega\hertz}. 
The qubit has a transition frequency of $\omega_{q}\approx$~\SI{5}{\giga\hertz} and can be tuned to access several phonon modes with longitudinal mode numbers around $n\approx$~400 through the AC-Stark effect. We note that there is a small uncertainty on the precise mode number of a specific mode ($\approx$ n$\pm$13), as the FSR slightly varies with frequency \cite{Garcia-Belles2025}. The device is housed in a superconducting aluminium 3D cavity for gate control and dispersive readout and is operated at \SI{10}{\milli\kelvin} in a dilution refrigerator to suppress thermal noise (Supplementary Information, Section \ref{app:setup}).

In this work, we use the population of a mechanical mode as a sensitive indicator for the presence of weak, resonant, external signals (Fig.~\ref{fig:protocol}(b)). Such signals could consist of, for example, high-frequency gravitational waves or dark matter interactions. They would couple energy into the mode, so that any excess population beyond the expected thermal background can be interpreted as a signature of new physics.  Our system is described by the Jaynes-Cummings Hamiltonian and a drive term caused by the potential external signal,
\begin{align}
\hat{H}/\hbar =\ & \omega_p \hat{a}^\dagger \hat{a} + \frac{1}{2} \omega_q \hat{\sigma}_z + g \left( \hat{a}^\dagger \hat{\sigma}_- + \hat{a} \hat{\sigma}_+ \right)\\ + & 2 \Omega_d \cos \left(\omega t \right) \left( \hat{a} + \hat{a}^\dagger  \right).
\label{eq:hbar-hamiltonian}
\end{align}
Here, $\omega_p$ is the frequency of the acoustic mode, $g$ is the qubit--phonon coupling strength, and $\hat{a}$ is the annihilation operator of the acoustic mode. The operators $\hat{\sigma}_z$ and $\hat{\sigma}_\pm$ describe the qubit. The external signal acts as a coherent drive with amplitude $\Omega_d$ and frequency $\omega$.

To characterize the excited-state population of the phonon modes, we adapt the protocol from Ref.~\cite{Geerlings2013, Jin2015} (Fig.~\ref{fig:protocol}(c)). We begin by cooling the qubit via an iSWAP operation (see End Matter) with a dedicated cold acoustic mode (gate C), realized by shifting the qubit into resonance with the cooling mode. We then perform another iSWAP operation between the qubit and target phonon mode (gate M). After this interaction, the population in the first excited state of the phonon mode, $P_p$, is transferred to the qubit's first excited state, $\ket{e}$. To probe this population, we drive Rabi oscillations between $\ket{e}$ and the second excited state $\ket{f}$ of the qubit (gate F), as illustrated in Fig.~\ref{fig:protocol}(d). To optimize measurement speed, we use only two pulse amplitudes, zero and that corresponding to a $\pi_{\text{ef}}$-pulse on the $\ket{e} \leftrightarrow \ket{f}$ transition (dashed lines in Fig.~\ref{fig:protocol}(d)). The qubit state is then mapped to the ground state by applying a $\pi_{\text{ge}}$ pulse (gate $\pi$), enhancing readout fidelity. We perform a standard dispersive readout on the qubit, followed by a wait time of at least six phonon relaxation times to allow the system to reach steady-state before the next repetition. The resulting signal contrast, $A_{\text{sig}}$, is defined as the difference in qubit response between the two drive amplitudes.
We then perform a reference measurement following the same sequence, but with an additional $\pi_{\text{ge}}$ pulse immediately after the qubit-phonon swaps (gate R in Fig.~\ref{fig:protocol}). This inverts the qubit population, so that the protocol effectively measures the ground state population, defining the reference contrast $A_{\text{ref}}$, as shown in Fig.~\ref{fig:protocol}(d).
The resulting phonon population is then extracted as \cite{Geerlings2013} 
\begin{equation}
    P_p = \frac{A_{\text{sig}}}{A_{\text{sig}} + A_{\text{ref}}}.
\end{equation}
The excited state population of the qubit itself, $P_q$, is characterized analogously, without applying the phonon-qubit swaps (gates C and M).

To reduce statistical uncertainty, each phonon population measurement is averaged over several million repetitions $n_{avg}$, typically accumulated over several days. We acquire data in $n_{blocks}$ blocks with $2 \times 10^6$ shots each. In between each block, we recalibrate key parameters such as the qubit frequency, the $\pi_{{ge}}$- and $\pi_{{ef}}$-pulse amplitudes, and the qubit-phonon iSWAP gate. This prevents long-term drifts in the measured populations caused by slow changes in the experimental setup (see also Supplementary Information, Section \ref{app:statistics}) .

\section*{Measured Populations}
\begin{figure}
  \begin{center} \includegraphics[width=1.0\columnwidth]{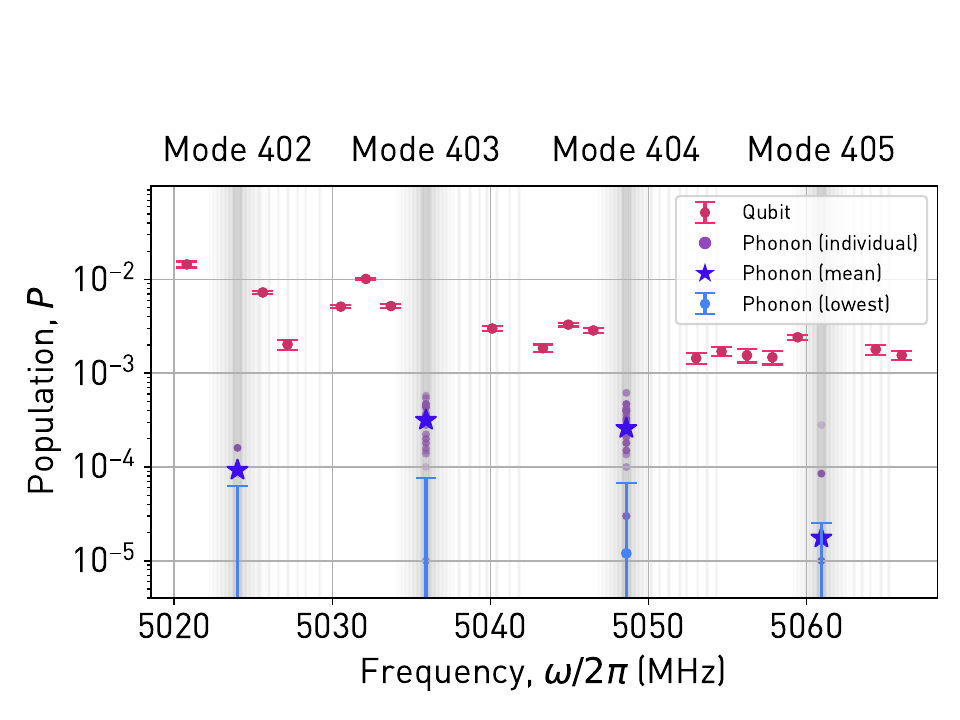}
  \caption{\textbf{Measured population.} First excited state population of the qubit (red) and four phonon modes (purple). Each mode was measured multiple times, ranging from 3 runs for mode 402 to 26 for mode 404. The opacity of each point reflects its uncertainty: faint points indicate larger standard deviations, while solid points represent more precise measurements. A star denotes the weighted average across all runs for a given mode, where weights are proportional to the inverse of the variance of each measurement. The blue markers highlight the measurement with the lowest value of mean plus standard deviation for each mode. Gray shaded lines indicate phonon modes: dark gray spans the finite linewidth of each Gaussian target mode, while lighter bars denote modes with higher transverse mode numbers.}
  \label{fig:pop_vs_freq}
  \end{center}
\end{figure}

\begin{figure}
  \begin{center} \includegraphics[width=1\columnwidth]{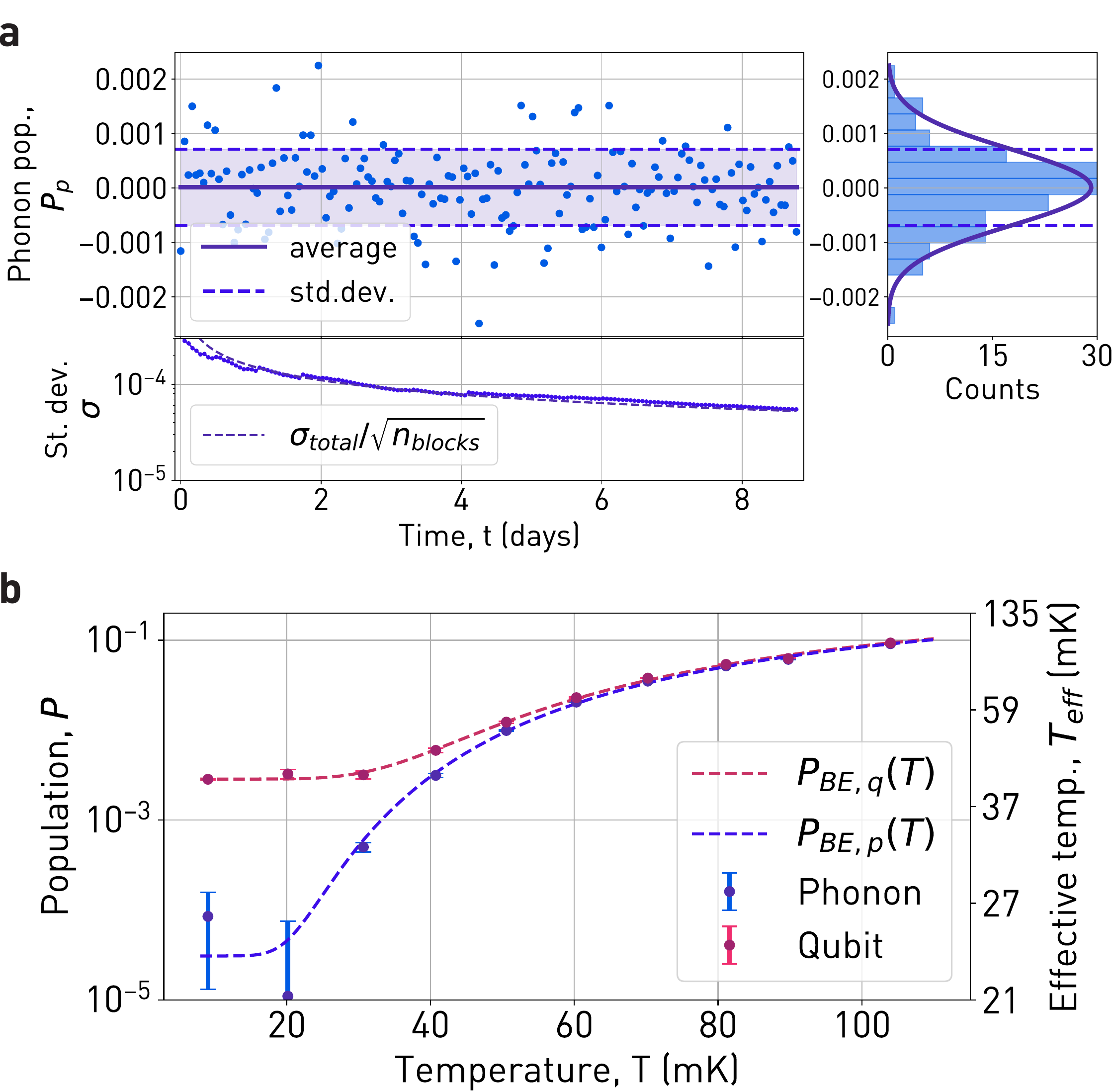}
  \caption{\textbf{Phonon population dynamics.} \textbf{a}, Phonon mode 404 population as a function of time, continuously measured for over a week with a total of $n_{avg}=324$ million averages. Each data point corresponds to the mean population of a measurement block. The negative values of $P_p$ are attributed to noise in the measurement protocol that lead to negative values of $A_{\text{sig}}$. The dashed lines correspond to the standard deviation taking into account all data, $\sigma_{total}$. The lower panel displays the standard error of the mean (SEM) up to the given point in time, $\sigma/\sqrt{n_{blocks}}$. The dashed line represents the expected behavior of the SEM, $\sigma_{total}/\sqrt{n_{blocks}}$. \textbf{b}, Population of the qubit and phonon mode 405 for elevated environmental temperatures. The points represent measurement data and the dashed line the Bose-Einstein distribution with an offset at low temperatures. Effective temperatures are calculated assuming a thermal distribution.}
  \label{fig:pop_vs_time}
  \end{center}
\end{figure}

We first characterize the excited-state population of the qubit by sweeping its frequency over a 45~MHz range. At higher frequencies, the qubit population is measured to be $P_q=1.5\times 10^{-3}$, as shown in Fig.~\ref{fig:pop_vs_freq}. As the qubit frequency is shifted to lower values, the excited-state population increases to approximately $P_q=10^{-2}$, which we attribute to excitations caused by the stronger AC-Stark-shift drive. We then measure the excited-state population of the four phonon modes within the qubit's tunable range. The phonon mode population, averaged over all measurement runs, is found to be between $\overline{P_p^{405}}$~=~(1.8$\pm$4.1)$\times10^{-5}$ (mode 405) and $\overline{P_p^{403}}$~=~(3.2$\pm$0.2)$\times10^{-4}$ (mode 403), indicating that the mechanical modes are one to two orders of magnitude colder than the qubit. Each phonon mode is measured across multiple cooldowns, yielding slightly different results. We attribute these variations to changes in the experimental setup and device parameters, as discussed later. The lowest populations obtained over a single measurement run are below $10^{-4}$ for all modes, when considering the upper bound given by the mean plus standard deviation (Fig.~\ref{fig:pop_vs_freq}). In the rest of this paper, we choose to quote the number obtained by the experiment with the most averaging - a measurement of the population of mode 404 spanning 9 days, see Fig.~\ref{fig:pop_vs_time}(a), yielding $P_p^{404}$~=~(1.2$\pm$5.5)$\times10^{-5}$. Below, when we base inferences about new physics on these numbers, we use the upper bound, $P_p^{404, \text{max}}$~=~6.7$\times10^{-5}$. 

As demonstrated in Fig.~\ref{fig:pop_vs_time}(a), the repeated recalibration of the experimental setup in between each block ensures long-term stability of the population dynamics over the course of more than a week. We recover a gaussian distribution of the measured population around the mean value $P_p^{404}$, and we observe the expected reduction of the standard deviation $\sigma$ with $1/\sqrt{n_{blocks}}$.

We also characterize the excited state populations of the qubit and phonon mode 405, and the corresponding effective temperatures, by applying a controlled heat load to the base stage of the dilution refrigerator. The cryostat temperature is monitored using a ruthenium oxide sensor mounted on the 3D cavity. As shown in Fig.~\ref{fig:pop_vs_time}(b), the measured first excited state populations follow the expected Bose-Einstein statistics, 
\begin{equation}
P_{BE}(T) = \left(1 - e^{-\beta \hbar\omega} \right) e^{-\beta \hbar \omega} + P_1^{\text{offset}},
\end{equation} with $\beta = 1 / k_B T$, where $k_B$ is the Boltzmann constant. At low temperatures, the population saturates at an offset $P_p^{\text{offset}}\approx 3\times 10^{-5}$, consistent with the population observed in Fig.~\ref{fig:pop_vs_freq}. We note that, while the populations are shown as effective temperatures in Fig.~\ref{fig:pop_vs_time}(b), we cannot determine whether the modes are in a thermal state.

The significantly lower population in the HBAR modes compared to the qubit is primarily attributed to their weaker coupling to electromagnetic noise, such as high-energy radiation that readily excite superconducting qubits but not mechanical resonators. Heating due to quasiparticles and measurement should also not affect the HBAR. Ultimately, the effective temperature of the HBAR modes is limited by the environmental temperature set by the dilution refrigerator, approximately \SI{10}{mK}. To identify the dominant mechanisms preventing us from measuring this limit, we simulate the full protocol (Fig.~\ref{fig:protocol}(c)) using QuTiP \cite{QuTip2024}, incorporating gate imperfections, thermalization, and decoherence of both qubit and phonon modes, as well as leakage to the qubit's second excited state (see Supplementary Information, Section \ref{app:statistics}, for details). 

\begin{figure}
  \begin{center} \includegraphics[width=1\columnwidth]{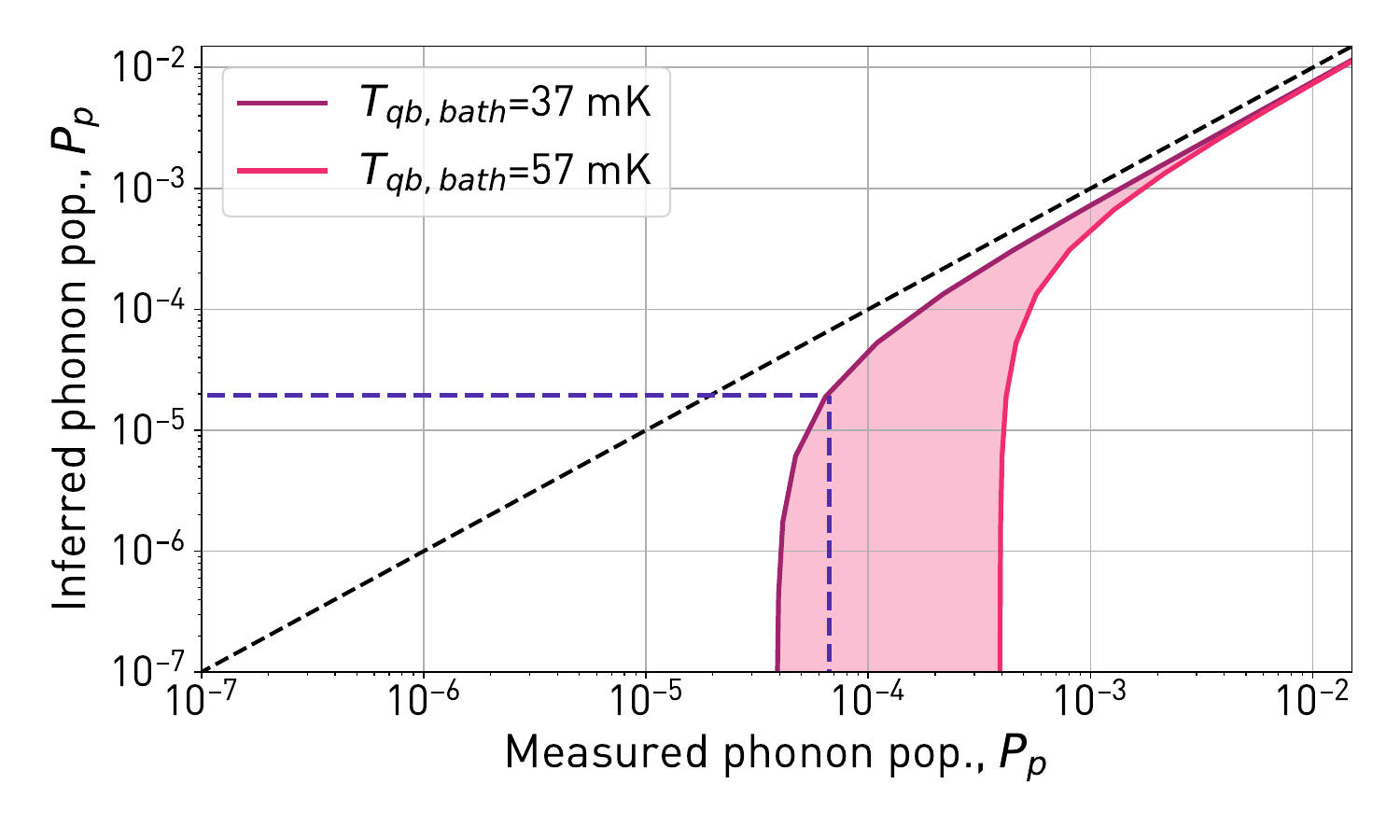}
  \caption{\textbf{Upper bounds of the measured and inferred phonon mode population.} The red region indicates simulated measured phonon populations for different inferred phonon populations, which is bounded by results for the maximum and minimum qubit bath temperature measured in  Fig.~\ref{fig:pop_vs_freq} (solid lines). The dashed vertical blue line shows the upper bound of the coldest measured phonon population of mode 404, $P_p^{404, \text{max}}$~=~6.7$\times10^{-5}$ (from Fig.~\ref{fig:pop_vs_freq}). Based on the simulations, the highest inferred phonon population consistent with the observed value is determined by the intersection of the vertical line with the upper boundary of the simulated region, defining the horizontal line at $P_p^{404, \text{inferred}}$~=~1.9$\times10^{-5}$.}
  \label{fig:simulation_results}
  \end{center}
\end{figure}

We find that the measured phonon population is limited primarily by errors introduced during the measurement procedure (Fig.~\ref{fig:protocol}(c)). The dominant source of error arises from the thermalization of the qubit to its bath during the phonon-qubit iSWAP operation (gate M, Fig.~\ref{fig:protocol}). We can therefore use the simulations to infer an upper bound on the actual phonon population based on the measured population of phonon and qubit. This bound is $P_p^{404, \text{inferred}}$~=~1.9$\times10^{-5}$, as illustrated in Fig.~\ref{fig:simulation_results}. Additional, less impactful error mechanisms, such as gate imperfections, finite coherence, and leakage, are analyzed in Supplementary Information, Section \ref{app:statistics}.

To the best of our knowledge, these excited-state populations are the lowest ever reported for any quantum system operating in the MHz or GHz range. This is relevant for quantum information processing tasks, as discussed in the End Matter.

\section*{Implications for Fundamental Physics}

By assuming that the measured excited-state population $P_p$ is entirely due to a particular external signal, we can provide an upper bound on its strength. Here we give two examples of signals relevant for fundamental physics.

\subsubsection*{Gravitational Waves} 

We first consider our macroscopic quantum device as a sensor for high-frequency gravitational waves (GWs), as described in Ref.~\cite{Linehan2024}. While gravitational waves in the Hz to kHz regime have been experimentally confirmed since 2015 \cite{Ligo2015}, no higher-frequency GWs have been observed so far - despite recent efforts with classical sensors \cite{Goryachev2021}. Gravitational waves in the GHz regime can originate from exotic processes predicted by theories beyond the Standard Model, including the motion of defects in spacetime called cosmic strings, collisions involving primordial black holes formed shortly after the Big Bang, or phase transitions in the state of the early universe \cite{Aggarwal2021,Aggarwal2025}.

Here, we consider a HBAR detector oriented along the z-direction, interacting with a monochromatic gravitational wave. The gravitational wave is described by $h_{ij}  = h_0 \epsilon_{ij} \cos \left( \omega t + \phi \right)$ with amplitude $h_0$, polarization tensor $ \epsilon_{ij}$, frequency $\omega$ and phase $\phi$. We assume that the wave is resonant with one of the acoustic modes and that it is plus polarized with respect to x- and z- direction, such that $\epsilon_{33}=1$. Under these assumptions, the gravitational wave acts as a classical coherent drive in equation~(\ref{eq:hbar-hamiltonian}) with the driving strength $\Omega_{GW} =
    - h_{0} \frac{\mu }{n^2}   \sqrt{ \frac{ \rho \omega^3 L^3}{2 \hbar  \pi^3 }  } $, 
where $\rho$ is the density of sapphire 
(derived in Supplementary Information, Section \ref{app:force_quant}). 

If the HBAR is coupled to an external bath at zero temperature with rate $\Gamma=1/T_{1,p}$, the presence of an external drive results in a non-zero steady-state population $\left< P_p \right> = 4 \Omega^2_d / \Gamma^2 $. Hence, by measuring the steady state population and the coupling rate, we can place an upper bound on gravitational wave amplitude
\begin{equation}
    h_0 =  \sqrt{\left< P_p \right>} \sqrt{\frac{\hbar \pi^3}{2 \rho \omega^3 L^3}}  \frac{\Gamma n^2}{\mu }.
   \label{eq:gw_main_0}
\end{equation}
For the device parameters described in Supplementary Information, Section~\ref{app:setup}, and $P_p^{404,\text{max}}$ we conclude that there are no resonant gravitational waves with amplitude larger than $h_0 = 5.5\times 10^{-18}$. Likewise, the inferred population $P_p^{\text{404,inferred}}$ yields an amplitude of $h_0 = 2.9 \times 10^{-18}$, while the other mode numbers give bounds with the same order of magnitude. This marks the first direct experimental search for gravitational waves in GHz frequency range \cite{Aggarwal2025}. Future generations of devices, outlined in Supplementary Information, Section \ref{app:improvements}, are expected to enable significantly higher sensitivities of up to $h_0\approx 8\times10^{-22}$ over a frequency band of several GHz.

\subsubsection*{Dark Matter} 
Similarly, our measurements allow us to calculate a bound on potential dark matter interactions \cite{Trickle2025}. Possible dark matter candidates with GHz-scale masses include dark photons - vector bosons that appear as oscillating electric fields, ultralight spin-0 scalars whose coherent oscillations modulate particle masses, and pseudoscalar axions that likewise form a classical microwave-frequency background \cite{Bozorgnia2024, carney2021}.

A monochromatic dark photon field at the HBAR mode frequency interacts with the electromagnetic field through a coupling parameter $\kappa$. This then leads to an oscillating electric field that resonantly drives the mode through piezoelectricity, with drive strength $\Omega_{DP}=-4\kappa e_{33}\frac{\mu}{\epsilon_r n}\sqrt{\frac{\rho_V\omega L}{\epsilon_0\hbar\pi c_{33}}}$. Here, $\rho_V=\SI{0.4}{GeV/cm^3}$ is the dark matter mass density, $\epsilon_0$ is the vacuum permittivity, $\epsilon_r = 10$ is the relative permittivity of AlN \cite{martin2004}, and $c_{33} = \SI{500}{GPa}$ is the stiffness tensor component of sapphire. The value of the piezoelectric tensor component of AlN, $e_{33}$, varies considerably in the literature, so we consider the range \SI{0.4}{\coulomb\per\square\meter}~\cite{Chu2017} to \SI{2.0}{\coulomb\per\square\meter}~\cite{martin2004}. 

As for the case of GWs, this then allows us to place an upper bound on the dark photon coupling parameter using the measured phonon mode population
\begin{equation}
    \kappa = \sqrt{\left< P_p \right>} \sqrt{\frac{\epsilon_0\hbar\pi c_{33}}{\rho_V \omega L}}  \frac{\Gamma \epsilon_r n}{8\mu e_{33}}.
\label{eq:dm_main}
\end{equation}
Using the measured value $P_p^{404,\text{max}}$, we rule out coupling parameters larger than $\kappa = 4.4\times 10^{-9}$ ($8.8\times 10^{-10}$) for the lower (higher) value of $e_{33}$, while using the inferred value $P_p^{\text{404,inferred}}$ gives $\kappa = 2.3\times 10^{-9}$ ($4.7\times 10^{-10}$). These values are currently not competitive with bounds set by cosmological and haloscope measurements~\cite{AxionLimits}. However, we note that while there are a large number of haloscope experiments around the mode frequency of $\sim 5$ GHz used in this work, they are more sparse in the 7 to 10 GHz range. Accessing HBAR modes in this range would only require modifying the qubit frequency and the thickness of the piezoelectric layer. Furthermore, improving our current bounds by one or two orders of magnitude would already reach the bounds set by cosmology in this frequency range. This can be achieved through longer integration times and improved device parameters, as discussed in the outlook section below and Section~\ref{app:improvements} of the Supplementary Information.


\section*{Outlook}
In this work, we have characterized the populations of HBAR phonon modes within a cQAD system. We observed remarkably low excited-state populations compared to other systems operating in the GHz regime. Using these results, we placed bounds on various predicted physical phenomena, such as high-frequency gravitational waves and dark matter. We further note that, unlike interferometric detectors such as LIGO, our device measures energy. As such, it is not subject to the usual standard quantum limit for quadrature measurements. It is analogous to a photon detector, except it is sensitive to any incoming signal that excites phonons in the HBAR.

There are many avenues for future improvements of our device. First, we note that our bounds were obtained with a one-week integration time, and using the results from a single mode. Fig.~\ref{fig:pop_vs_time}(a) indicates that longer integration times will continue to improve the statistical uncertainty. The measurement time can also be used more efficiently and the bandwidth effectively increased by measuring other modes during the wait period in our protocol, which currently takes up most of the time. Second, the frequency range of our sensor can be expanded to 3-10 GHz by incorporating flux-tunable transmons and down to MHz frequencies by using a fluxonium qubit instead \cite{lee2023, najera2024}. Third, HBAR resonators with much higher quality factors have been demonstrated \cite{galliou2013, Garcia-Belles2025}, especially at lower frequencies. Together with increasing the beam waist and the favorable scaling with decreasing mode number in Equations \ref{eq:gw_main_0} and \ref{eq:dm_main}, we expect significant gains in sensitivity. Finally, well-established techniques for quantum control of HBAR modes allow us to explore using non-classical mechanical states, such as squeezed \cite{Marti2024} or cat \cite{Bild2023} states, for enhanced sensitivity.

The fundamental physics implications that we have discussed here are just a few examples of how our results can be applied. Additionally, in the Supplementary Information, Section \ref{app:collapse}, we calculate parameter bounds for spontaneous collapse model theories based on our results \cite{GhirardiPRD86, Schrinski2023}, reaching sensitivities comparable with ultra-cold cantilever experiments \cite{VinantePRL20}. We also discuss interesting prospects for improved $\hbar$BAR devices. Furthermore, low populations of mechanical modes allow us to calculate bounds on other physical effects such as generalized uncertainty principles \cite{Bosso2023}, spacetime fluctuations \cite{Porcino95,DonadiPRD25}, topological defects \cite{Roberts2017}, or holographic noise models \cite{Hogan2010}. More generally, the ability to perform highly precise measurements on mechanical modes also positions $\hbar$BAR devices as promising tools for measuring weak gravitational forces, including precision tests of the gravitational constant \cite{Bose2024}, tests of Newton's law at short distances \cite{Clifton2012}, direct graviton detection \cite{Tobar2024}, probing gravitational interactions between quantum systems \cite{Bose2017,Marletto2017}, and investigating gravitationally induced decoherence mechanisms \cite{Bassi2017}. Ultimately, cQAD platforms hold the potential to address fundamental, unresolved questions about the interplay between gravity and quantum mechanics \cite{ChapelHill2011}.

\section*{Data availability}
All data are available from the corresponding authors upon reasonable request.

\section*{Acknowledgments}
We thank Yu Yang, Igor Kladari\'{c}, Raquel Garc\'{\i}a Bell\'{e}s, and Otto Schmid for useful discussions and Ryan Linehan, Tanner Trickle and Diego Blas for comments on the manuscript. Fabrication of the quantum devices was performed at ETH Zurich and at IBM Zurich. This work was funded by the QuantERA II Program that has received funding from the European Union’s Horizon 2020 research and innovation program under grant agreement no 101017733, and with the Swiss National Science Foundation.
M.F. was supported by the Swiss National Science Foundation Ambizione Grant No. 208886, and by The Branco Weiss Fellowship -- Society in Science, administered by the ETH Z\"{u}rich.

\section*{Author Contributions}
A.O., S.S. and M.B. planned and performed the experiment, analyzed the data and derived the results. S.S. and A.O. performed numerical simulations of the experiment. A.O., M.F., D.S., M.B. and Y.C. derived the theoretical models. S.S. created the figures and S.S., A.O., Y.C. and M.F. wrote the manuscript with input from all authors. Y.C. supervised the project.

\onecolumngrid
\ \\ \newpage
\section*{End Matter}
\twocolumngrid

\paragraph*{Implications for Quantum Information Processing} -- 
\label{app:comparison}

Proper initialization of a quantum system in its ground state is one of the core prerequisites for quantum computing and simulations, long-lived quantum memories, or quantum sensors. In order to reach the required regime where $k_B T \ll \hbar\omega$, systems operating in the MHz to GHz regime are typically cooled passively using dilution refrigerators. Additionally, active cooling schemes can be used to overcome limitations imposed by electromagnetic and thermal noise.

\begin{figure}
  \begin{center} \includegraphics[width=0.9\columnwidth]{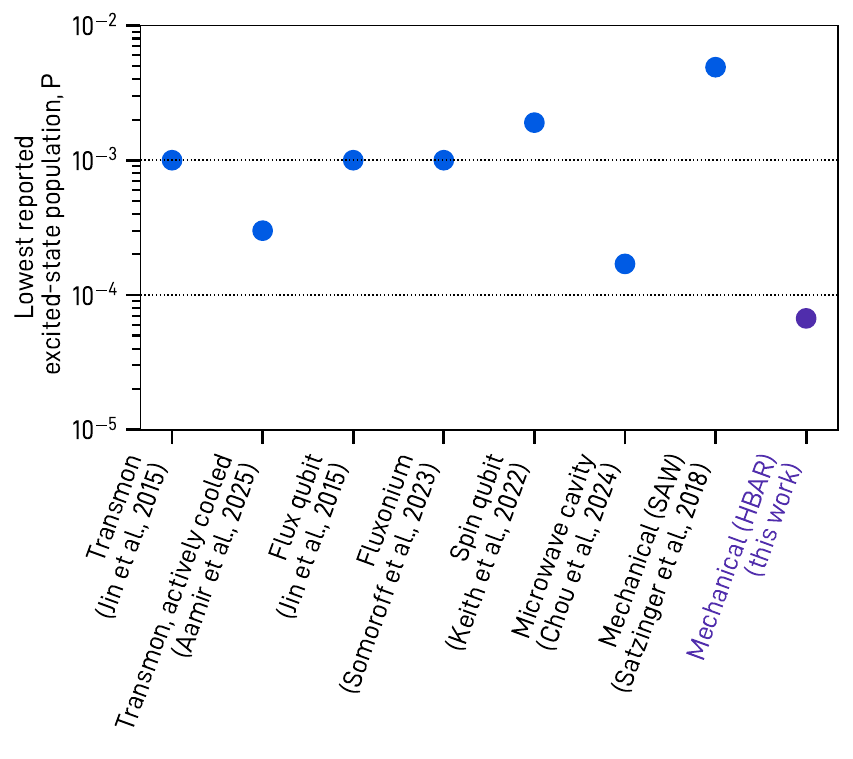}
  \caption{\textbf{Lowest reported excited-state populations of quantum systems operating in the MHz to GHz range.} For each type of system, we list the lowest value, or upper bound, reported in literature to date: Transmon without \cite{Jin2015} and with active cooling \cite{Aamir2025}; Flux \cite{Jin2015}, Fluxonium \cite{Somoroff2023} and spin qubits \cite{Keith2022}; microwave cavities \cite{Chou2024}; and the coldest mechanical mode reported so far, a surface acoustic wave resonator \cite{Satzinger2018}, together with the current work based on an $\hbar$BAR.}
  \label{fig:cold_comparison}
  \end{center}
\end{figure}

In Fig.~\ref{fig:cold_comparison}, we compare the lowest reported excited-state populations for quantum systems operating in the MHz and GHz regime. To the best of our knowledge, the excited-state population for HBARs presented in this manuscript is the lowest currently reported.

This positions $\hbar$BARs as not only a quantum resource with high-fidelity state initialization, but also a promising platform for active cooling of other quantum systems, such as transmons. In fact, previous cQAD implementations routinely utilized acoustic modes of coupled HBARs for transmon qubit cooling \cite{Bild2023, Yang2024, Marti2024}.

Using a sequence of iSWAP gates (see Fig.~\ref{fig:protocol}c in the main text, gate C), we can suppress the qubit's excited-state population to a value more than three times lower compared to the lowest values reported to date \cite{Aamir2025}. 

Our current device accesses $\mathcal{O}(10)$ acoustic modes with the AC-stark tunable qubit, with next-generation devices expected to reach $\mathcal{O}(100)$ modes using flux-tunable qubits. As suggested by simulations, two iSWAP gates ($\sim$850 ns each) would be sufficient to cool the qubit substantially below 0.1$\%$ in only about 1.8$\mu$s, comparable to other active cooling approaches \cite{Salathe2018, Riste2012, Geerlings2013}. While faster active cooling protocols exist for superconducting circuits \cite{Reed2010, Magnard2018, Egger2018, Zhou2021, Ding2025}, they typically exhibit higher excited-state populations. 
While our protocol requires the hardware overhead of a HBAR, it comes with the advantage of requiring no external feedback, additional control lines, flux pulses, nor strong drives. This makes it attractive for larger devices where wiring density is a bottleneck.

\paragraph*{Description of the iSWAP gate} -- 
\label{app:comparison}
Broadly speaking, an iSWAP gate exchanges the $\ket{10}$ and $\ket{01}$ states and multiplies each exchanged amplitude by a 90-degree phase factor $i$, while leaving the $\ket{00}$ and $\ket{11}$ states unaffected. This corresponds to a coherent swap of a single excitation between two quantum systems combined with a phase rotation.
More precisely, the iSWAP unitary is written in the computational basis ${\ket{00},\ket{01},\ket{10},\ket{11}}$ as

\begin{equation}
\mathrm{iSWAP} =
\begin{pmatrix}
1 & 0 & 0 & 0 \\
0 & 0 & i & 0 \\
0 & i & 0 & 0 \\
0 & 0 & 0 & 1
\end{pmatrix}.
\end{equation}

\appendix
\onecolumngrid
\ \\ \newpage
\section*{Supplementary Information}
\twocolumngrid

\section{Experimental Setup}\label{app:setup}

\begin{figure}[ht!]
  \begin{center} \includegraphics[width=0.76\columnwidth]{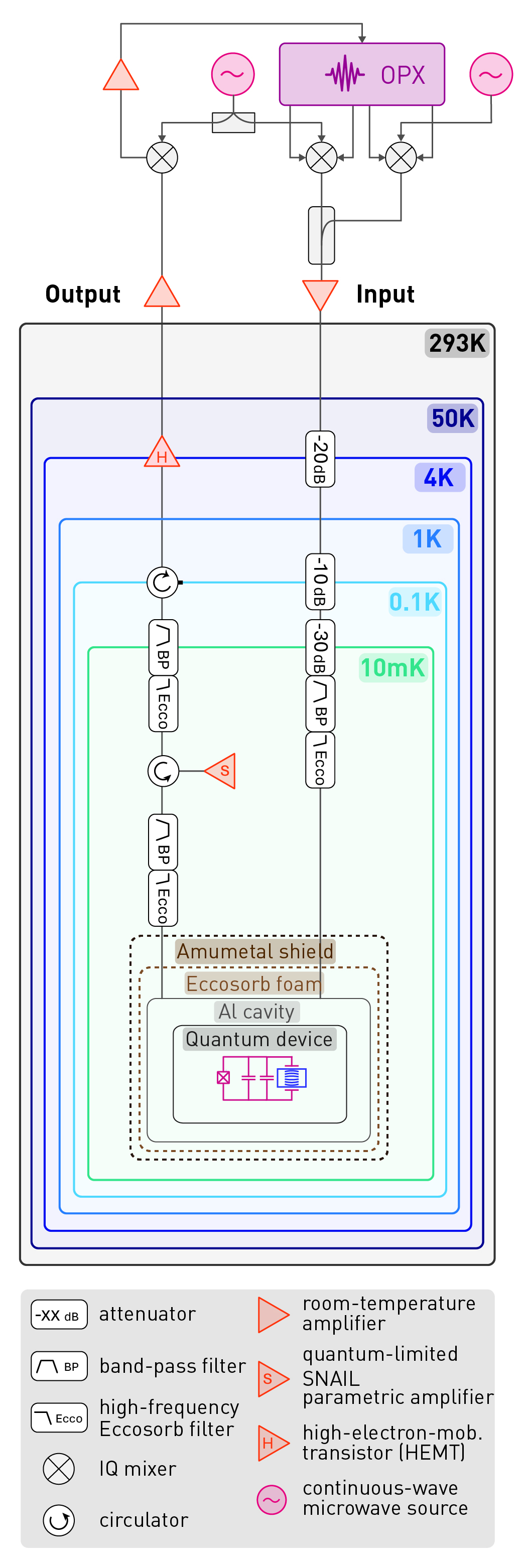}
  \caption{\textbf{Simplified schematic of the experimental setup.} The quantum device is operated at the base temperature stage of a dilution refrigerator, and controlled and read out using microwave electronics, see text for details.}
  \label{fig:cabling_diagram}
  \end{center}
\end{figure}

Our sample (Fig.~\ref{fig:protocol}(a)) is operated at a temperature of approximately \SI{10}{mK} in a dilution refrigerator. Control and readout of the superconducting qubit is realized using an aluminum 3D cavity that houses the quantum device (Fig.~\ref{fig:cabling_diagram}). The packaged sample is shielded from external electromagnetic radiation by high-frequency absorbing Eccosorb foam and a magnetic Amumetal shield. At room temperature, qubit rotation and readout pulses are generated using a Quantum Machines OPX arbitrary waveform generator. The pulses are subsequently up-converted to GHz frequencies via continuous-wave microwave sources and IQ mixers, combined to a single control line using a directional coupler, and pre-amplified before entering the cryogenic system. Inside the dilution refrigerator, signals undergo attenuation and filtering before reaching the quantum device. Upon leaving the sample, the readout response is filtered and amplified using a quantum-limited superconducting nonlinear asymmetric inductive element parametric amplifier (SNAIL parametric amplifier), a high-electron-mobility transistor (HEMT), and additional room-temperature amplifiers. Finally, after down-conversion to \SI{250}{kHz}, the readout signal is digitized and recorded using an FPGA integrated in the OPX controller. Typical parameters of our cQAD quantum device are summarized in Table~\ref{tab:hbar_qubit_params}. These values are used for the fundamental physics calculations presented in Supplementary Information, Sections \ref{app:force_quant}, \ref{sec:dark_mat} and \ref{app:collapse}.

\begin{table}
\centering
\renewcommand{\arraystretch}{1.2}   
\begin{tabularx}{\linewidth}{|>{\raggedright\arraybackslash}X|c|c|}
\hline
\textbf{Parameter} & \textbf{Symbol} & \textbf{Value} \\ \hline
\multicolumn{3}{|c|}{\textbf{HBAR properties}} \\ \hline
Phonon mode frequency          & $\omega_p / 2\pi$              & \SI{5048.630\pm0.001}{\mega\hertz}      \\ \hline
Phonon mode FSR                     & $\omega_{\mathrm{FSR}} / 2\pi$ & \SI{12.5\pm0.2}{\mega\hertz}       \\ \hline
Phonon mode number             & $n$                            &\SI{404\pm13}{} \\ \hline
Phonon mode 404 lifetime                & $T_{1,p}^{404}$                       & \SI{112\pm9}{\micro\second}         \\ \hline
Phonon mode 404 coherence time          & $T_{2,p}^{404}$                       & \SI{200\pm25}{\micro\second}         \\ \hline
Phonon quality factor          & $Q_{p}^{404}$                       & 3.6$\times10^6$         \\ \hline
Dielectric constant (sapphire) & $\epsilon_{r}$ & 10            \\ \hline
Stiffness $c-$axis (sapphire)  & $c_{33}$                       & \SI{500}{\giga\pascal}       \\ \hline
Piezoelectric coefficient $c-$axis (AlN)  & $e_{33}$                       & \SI{0.4}{\coulomb\per\square\meter} to \SI{2.0}{\coulomb\per\square\meter}   \\ \hline
sapphire density                   & $\rho$                            & \SI{3980}{\kilo\gram\per\cubic\meter}       \\ \hline    
HBAR height                    & $L$                            & \SI{435}{\micro\meter}       \\ \hline
Acoustic mode waist \newline (1/e radius)                     & $\mu$                            & \SI{27}{\micro\meter}          \\ \hline
Effective mass                 & $m_{\mathrm{eff}}$             & \SI{1}{\micro\gram}           \\ \hline
Number of atoms                & $N_a$                          & $5\times10^{17}$ \\ \hline
Effective delocalization       & $x_{\text{del}}$                            & $1\times10^{-18}$~m \\ \hline
\multicolumn{3}{|c|}{\textbf{Qubit properties}} \\ \hline
Qubit frequency                & $\omega_q / 2\pi$              & \SI{5068.81\pm0.02}{\mega\hertz}       \\ \hline
Qubit anharmonicity            & $\alpha / 2\pi$                & \SI{-185.12\pm0.02}{\mega\hertz}       \\ \hline
Qubit ge lifetime                 & $T_{{1,ge}}$                       & \SI{28\pm3}{\micro\second}        \\ \hline
Qubit ge dephasing time           & $T_{\phi}$                       & \SI{20\pm6}{\micro\second}          \\ \hline
Qubit ef lifetime           & $T_{{1,ef}}$                       & \SI{20\pm4}{\micro\second}          \\ \hline
Qubit - phonon coupling          & $g / 2\pi$                     & \SI{280}{\kilo\hertz}        \\ \hline
Initial qubit temperature, after cooling     & $T_{{qb,0}}$            & \SI{30}{\milli\kelvin} \\ \hline
Qubit bath temperature    & $T_{{qb,bath}}$            & \SI{40}{\milli\kelvin} \\ \hline
\end{tabularx}
\caption{\textbf{Parameters of the quantum device employed in the presented experiments.}
For $\omega_p$, $\omega_{\mathrm{FSR}}$, and $n$ we are using the parameters for mode $n=404$, with typical values for $T_{1,p}$ and $T_{2,p}$. $Q_{p}^{404}$ is calculated therefrom. $\varepsilon_{\mathrm{r}}$ is an approximation based on Ref.~\cite{Floch2008} and references therein, $c_{33}$ is based on Refs.~\cite{Hovis2006,Vodenitcharova2007}, $e_{33}$ represents the range of values listed in Refs.~\cite{Chu2017,martin2004}, and $\rho$ is taken from Ref.~\cite{roditi_2025}. The listed qubit frequency $\omega_q$ corresponds to the value without Stark-shift drive.
$L$, $\mu$, $m_{\mathrm{eff}}$, $N_a$ and $x_{\text{del}}$ are estimates calculated in Ref.~\cite{Bild2023}. $T_{{qb,0}}$, $T_{{1,ef}}$ and $T_{\text{qb,bath}}$ represent typical values, and $T_{{1,ge}}$ and $T_{\phi}$ are the average values of the main data set presented in Figure~3 in the main text. The values listed in this table serve as the basis for the calculations presented in Supplementary Information, Sections \ref{app:force_quant}, \ref{sec:dark_mat} and \ref{app:collapse}.}
\label{tab:hbar_qubit_params}
\end{table}

\section{Limitations on the Measured Populations}
\label{app:statistics}

\begin{figure}
  \begin{center} \includegraphics[width=0.9\columnwidth]{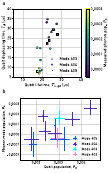}
  \caption{\textbf{Correlations between measured phonon mode populations and qubit properties.} \textbf{a}, Correlation between measured phonon mode population and qubit lifetime and dephasing time. Each point corresponds to an individual phonon mode population measurement, color coded with the mean of the extracted population, and placed according to the measured qubit lifetime and dephasing time. \textbf{b}, Correlation between the measured phonon mode population and the measured qubit population. Points denote the mean of a specific measurement, lines correspond to the standard deviation.}
  \label{fig:simulation_data}
  \end{center}
\end{figure}

The extracted phonon mode populations are limited by errors induced by the experimental protocol. We find that two factors dominate: the thermalisation of the qubit to its thermal bath, characterized by the qubit lifetime $T_1$ and the effective temperature of the bath $T_{{qb, bath}}$, and the qubit dephasing time $T_{\phi}$, which predominantly affects the fidelity of the iSWAP operation. We experimentally observe that a higher qubit lifetime $T_1$ and coherence time $T_2$ is correlated with lower measured phonon mode populations, see Fig.~\ref{fig:simulation_data}(a). At the same time, higher qubit populations tend to limit the phonon mode population, as shown in Fig.~\ref{fig:simulation_data}(b). These results are consistent with the two limiting factors above.

To study these effects quantitatively, we perform master equation simulations to model the experimental protocol described in Fig.~\ref{fig:protocol} in the main text, using the parameters in Table~\ref{tab:hbar_qubit_params}. Specifically, we simulate a Jaynes-Cummings interaction between a three-level superconducting qubit coupled to a single-mode mechanical resonator, both initialized in their respective thermal states at a given temperature. We explicitly model the pulse sequence using time-dependent drives, waiting periods, and a measurement window. Dissipative processes are included through Lindblad collapse operators. 

\begin{figure*}
  \begin{center} \includegraphics[width=2.\columnwidth]{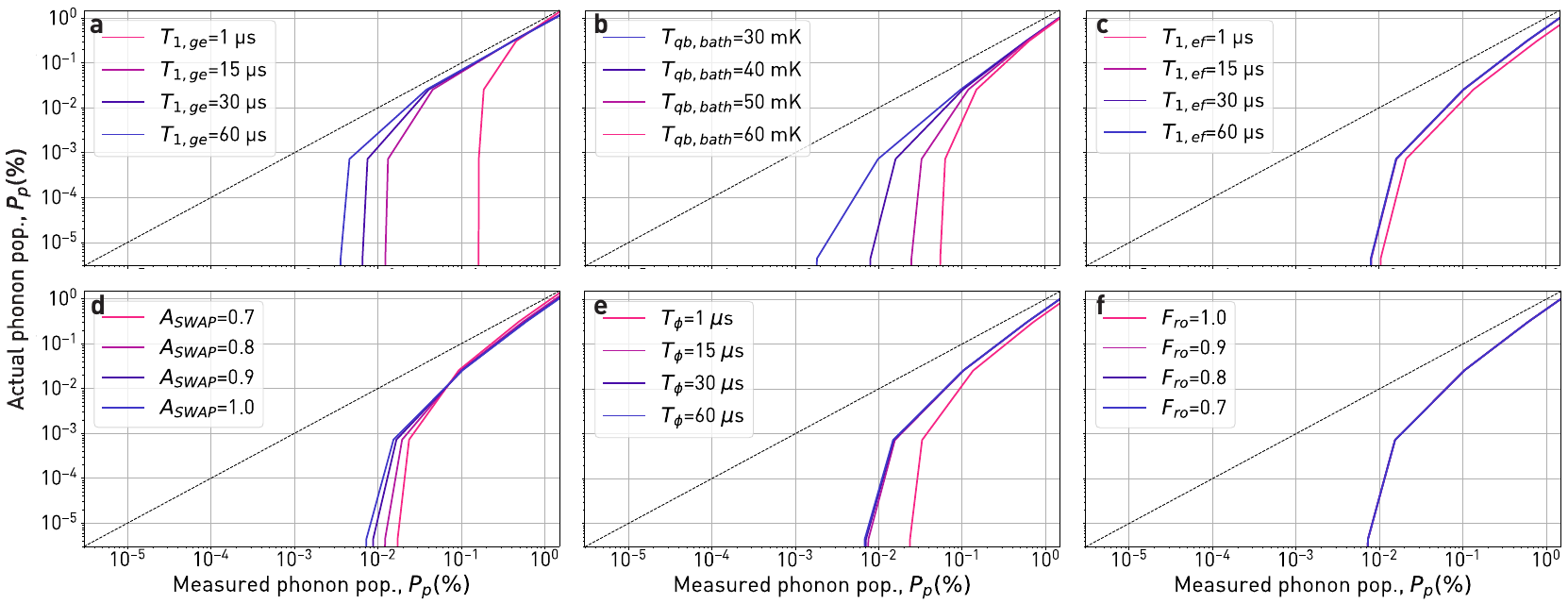}
  \caption{\textbf{Parameter sweeps in numerical simulations.} \textbf{a,} Sweeping the qubit ge relaxation time $T_{\text{1,ge}}$. \textbf{b,} Sweep of the qubit bath temperature. \textbf{c,} Sweep of the qubit ef relaxation time $T_{\text{1,ef}}$. \textbf{d,} Sweeping the relative amount of energy that gets swapped between qubit and phonon during the iSWAP operation, $A_{\text{iSWAP}}$, regardless of other experimental conditions. This quantifies pulse calibration errors, where $A_{\text{iSWAP}}$=1.0 corresponds to a perfectly calibrated iSWAP operation, without taking into account other device properties such as coherences. \textbf{e,} Sweep of the qubit ge dephasing time, $T_{\phi}$. \textbf{f,} Sweeping the readout fidelity $F_{{ro}}$.}
  \label{fig:simulations_all}
  \end{center}
\end{figure*}

This framework enables parameter sweeps that provide insights into limitations of the presented scheme. As can be seen in Fig.~\ref{fig:simulations_all}, the simulations suggest that the measured phonon population is predominantly limited by the thermalization of the qubit to its bath during the protocol, characterized by the qubit relaxation time $T_{\text{1,ge}}$ (Fig.~\ref{fig:simulations_all}(a)) and the temperature of the qubit bath $T_{\text{qb, bath}}$ (Fig.~\ref{fig:simulations_all}(b)), and by the qubit dephasing ($T_{{\phi}}$, panel e). For a transmon in a 3D microwave cavity, reducing the qubit's effective bath temperature is primarily achieved by suppressing external heating mechanisms. The dominant mechanisms are stray infrared radiation, residual thermal photons, and quasiparticles. These can be mitigated through optimized electromagnetic shielding and improved filtering on all microwave lines. In addition, ensuring robust thermal anchoring of both the cavity and the quantum device to the base plate of the dilution refrigerator helps bring the system closer to the physical temperature of the refrigerator stage.

Other factors such as ef-level decay (panel c) or phonon-qubit iSWAP infidelities (panel d) only increase the measured phonon population marginally compared to the inferred value. Given the measured e-level population of the qubit and assuming a thermal distribution, we expect the f-level population to be negligible. Because the protocol always takes the ratio of the signal and its reference, any noise or systematic error that affects both quantities equally is canceled out in the extracted mean population. This particularly holds for readout infidelities, to which the extracted mean population is insensitive (Fig.~\ref{fig:simulations_all}(f)), as long as the readout fidelity remains the same for both signal and reference measurements. However, improving readout fidelity still helps to increase the signal-to-noise ratio of the measurement process, and thereby accelerates the reduction of measurement variance with averaging.

\begin{figure*}
  \begin{center} \includegraphics[width=1.8\columnwidth]{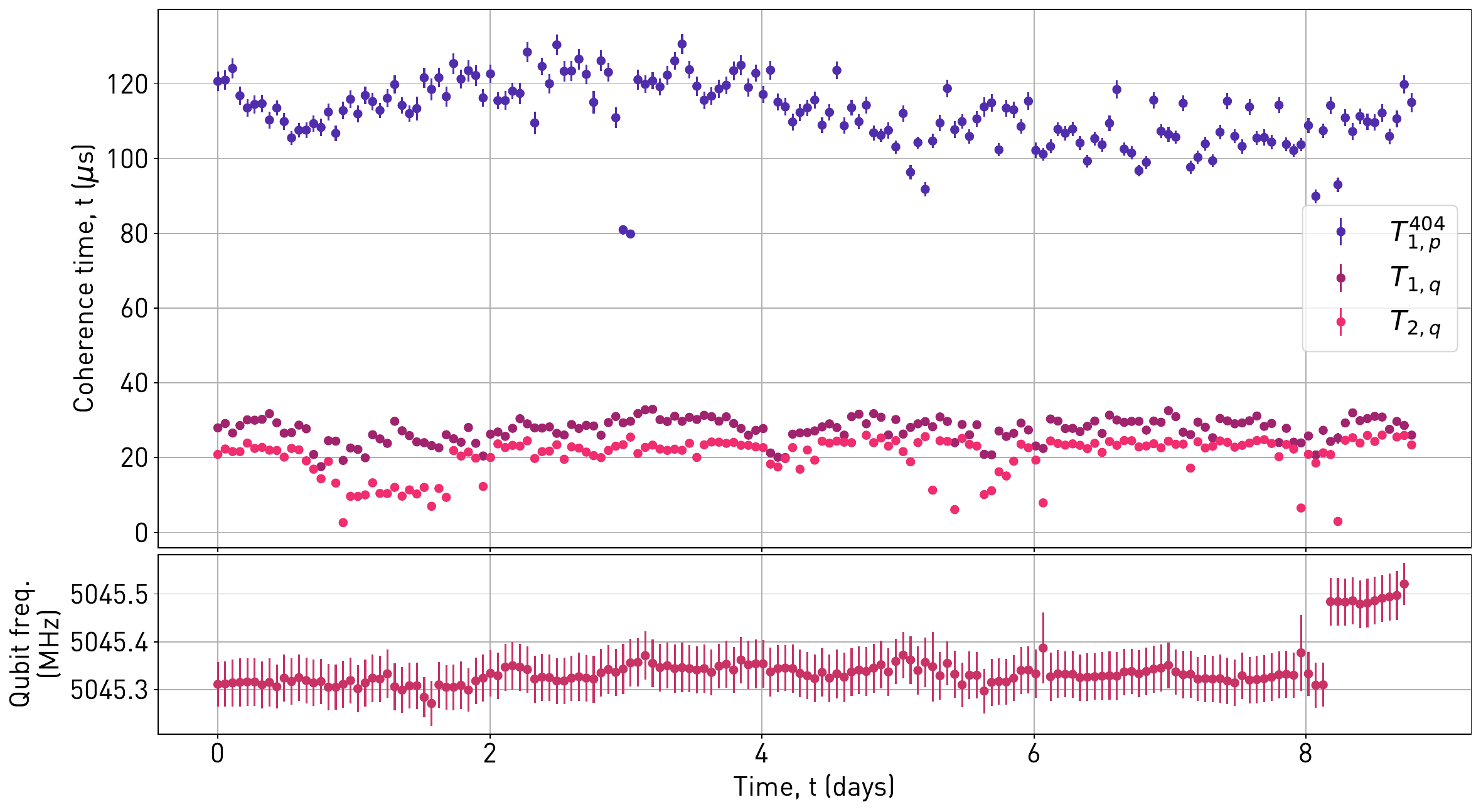}
  \caption{\textbf{Device parameter stability for the main dataset with a total experimental duration of 9 days.} Each point corresponds to a characterization measurement in between two data acquisition blocks. We display the lifetime of phonon mode 404 and the qubit, as well as the coherence time and the frequency of the qubit.}
  \label{fig:1week_stability}
  \end{center}
\end{figure*}

Moreover, we notice that the stability of the lifetime and coherence time of phonon and qubit, as well as the qubit frequency, are comparable to what is typically observed in the literature over the course of more than a week (corresponding to the main dataset presented in the main text), see Figure~\ref{fig:1week_stability}. 

\section{Sensitivity to High-Frequency Gravitational Waves} \label{app:force_quant}

In the following Section, we derive a bound on the amplitude of a gravitational wave impinging on our device, inferred from the occupation number of a resonant acoustic mode.
We start with a description of the acoustic eigenmodes in a $\hbar$BAR device, followed by the derivation of a quantized Hamiltonian describing the interaction of such a mode with a gravitational wave. In the third part, we find an expression for the number of excitations in an acoustic mode, excited by a resonant gravitational wave with a specific amplitude.

\subsection{Phonon Modes in $\hbar$BARs}

The displacement of atoms $\Vec{u}$ from their equilibrium position in a crystal is described by the Christoffel equation
\begin{equation}
    \rho \frac{\partial^2 u_i}{\partial t^2} - c_{ijlm}\frac{\partial^2 u_l}{\partial x_j \partial x_m} = 0 , 
    \label{eq:christoffel}
\end{equation}
where $\rho$ is the density and $c_{ijlm}$ the stiffness tensor of the crystal material (sapphire). We have assumed Einstein notation with indices $i,j,l,m \in \{x,y,z\}$ denoting vectors and tensors. 

The solutions of this wave equation are the eigenmodes of the acoustic resonator $ \Vec{u}_\lambda(\Vec{x}, t)$. We can decompose them into time and position-dependent contributions
\begin{equation}
    \Vec{u}_\lambda(\Vec{x}, t) = \Vec{f}_\lambda(\Vec{x}) \widetilde{b}_\lambda(t),  
    \label{eq:normal_modes_f}
\end{equation}
where the time-dependent part $\widetilde{b}_\lambda(t) = A_{\lambda} \exp\left( -\mathrm{i} \omega_\lambda t \right)  $ describes oscillations with the eigenmode frequency $\omega_{\lambda}$ and the oscillation amplitude $A_{\lambda}$. 
We can describe the spatial profile of a mode in the center of mass reference frame with a Laguerre-Gaussian beam propagating in the z-direction,
\begin{equation}
    \Vec{f}_{\lambda} \left(r,\phi,z \right) = \sqrt{\frac{2}{L \mu^2 } }  \cdot \text{LG}_{pl} \left(r, \phi \right) \sin \left( \frac{n \pi z}{L}   \right)   \hat{\mathbf{z}}.
    \label{eq:appendix-mode-shape}
\end{equation}
Here, $n$ denotes odd longitudinal mode numbers (see Ref.~\cite{Bild2023}), while for even mode numbers we replace the sine-term in equation~(\ref{eq:appendix-mode-shape}) with its cosine counterpart. $L$ represents the length of the resonator in z-direction, and $\mu$ the beam waist of the Laguerre-Gaussian mode. The piezoelectric material in our device couples the electric field of the qubit only to longitudinal sound waves, and therefore we only consider displacements in the $\hat{\mathbf{z}}$  direction. The transverse profile of the beam is described by the Laguerre-Gaussian mode $\text{LG}_{pl}$ with radial index $p$ and azimuthal index $l$

\begin{align}
    \text{LG}_{pl}(r, \phi) 
    &= \sqrt{\frac{2p!}{\pi (p + |l|)!}} \left( \frac{r \sqrt{2}}{\mu} \right)^{|l|} 
       \exp \left( - \left( \frac{r}{\mu} \right)^2 \right) \notag \\
    &\quad \cdot L_p^{|l|} \left( \frac{2 r^2}{\mu^2} \right) 
       \exp \left( - \mathrm{i} l \phi \right) \,,
\end{align}
where $L_p^{|l|}$ denotes generalized Laguerre polynomial. 
In our device, we only strongly couple to the fundamental mode with $p=0$, $l=0$. The prefactors $\sqrt{\frac{2}{L \mu^2 } }$ in equation~(\ref{eq:appendix-mode-shape}) represent a normalization, such that 
$
    \int_V dV \Vec{f}_\lambda\cdot\Vec{f}_{\lambda^\prime} = \delta_{\lambda\lambda^\prime}\, 
$.

\subsection{Coupling to a Gravitational Wave}

In this section we follow the approach described in \cite{Maggiore2007, goryachev_gravitational_2014}. A gravitational wave $h_{\alpha \beta}$ can be described as a small perturbation to the flat Minkowski metric $g_{\alpha \beta} = \eta_{\alpha \beta} + h_{\alpha \beta}$, and can therefore be written as $h_{ij}  = h_0 \epsilon_{ij} \cdot \cos \left( \omega \left( t - x/c \right) \right)$, with the wave amplitude $h_0$, the polarization tensor $\epsilon_{ij}$, and its frequency $\omega$. We use Einstein notation, where indices $ij$ stand for spatial coordinates: $x=1$, $y=2$, $z=3$. 

When a gravitational wave interacts with an acoustic resonator, its action can be described by an external Newtonian force density $F_i = \Ddot{h}_{ij} \left( t \right) x_j \rho / 2 $ \cite{Maggiore2007}, which depends on the position $x_j$ of the atoms inside the resonator. Here, we have assumed that the wavelength of gravitational wave is much larger than the detector, such that the effect of gravitational wave is uniform within our device. This leads to the total equation of motion, 
\begin{equation}
    \rho \frac{\partial^2 u_i}{\partial t^2} - c_{ijlm}\frac{\partial^2 u_l}{\partial x_j \partial x_m} =  \frac{\rho}{2}\Ddot{h}_{ij}x_j\, .
    \label{eq:christoffel-GW}
\end{equation}
The general solution of equation~(\ref{eq:christoffel-GW}) is a linear combination of eigenmodes of the acoustic resonator, $  \Vec{u} = \sum_\lambda c_\lambda (t) \Vec{u}_{\lambda} = \sum_\lambda b_\lambda (t) \Vec{f}_{\lambda} $, where we introduced the new label $b_\lambda (t) = c_\lambda (t)  A_{\lambda} \exp\left( -\mathrm{i} \omega_\lambda t \right) $. Inserting this ansatz in equation~(\ref{eq:christoffel-GW}), multiplying both sides with $\Vec{f}_{\lambda^\prime}$ and integrating over the full mode volume of the resonator gives the differential equation  
\begin{equation}
    \Ddot{b}_\lambda + \gamma \Dot{b}_\lambda + \omega^2_\lambda b_\lambda = \frac{1}{2} \Ddot{h}_{ij} \int_V dV f_{\lambda i} x_j,
    \label{eq:driven-harmonic-oscillator}
\end{equation}
where we have introduced the term $\gamma \Dot{b_\lambda}$ to account for dissipation of the mechanical mode. Equation~(\ref{eq:driven-harmonic-oscillator}) describes the acoustic mode as a damped harmonic oscillator, driven by a gravitational wave. 
We call the integral on the right hand side of equation~(\ref{eq:driven-harmonic-oscillator}) the coupling to the gravitational wave $\xi_{ij}$. It is proportional to the overlap of the force exerted by the gravitational wave and the mode shape given in equation~(\ref{eq:appendix-mode-shape}). Evaluating the integral gives  
\begin{equation}
    \xi_{ij} \coloneqq \int_V dV f_{\lambda i} x_j = \frac{4 L^{\frac{3}{2}} \mu}{\pi^{\frac{3}{2}} n^2} \delta_{i3} \delta_{j3}. 
    \label{eq:GW-coupling}
\end{equation}
The coupling $\xi_{ij}$ is nonzero only if the gravitational wave induces vibrations in the resonator. Specifically, the coupling is nonzero only in the direction of the main detection axis, as expressed by the Kronecker delta symbols in equation~(\ref{eq:GW-coupling}). Moreover, the integral is zero for all even mode numbers $n$. For this reason, we assume odd mode numbers from now on. 
The final equation of motion is then
\begin{equation}
    \Ddot{b}_\lambda + \gamma \Dot{b}_\lambda + \omega^2_\lambda b_\lambda = \frac{1}{2} \Ddot{h}_{33} \xi_{33}
    \label{eq:driven-harmonic-oscillator-2}
\end{equation}

We proceed by quantizing $b_\lambda$ as it is done for the position of a harmonic oscillator, i.e. $\hat{b}_\lambda = \sqrt{\frac{\hbar}{2 \rho \omega_\lambda}} \left( \hat{a} + \hat{a}^\dagger \right)$. 
With this, we arrive at the Hamiltonian 
\begin{equation}
    H = \hbar \omega_{\lambda}  \left(\hat{a}^\dagger\hat{a} + \frac{1}{2}\right)
    - \frac{1}{2}  \sqrt{ \frac{\hbar \rho}{2 \omega_{\lambda} }  }  \Ddot{h}_{33}  \xi_{33} \left( \hat{a}^\dagger + \hat{a} \right),
    \label{eq:Hamiltonian-with-GW-drive-general}
\end{equation}
describing the coupling between our acoustic mode and a gravitational wave.

\subsection{Steady-state Solution of a Driven Dissipative Oscillator}

We consider a gravitational wave resonant with the phonon mode $\lambda$, $h_{33}=h_0 \epsilon_{33} \cos \left( \omega t \right)$, where $\epsilon_{33}$ is the polarization tensor projected onto the z-direction and $\omega = \omega_{\lambda}$. 
The Hamiltonian in equation~(\ref{eq:Hamiltonian-with-GW-drive-general}) simplifies to
\begin{equation}
    \hat{H}/\hbar =  \omega  \left(\hat{a}^\dagger\hat{a} + \frac{1}{2}\right)
    + 2  \Omega_{GW} \cos \left( \omega t \right) \left( \hat{a}^\dagger + \hat{a} \right),
\end{equation}
with the driving strength  
\begin{align} 
    \Omega_{GW} &=
    - h_{0} \frac{\epsilon_{33} \mu }{n^2}   \sqrt{ \frac{ \rho \omega^3 L^3}{2 \hbar \pi^3 }  } 
    \label{eq:app-gw-driving-strength}
\end{align}
which corresponds to equation~(\ref{eq:hbar-hamiltonian}) in the main text, with $\Omega_d = \Omega_{GW}$. 
We proceed by moving into the interaction picture, which corresponds to the unitary transformation $U = \exp \left(\mathrm{i} \omega t \left(\hat{a}^\dagger\hat{a} + \frac{1}{2}\right) \right)$. In the rotating wave approximation, the Hamiltonian becomes
\begin{equation}
    \hat{H}/\hbar = 
   \Omega_d \left(\hat{a}^\dagger + \hat{a} \right).
    \label{eq:E_GW}
\end{equation}

We now want to link the amplitude of the gravitational wave with the excitation number of the acoustic resonator.
For this reason, we consider the Lindblad master equation, which describes the full evolution of the acoustic resonator coupled to an external bath at zero temperature 
\begin{equation}
        \dot{\rho} = - \frac{\mathrm{i}}{\hbar} \left[\hat{H}, \rho\right] 
        + \Gamma \left( \hat{a}\rho \hat{a}^\dagger - \frac{1}{2}\hat{a}^\dagger\hat{a}\rho
        - \frac{1}{2}\rho \hat{a}^\dagger\hat{a} \right)
        \label{eq:master-equation}.
\end{equation}
The master equation contains only one Lindblad collapse operator $L = \sqrt{\Gamma} \hat{a}$, which describes thermalization of the acoustic mode to an external bath at zero temperature with rate $\Gamma = 1/ T_{1,p}$. Therefore, any population of the mode in the steady-state is a result of a coherent drive, which, in this case, we assume is induced by the gravitational wave. To derive the occupation number in the steady-state one can either solve equation~(\ref{eq:master-equation}) with the condition $\dot{\rho}=0$, or directly consider the evolution of the occupation number
\begin{equation}
    \frac{\mathrm{d} \left< P_p \right>}{\mathrm{d}t} =  \frac{\mathrm{d} }{\mathrm{d}t}  \mathrm{Tr} \left( \hat{n} \rho \right) =   \mathrm{Tr} \left( \hat{n} \frac{\mathrm{d} \rho}{\mathrm{d}t}  \right) = 0 .
    \label{eq:occupation-steady-state-condition}
\end{equation}

The solution to equation~(\ref{eq:occupation-steady-state-condition}) is the occupation number in the steady state, 
\begin{equation}
    \left< P_p \right> = \frac{4 \Omega^2_d}{ \Gamma^2}.
\end{equation}
Using equation~(\ref{eq:app-gw-driving-strength}), we extract the bound on gravitational wave amplitude
\begin{equation}
   h_0 =  \sqrt{\left< P_p \right>} \sqrt{\frac{\hbar \pi^3}{2 \rho \omega^3 L^3}}  \frac{\Gamma n^2}{\mu \epsilon_{33}}.
   \label{eq:app-h0-bound}
\end{equation}

To determine the smallest measurable amplitude, we assume $\epsilon_{33} = 1$ (the maximal value for the polarization tensor projection), which leads to equation \ref{eq:gw_main_0} in the main text. 
This amplitude bound can be interpreted the following way: 
in the experiment, we measure the steady-state occupation number of an acoustic mode, which is a result of coupling to a thermal bath (e.g. dilution refrigerator at 10 mK) as well as coherent driving coming from external signals (e.g. gravitational waves). 
With the aim of placing the most conservative upper bound on the gravitational wave amplitude, we assume that the external bath is at zero temperature. We further assume that the population is induced solely by a coherent gravitational wave drive according to equation~(\ref{eq:app-h0-bound}). 
To tighten the bound on the amplitude, we need a low excited-state population $\left< P_p \right>$ and a small decoherence rate $\Gamma$. Furthermore, it is beneficial to design the detector such that the mode number $n$ is small while keeping frequency $\omega$ and length $L$ large. Finally, an effective detector should have a large beam waist $\mu$ and density $\rho$.

\begin{figure*}
  \begin{center} \includegraphics[width=2\columnwidth]{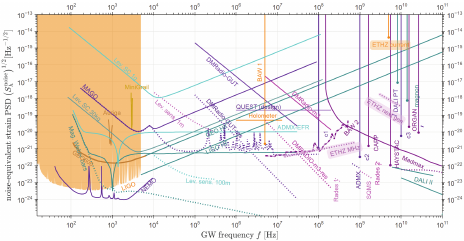}
  \caption{\textbf{Overview of characteristic strain sensitivities of existing and proposed experiments.} The y-axis shows the strain power spectral density (PSD), $\sqrt{S_h(\omega)}=h_c/\sqrt{\omega}$, a metric describing detector sensitivity across frequency. Solid lines represent broadband detectors, dashed lines resonant ones. Published results are marked in orange, experiments in active development in purple, and proposed setups in cyan. "ETHZ" indicates the result in this work and proposed future devices in Table~\ref{tab:detector-comparison}. This plot is based on the open database from Ref.~\cite{HFGWPlotter_Omega}. We refer to Ref.~\cite{Aggarwal2025} for details on the listed experiments.}
  \label{fig:gw_projection}
  \end{center}
\end{figure*}

\section{Future Device Improvements} \label{app:improvements}
In this Section, we discuss potential upgrades to our experimental setup aimed at enhancing its sensitivity to gravitational waves, dark matter signals, or collapse model parameters. 

For a next-generation device (see Table~\ref{tab:detector-comparison}), we anticipate coupling the HBAR to a flux-tunable qubit, thereby enabling access to hundreds of acoustic modes in the 3-10~GHz frequency range, constrained mainly by microwave electronics. Utilizing a larger mode set, particularly at lower frequencies, allows the selection of modes with higher quality factors. Preliminary measurements of test devices have demonstrated mode lifetimes on the order of 1~ms at 3~GHz. With additional improvements in materials and fabrication processes \cite{Garcia-Belles2025}, we expect to reduce the dominating surface scattering losses in HBARs and thereby achieve comparable quality factors even at higher frequencies. Furthermore, enhancements in material and fabrication quality are expected to extend qubit lifetimes. Collectively, these advancements, as detailed in Table~\ref{tab:detector-comparison}, promise substantial gains in the sensitivity to gravitational wave or dark matter signals, and collapse model parameters.

\begin{figure}
  \begin{center} \includegraphics[width=1\columnwidth]{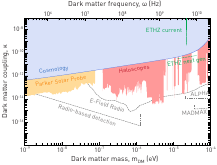}
  \caption{\textbf{Searches for dark photon dark matter.} Experimentally excluded regions are shown by solid areas: the blue region is excluded by cosmological observations \cite{PaolaArias_2012}, the yellow region by results from the Parker Solar Probe mission \cite{An2025}, and the red region by a variety of haloscope experiments, see e.g. Ref.~\cite{Semertzidis2022} for an overview. Dashed lines mark proposed or planned experiments - radio-based dark matter detection \cite{Irwin_DMRadio}, the dark E-field experiment \cite{Godfrey2021}, the plasma haloscope-based ALPHA experiment \cite{Millar2023}, and the dielectric haloscope-based MADMAX collaboration effort \cite{Brun2019}. The results discussed in this manuscript are colored green. This plot is based on data from the open dark matter limit library in Ref.~\cite{AxionLimits}.}
  \label{fig:dm_projection}
  \end{center}
\end{figure}

In a subsequent step, we intend to investigate sensing protocols using HBAR-modes with lower mode numbers ($n\approx1-70$). This approach is motivated by the favourable scaling of the projected bounds (Eqs.~\ref{eq:gw_main_0}, \ref{eq:dm_main}) with lower phonon mode frequencies and mode numbers. To control and read out the low-frequency modes, we anticipate coupling the mechanical resonator to a fluxonium qubit operating at MHz frequencies~\cite{lee2023, najera2024}. High acoustic quality factors in the MHz regime are expected by manufacturing the HBAR out of a single, piezoelectric material such as quartz (SiO$_2$). The upper frequency limit of this device will be limited by the coupling between the acoustic modes and the superconducting qubit, which is expected to decrease with higher frequencies \cite{Garcia-Belles2025}. In addition, MHz frequency devices placed inside a dilution refrigerator are not passively cooled into their ground state.  To overcome this limitation, various active cooling protocols have been developed with fluxonium qubits \cite{najera2024,  Zhang_2021}. These methods could be adapted to our system by first cooling the qubit and then transferring its state to the mechanical mode. To avoid thermal noise, the detection of external signals must then happen before the mechanical mode returns to thermal equilibrium, which can still result in a large duty cycle due to the high-Q and low frequency of the modes. As shown in Table~\ref{tab:detector-comparison} and Figures~\ref{fig:gw_projection} and \ref{fig:dm_projection}, such a device not only enables access to previously unexplored parameter regimes, but also further enhances the overall sensitivity of our detection strategy.

\begin{table*}[ht]
  \centering
  \begin{tabular*}{\textwidth}{@{\extracolsep{\fill}} l c c c @{}}
    \toprule
    \textbf{Quantity} 
      & \textbf{Current} 
      & \textbf{Next generation} 
      & \textbf{MHz device} \\
    \midrule
    Frequency
      & \SI{5}{\giga\hertz}
      & 3-10~\SI{}{\giga\hertz}
      & 15-800~\SI{}{\mega\hertz} \\
    Qubit \(T_1\)
      & \SI{25}{\micro\second}
      & \SI{50}{\micro\second}
      & \SI{50}{\micro\second} \\
    Phonon \(T_1\)
      & \SI{100}{\micro\second}
      & \SI{1}{\milli\second}
      & \SI{10}{\milli\second} \\
    HBAR height $L$                            
      & \SI{435}{\micro\meter}    
      & \SI{435}{\micro\meter} 
      & \SI{1}{\milli\meter} \\
    Beam waist $\mu$                            
      & \SI{27}{\micro\meter}
      & \SI{27}{\micro\meter}
      & \SI{700}{\micro\meter}\\ 
    Material 
      & sapphire 
      & sapphire 
      & quartz \\
    Integration time
      & 1~week
      & 1~year
      & 1~year \\
    Excited state population 
      & \(6.7 \times 10^{-5}\)
      & \(1.0\times10^{-5}\)
      & \(1.0\times10^{-5}\)   \\
      \midrule
    Estimated GW amplitude bound \(h_0\)
      & \(5.5 \times 10^{-18}\)
       & \(1.8 \times 10^{-19}\)
      & \(8.6 \times 10^{-22}\)    \\
    Estimated DM coupling bound \(\kappa\)
      & \(8.8\times10^{-10}\)
      & \(3.0\times10^{-11}\)
      & - \\
    Estimated CSL rate \(\lambda_{\mathrm{CSL}}\)
      & \(5.7\times10^{-8}\,\mathrm{s}^{-1}\)
      & \(10^{-10}\,\mathrm{s}^{-1}\)
      & \(10^{-13}\,\mathrm{s}^{-1}\) \\
    Estimated CSL radius \(r_{\mathrm{CSL}}\)
      & \(3.0\times10^{-7}\,\mathrm{m}\)
      & \(3.0\times10^{-7}\,\mathrm{m}\)
      & \(3.0\times10^{-7}\,\mathrm{m}\) \\
    \bottomrule
  \end{tabular*}
  \caption{\textbf{Comparison of current and future device parameters.} For the estimated bounds, we assume the same excited-state population as measured in the current experiment. For the MHz device, we assume that we can sufficiently well cool the system in order to reach the same measured populations as for higher-frequency modes. We do not calculate a DM coupling bound $\kappa$ for the MHz device, because the suppression of the kinetic mixing due to electromagnetic shielding of the device \cite{Trickle2025} is expected to substantially lower the device's sensitivity at these frequencies. This does not affect the detection of gravitational waves.}
  \label{tab:detector-comparison}
\end{table*}

\section{Dark Matter Interaction} \label{sec:dark_mat}
Dark photons are ultra-light dark matter candidates that, for the purposes of coupling to $\hbar$BAR devices, can be treated as a classical oscillating electric field $\vec{E}(\vec{x}, t)$ that interacts with the strain field $\overset{\leftrightarrow}{S}(\vec{x}, t)$ of the mechanical modes through the piezoelectric tensor $\overset{\leftrightarrow}{e}(\vec{x})$. The interaction Hamiltonian can be written as
\begin{equation}
   H_{DP} = -\int{\text{d}^3x\,  e_{ijk}E_i S_{jk}}.
   \label{eq:HDP}
\end{equation}
As in the previous section, we only consider the dominant strain component $S_{33}$, which, for AlN, makes $e_{33}$ (in Voigt notation) and $E_3$ the only relevant components. We also note that the integral is essentially over the volume of the piezoelectric AlN layer, since $e_{33}=0$ elsewhere. 

Assuming that the dark photon electric field is polarized in the 3 direction, its amplitude, in SI units, is given by~\cite{Linehan2024}
\begin{equation}
   E_3 = \kappa\frac{\sqrt{2\rho_V}}{\epsilon_r\sqrt{\epsilon_0}}\cos\left(\omega t\right).
   \label{eq:EMfield}
\end{equation}

The quantized strain field of the HBAR mode can be written as \cite{Bild2023}
\begin{equation}
   S_{33} = S_0 \, LG_{00}(r,\phi) \, \sin\left( \dfrac{n \pi z}{L}\right) (a+a^{\dagger}),
   \label{eq:strain}
\end{equation}
where we consider the fundamental transverse Laguerre-Gaussian mode as before, and $S_0 = \sqrt{4\hbar\omega/(L \mu^2c_{33})}$ is the strain per phonon. 

Inserting equation~(\ref{eq:EMfield}) and (\ref{eq:strain}) into equation~(\ref{eq:HDP}), we find
\begin{eqnarray}
   H_{DP} &=& -8\kappa e_{33} \cos\left(\omega t\right)\frac{\mu}{\epsilon_r n}\sqrt{\frac{\rho_V \omega L}{\epsilon_0\hbar\pi c_{33}}} (a+a^{\dagger}).
   \label{HDP}
\end{eqnarray}
Identifying the result with the drive Hamiltonian of equation~(\ref{eq:hbar-hamiltonian}), we find a driving strength due to dark photons of 
\begin{equation}
    \Omega_{DP} = -4\kappa e_{33}\frac{\mu}{\epsilon_r n}\sqrt{\frac{\rho_V\omega L}{\epsilon_0\hbar\pi c_{33}}},
\end{equation}
where we have also made the rotating wave approximation as in the previous section.

\section{Testing Wavefunction Collapse Models} \label{app:collapse}
Collapse models are modified versions of the Schr\"odinger equation designed to explain the collapse of the wavefunction and the disappearance of quantum superpositions in macroscopic objects, regardless of their isolation from the environment. 
Collapse models address this issue by introducing stochastic modifications to the Schr{\"o}dinger equation through the addition of noise terms that are intrinsic to the system. 
Fundamentally, these modifications can be interpreted as arising from stochastic forces that constitute an intrinsic noise term in the system.
Recently, the possibility of performing high-precision experiments has allowed for bounding the parameter space of these possible modifications.
Although a direct test of collapse models would require the preparation of a macroscopic superposition state and observation of its decoherence, the possible existence of a stochastic noise would result in effects that are observable even without preparing nonclassical states.
Striking examples that have been widely explored experimentally are the emission of electromagnetic waves \cite{Donadi_Xray_21} and the heating of bulk matter \cite{GhirardiPRD86}. 
Clearly, the latter case is directly related to our experiment and will thus be discussed below.

Following the approach of Ref.~\cite{Schrinski2023}, a measurement of the HBAR occupation number can be used to bound possible nonlinear modifications to the Schr\"odinger equation.
The latter results in a diffusion in phase-space at rate $\Gamma_D$, which is compensated by the oscillator relaxation at rate $\gamma_\downarrow = 1/T_1$. In the steady state, the energy of the system is thus $E=\hbar\omega(1/2+\Gamma_D/\gamma_\downarrow)$, which allows us to write $\Gamma_D = \overline{n} \gamma_\downarrow$ with $\overline{n}$ the oscillator's occupation number. 

\begin{figure}
    \centering
    \includegraphics[width=0.9\linewidth]{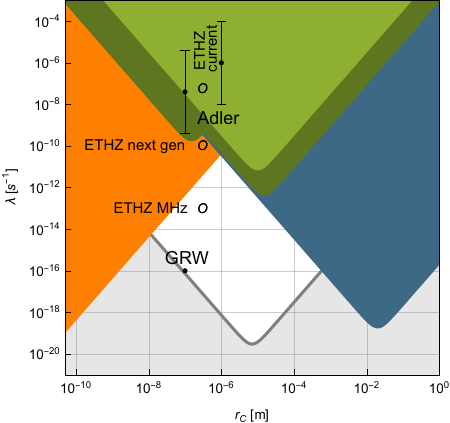}
    \caption{\textbf{Bounds on CSL parameters.} Bounds on the free parameters of the CSL model coming from non-interferometric tests using LISA Pathfinder (blue), X-ray emission tests (orange) and cantilever experiments (green).  Grey areas are theoretically excluded regions \cite{Donadi_Xray_21,VinantePRL20}. Filled black dots are different theoretical predictions for the parameters \cite{GhirardiPRD86,Adler_2007}. Circles are the exclusion values achieved by the current experiment (top) and next generation devices (bottom), see Tab.~\ref{tab:detector-comparison}.}
    \label{figCSL}
\end{figure}

Importantly, the diffusion rate $\Gamma_D$ depends on both the modification strength $\tau_e$, as well as on the details of the experimental platform. For our $\hbar$BAR device we have $\Gamma_D = 3.5\times 10^{13}/\tau_e$ \cite{Schrinski2023}, which, together with the relation $\Gamma_D = \overline{n} \gamma_\downarrow$, implies $\tau_e = 3.5\times 10^{13} T_1 / \overline{n}$. Therefore, a measurement of $T_1$ and $\overline{n}$ results in an upper bound to the possible modification parameter $\tau_e$.

In our experiment, we can estimate the occupation number to coincide with the measured Fock $\ket{1}$ population upper bound $P_p^{404,\text{max}}$ or $P_p^{404,\text{inferred}}$, since we assume the population of higher energy levels to constitute a negligible contribution. This results in an exclusion of modification values $\tau_e^{\text{max}} <  \SI{5.9 e13}{\second}$ or $\tau_e^{\text{inferred}} <  \SI{2.1 e14}{\second}$, respectively. 

We can convert these numbers into bounds for the free parameters of the continuous spontaneous localization (CSL) model. These are the collapse rate $\lambda_\text{CSL} = (\SI{1}{\atomicmassunit}/m_e)^2/\tau_e$, giving $\lambda_\text{CSL}^\text{max} = \SI{ 5.7 e-8}{\second^{-1}}$ and $\lambda_\text{CSL}^\text{inferred} = \SI{1.6 e-8}{\second^{-1}}$, and localization length scale $r_\text{CSL}=\hbar/\sqrt{2}\sigma_q=\SI{ 3.0 e-7}{\meter}$.

Although these bounds are on the same level as the one obtained from ultracold cantilever experiments \cite{VinantePRL20}, the estimated parameters of future $\hbar$BAR devices are expected to improve the bound on $\lambda_\text{CSL}$ by up to five orders of magnitude, see Tab.~\ref{tab:detector-comparison} and Fig.~\ref{figCSL}.
This would result in testing parameter regimes still unexplored by current experiments and observations.

\bibliography{references}

\begin{thebibliography}{73}%
\makeatletter
\providecommand \@ifxundefined [1]{%
 \@ifx{#1\undefined}
}%
\providecommand \@ifnum [1]{%
 \ifnum #1\expandafter \@firstoftwo
 \else \expandafter \@secondoftwo
 \fi
}%
\providecommand \@ifx [1]{%
 \ifx #1\expandafter \@firstoftwo
 \else \expandafter \@secondoftwo
 \fi
}%
\providecommand \natexlab [1]{#1}%
\providecommand \enquote  [1]{``#1''}%
\providecommand \bibnamefont  [1]{#1}%
\providecommand \bibfnamefont [1]{#1}%
\providecommand \citenamefont [1]{#1}%
\providecommand \href@noop [0]{\@secondoftwo}%
\providecommand \href [0]{\begingroup \@sanitize@url \@href}%
\providecommand \@href[1]{\@@startlink{#1}\@@href}%
\providecommand \@@href[1]{\endgroup#1\@@endlink}%
\providecommand \@sanitize@url [0]{\catcode `\\12\catcode `\$12\catcode `\&12\catcode `\#12\catcode `\^12\catcode `\_12\catcode `\%12\relax}%
\providecommand \@@startlink[1]{}%
\providecommand \@@endlink[0]{}%
\providecommand \url  [0]{\begingroup\@sanitize@url \@url }%
\providecommand \@url [1]{\endgroup\@href {#1}{\urlprefix }}%
\providecommand \urlprefix  [0]{URL }%
\providecommand \Eprint [0]{\href }%
\providecommand \doibase [0]{https://doi.org/}%
\providecommand \selectlanguage [0]{\@gobble}%
\providecommand \bibinfo  [0]{\@secondoftwo}%
\providecommand \bibfield  [0]{\@secondoftwo}%
\providecommand \translation [1]{[#1]}%
\providecommand \BibitemOpen [0]{}%
\providecommand \bibitemStop [0]{}%
\providecommand \bibitemNoStop [0]{.\EOS\space}%
\providecommand \EOS [0]{\spacefactor3000\relax}%
\providecommand \BibitemShut  [1]{\csname bibitem#1\endcsname}%
\let\auto@bib@innerbib\@empty
\bibitem [{\citenamefont {Chu}\ \emph {et~al.}(2017)\citenamefont {Chu}, \citenamefont {Kharel}, \citenamefont {Renninger}, \citenamefont {Burkhart}, \citenamefont {Frunzio}, \citenamefont {Rakich},\ and\ \citenamefont {Schoelkopf}}]{Chu2017}%
  \BibitemOpen
  \bibfield  {author} {\bibinfo {author} {\bibfnamefont {Y.}~\bibnamefont {Chu}}, \bibinfo {author} {\bibfnamefont {P.}~\bibnamefont {Kharel}}, \bibinfo {author} {\bibfnamefont {W.~H.}\ \bibnamefont {Renninger}}, \bibinfo {author} {\bibfnamefont {L.~D.}\ \bibnamefont {Burkhart}}, \bibinfo {author} {\bibfnamefont {L.}~\bibnamefont {Frunzio}}, \bibinfo {author} {\bibfnamefont {P.~T.}\ \bibnamefont {Rakich}},\ and\ \bibinfo {author} {\bibfnamefont {R.~J.}\ \bibnamefont {Schoelkopf}},\ }\bibfield  {title} {\bibinfo {title} {Quantum acoustics with superconducting qubits},\ }\href {https://doi.org/10.1126/science.aao1511} {\bibfield  {journal} {\bibinfo  {journal} {Science}\ }\textbf {\bibinfo {volume} {358}},\ \bibinfo {pages} {199} (\bibinfo {year} {2017})}\BibitemShut {NoStop}%
\bibitem [{\citenamefont {Satzinger}\ \emph {et~al.}(2018)\citenamefont {Satzinger}, \citenamefont {Zhong}, \citenamefont {Chang}, \citenamefont {Peairs}, \citenamefont {Bienfait}, \citenamefont {Chou}, \citenamefont {Cleland}, \citenamefont {Conner}, \citenamefont {Dumur}, \citenamefont {Grebel}, \citenamefont {Gutierrez}, \citenamefont {November}, \citenamefont {Povey}, \citenamefont {Whiteley}, \citenamefont {Awschalom}, \citenamefont {Schuster},\ and\ \citenamefont {Cleland}}]{Satzinger2018}%
  \BibitemOpen
  \bibfield  {author} {\bibinfo {author} {\bibfnamefont {K.~J.}\ \bibnamefont {Satzinger}}, \bibinfo {author} {\bibfnamefont {Y.~P.}\ \bibnamefont {Zhong}}, \bibinfo {author} {\bibfnamefont {H.-S.}\ \bibnamefont {Chang}}, \bibinfo {author} {\bibfnamefont {G.~A.}\ \bibnamefont {Peairs}}, \bibinfo {author} {\bibfnamefont {A.}~\bibnamefont {Bienfait}}, \bibinfo {author} {\bibfnamefont {M.-H.}\ \bibnamefont {Chou}}, \bibinfo {author} {\bibfnamefont {A.~Y.}\ \bibnamefont {Cleland}}, \bibinfo {author} {\bibfnamefont {C.~R.}\ \bibnamefont {Conner}}, \bibinfo {author} {\bibfnamefont {{\'E}.}~\bibnamefont {Dumur}}, \bibinfo {author} {\bibfnamefont {J.}~\bibnamefont {Grebel}}, \bibinfo {author} {\bibfnamefont {I.}~\bibnamefont {Gutierrez}}, \bibinfo {author} {\bibfnamefont {B.~H.}\ \bibnamefont {November}}, \bibinfo {author} {\bibfnamefont {R.~G.}\ \bibnamefont {Povey}}, \bibinfo {author} {\bibfnamefont {S.~J.}\ \bibnamefont {Whiteley}}, \bibinfo {author} {\bibfnamefont {D.~D.}\ \bibnamefont {Awschalom}}, \bibinfo
  {author} {\bibfnamefont {D.~I.}\ \bibnamefont {Schuster}},\ and\ \bibinfo {author} {\bibfnamefont {A.~N.}\ \bibnamefont {Cleland}},\ }\bibfield  {title} {\bibinfo {title} {Quantum control of surface acoustic‐wave phonons},\ }\href {https://doi.org/10.1038/s41586-018-0719-5} {\bibfield  {journal} {\bibinfo  {journal} {Nature}\ }\textbf {\bibinfo {volume} {563}},\ \bibinfo {pages} {661} (\bibinfo {year} {2018})}\BibitemShut {NoStop}%
\bibitem [{\citenamefont {Arrangoiz-Arriola}\ \emph {et~al.}(2019)\citenamefont {Arrangoiz-Arriola}, \citenamefont {Wollack}, \citenamefont {Wang}, \citenamefont {Pechal}, \citenamefont {Jiang}, \citenamefont {McKenna}, \citenamefont {Witmer}, \citenamefont {Van~Laer},\ and\ \citenamefont {Safavi-Naeini}}]{Arrangoiz-Arriola2019}%
  \BibitemOpen
  \bibfield  {author} {\bibinfo {author} {\bibfnamefont {P.}~\bibnamefont {Arrangoiz-Arriola}}, \bibinfo {author} {\bibfnamefont {E.~A.}\ \bibnamefont {Wollack}}, \bibinfo {author} {\bibfnamefont {Z.}~\bibnamefont {Wang}}, \bibinfo {author} {\bibfnamefont {M.}~\bibnamefont {Pechal}}, \bibinfo {author} {\bibfnamefont {W.}~\bibnamefont {Jiang}}, \bibinfo {author} {\bibfnamefont {T.~P.}\ \bibnamefont {McKenna}}, \bibinfo {author} {\bibfnamefont {J.~D.}\ \bibnamefont {Witmer}}, \bibinfo {author} {\bibfnamefont {R.}~\bibnamefont {Van~Laer}},\ and\ \bibinfo {author} {\bibfnamefont {A.~H.}\ \bibnamefont {Safavi-Naeini}},\ }\bibfield  {title} {\bibinfo {title} {Resolving the energy levels of a nanomechanical oscillator},\ }\href {https://doi.org/10.1038/s41586-019-1386-x} {\bibfield  {journal} {\bibinfo  {journal} {Nature}\ }\textbf {\bibinfo {volume} {571}},\ \bibinfo {pages} {537} (\bibinfo {year} {2019})}\BibitemShut {NoStop}%
\bibitem [{\citenamefont {Qiao}\ \emph {et~al.}(2023)\citenamefont {Qiao}, \citenamefont {Dumur}, \citenamefont {Andersson}, \citenamefont {Yan}, \citenamefont {Chou}, \citenamefont {Grebel}, \citenamefont {Conner}, \citenamefont {Joshi}, \citenamefont {Miller}, \citenamefont {Povey}, \citenamefont {Wu},\ and\ \citenamefont {Cleland}}]{Qiao2023}%
  \BibitemOpen
  \bibfield  {author} {\bibinfo {author} {\bibfnamefont {H.}~\bibnamefont {Qiao}}, \bibinfo {author} {\bibfnamefont {E.}~\bibnamefont {Dumur}}, \bibinfo {author} {\bibfnamefont {G.}~\bibnamefont {Andersson}}, \bibinfo {author} {\bibfnamefont {H.}~\bibnamefont {Yan}}, \bibinfo {author} {\bibfnamefont {M.-H.}\ \bibnamefont {Chou}}, \bibinfo {author} {\bibfnamefont {J.}~\bibnamefont {Grebel}}, \bibinfo {author} {\bibfnamefont {C.~R.}\ \bibnamefont {Conner}}, \bibinfo {author} {\bibfnamefont {Y.~J.}\ \bibnamefont {Joshi}}, \bibinfo {author} {\bibfnamefont {J.~M.}\ \bibnamefont {Miller}}, \bibinfo {author} {\bibfnamefont {R.~G.}\ \bibnamefont {Povey}}, \bibinfo {author} {\bibfnamefont {X.}~\bibnamefont {Wu}},\ and\ \bibinfo {author} {\bibfnamefont {A.~N.}\ \bibnamefont {Cleland}},\ }\bibfield  {title} {\bibinfo {title} {Splitting phonons: Building a platform for linear mechanical quantum computing},\ }\href {https://doi.org/10.1126/science.adg8715} {\bibfield  {journal} {\bibinfo  {journal} {Science}\ }\textbf
  {\bibinfo {volume} {380}},\ \bibinfo {pages} {1030} (\bibinfo {year} {2023})}\BibitemShut {NoStop}%
\bibitem [{\citenamefont {Wollack}\ \emph {et~al.}(2022)\citenamefont {Wollack}, \citenamefont {Cleland}, \citenamefont {Gruenke}, \citenamefont {Wang}, \citenamefont {Arrangoiz-Arriola},\ and\ \citenamefont {Safavi-Naeini}}]{Wollack2022}%
  \BibitemOpen
  \bibfield  {author} {\bibinfo {author} {\bibfnamefont {E.~A.}\ \bibnamefont {Wollack}}, \bibinfo {author} {\bibfnamefont {A.~Y.}\ \bibnamefont {Cleland}}, \bibinfo {author} {\bibfnamefont {R.~G.}\ \bibnamefont {Gruenke}}, \bibinfo {author} {\bibfnamefont {Z.}~\bibnamefont {Wang}}, \bibinfo {author} {\bibfnamefont {P.}~\bibnamefont {Arrangoiz-Arriola}},\ and\ \bibinfo {author} {\bibfnamefont {A.~H.}\ \bibnamefont {Safavi-Naeini}},\ }\bibfield  {title} {\bibinfo {title} {Quantum state preparation and tomography of entangled mechanical resonators},\ }\href {https://doi.org/10.1038/s41586-022-04500-y} {\bibfield  {journal} {\bibinfo  {journal} {Nature}\ }\textbf {\bibinfo {volume} {604}},\ \bibinfo {pages} {463} (\bibinfo {year} {2022})}\BibitemShut {NoStop}%
\bibitem [{\citenamefont {van Thiel}\ \emph {et~al.}(2025)\citenamefont {van Thiel}, \citenamefont {Weaver}, \citenamefont {Berto}, \citenamefont {Duivestein}, \citenamefont {Lemang}, \citenamefont {Schuurman}, \citenamefont {{\v{Z}}emli{\v{c}}ka}, \citenamefont {Hijazi}, \citenamefont {Bernasconi}, \citenamefont {Ferrer}, \citenamefont {Cataldo}, \citenamefont {Lachman}, \citenamefont {Field}, \citenamefont {Mohan}, \citenamefont {de~Vries}, \citenamefont {Bultink}, \citenamefont {van Oven}, \citenamefont {Mutus}, \citenamefont {Stockill},\ and\ \citenamefont {Gr{\"o}blacher}}]{vanThiel2025}%
  \BibitemOpen
  \bibfield  {author} {\bibinfo {author} {\bibfnamefont {T.~C.}\ \bibnamefont {van Thiel}}, \bibinfo {author} {\bibfnamefont {M.~J.}\ \bibnamefont {Weaver}}, \bibinfo {author} {\bibfnamefont {F.}~\bibnamefont {Berto}}, \bibinfo {author} {\bibfnamefont {P.}~\bibnamefont {Duivestein}}, \bibinfo {author} {\bibfnamefont {M.}~\bibnamefont {Lemang}}, \bibinfo {author} {\bibfnamefont {K.~L.}\ \bibnamefont {Schuurman}}, \bibinfo {author} {\bibfnamefont {M.}~\bibnamefont {{\v{Z}}emli{\v{c}}ka}}, \bibinfo {author} {\bibfnamefont {F.}~\bibnamefont {Hijazi}}, \bibinfo {author} {\bibfnamefont {A.~C.}\ \bibnamefont {Bernasconi}}, \bibinfo {author} {\bibfnamefont {C.}~\bibnamefont {Ferrer}}, \bibinfo {author} {\bibfnamefont {E.}~\bibnamefont {Cataldo}}, \bibinfo {author} {\bibfnamefont {E.}~\bibnamefont {Lachman}}, \bibinfo {author} {\bibfnamefont {M.}~\bibnamefont {Field}}, \bibinfo {author} {\bibfnamefont {Y.}~\bibnamefont {Mohan}}, \bibinfo {author} {\bibfnamefont {F.~K.}\ \bibnamefont {de~Vries}}, \bibinfo {author}
  {\bibfnamefont {C.~C.}\ \bibnamefont {Bultink}}, \bibinfo {author} {\bibfnamefont {J.~C.}\ \bibnamefont {van Oven}}, \bibinfo {author} {\bibfnamefont {J.~Y.}\ \bibnamefont {Mutus}}, \bibinfo {author} {\bibfnamefont {R.}~\bibnamefont {Stockill}},\ and\ \bibinfo {author} {\bibfnamefont {S.}~\bibnamefont {Gr{\"o}blacher}},\ }\bibfield  {title} {\bibinfo {title} {Optical readout of a superconducting qubit using a piezo-optomechanical transducer},\ }\href {https://doi.org/10.1038/s41567-024-02742-3} {\bibfield  {journal} {\bibinfo  {journal} {Nature Physics}\ }\textbf {\bibinfo {volume} {21}},\ \bibinfo {pages} {401} (\bibinfo {year} {2025})}\BibitemShut {NoStop}%
\bibitem [{\citenamefont {Linehan}\ \emph {et~al.}(2025)\citenamefont {Linehan}, \citenamefont {Trickle}, \citenamefont {Conner}, \citenamefont {Ghosh}, \citenamefont {Lin}, \citenamefont {Sholapurkar},\ and\ \citenamefont {Cleland}}]{Linehan2024}%
  \BibitemOpen
  \bibfield  {author} {\bibinfo {author} {\bibfnamefont {R.}~\bibnamefont {Linehan}}, \bibinfo {author} {\bibfnamefont {T.}~\bibnamefont {Trickle}}, \bibinfo {author} {\bibfnamefont {C.~R.}\ \bibnamefont {Conner}}, \bibinfo {author} {\bibfnamefont {S.}~\bibnamefont {Ghosh}}, \bibinfo {author} {\bibfnamefont {T.}~\bibnamefont {Lin}}, \bibinfo {author} {\bibfnamefont {M.}~\bibnamefont {Sholapurkar}},\ and\ \bibinfo {author} {\bibfnamefont {A.~N.}\ \bibnamefont {Cleland}},\ }\bibfield  {title} {\bibinfo {title} {Listening for new physics with quantum acoustics},\ }\href {https://doi.org/10.1103/63zj-d8z4} {\bibfield  {journal} {\bibinfo  {journal} {Phys. Rev. D}\ }\textbf {\bibinfo {volume} {112}},\ \bibinfo {pages} {115005} (\bibinfo {year} {2025})}\BibitemShut {NoStop}%
\bibitem [{\citenamefont {Trickle}(2025)}]{Trickle2025}%
  \BibitemOpen
  \bibfield  {author} {\bibinfo {author} {\bibfnamefont {T.}~\bibnamefont {Trickle}},\ }\bibfield  {title} {\bibinfo {title} {Piezoelectric bulk acoustic resonators for dark photon detection},\ }\href {https://doi.org/10.1103/yq4n-xq9y} {\bibfield  {journal} {\bibinfo  {journal} {Phys. Rev. D}\ }\textbf {\bibinfo {volume} {112}},\ \bibinfo {pages} {055043} (\bibinfo {year} {2025})}\BibitemShut {NoStop}%
\bibitem [{\citenamefont {Aggarwal}\ \emph {et~al.}(2025)\citenamefont {Aggarwal}, \citenamefont {Aguiar}, \citenamefont {Blas}, \citenamefont {Bauswein}, \citenamefont {Cella}, \citenamefont {Clesse}, \citenamefont {Cruise}, \citenamefont {Domcke}, \citenamefont {Ellis}, \citenamefont {Figueroa}, \citenamefont {Franciolini}, \citenamefont {García-Cely}, \citenamefont {Geraci}, \citenamefont {Goryachev}, \citenamefont {Grote}, \citenamefont {Hindmarsh}, \citenamefont {Ito}, \citenamefont {Kopp}, \citenamefont {Lee}, \citenamefont {Martineau}, \citenamefont {McDonald}, \citenamefont {Muia}, \citenamefont {Mukund}, \citenamefont {Ottaway}, \citenamefont {Peloso}, \citenamefont {Peters}, \citenamefont {Quevedo}, \citenamefont {Ricciardone}, \citenamefont {Ringwald}, \citenamefont {Steinlechner}, \citenamefont {Steinlechner}, \citenamefont {Sun}, \citenamefont {Tamarit}, \citenamefont {Tobar}, \citenamefont {Torrenti}, \citenamefont {Ünal},\ and\ \citenamefont {White}}]{Aggarwal2025}%
  \BibitemOpen
  \bibfield  {author} {\bibinfo {author} {\bibfnamefont {N.}~\bibnamefont {Aggarwal}}, \bibinfo {author} {\bibfnamefont {O.~D.}\ \bibnamefont {Aguiar}}, \bibinfo {author} {\bibfnamefont {D.}~\bibnamefont {Blas}}, \bibinfo {author} {\bibfnamefont {A.}~\bibnamefont {Bauswein}}, \bibinfo {author} {\bibfnamefont {G.}~\bibnamefont {Cella}}, \bibinfo {author} {\bibfnamefont {S.}~\bibnamefont {Clesse}}, \bibinfo {author} {\bibfnamefont {A.~M.}\ \bibnamefont {Cruise}}, \bibinfo {author} {\bibfnamefont {V.}~\bibnamefont {Domcke}}, \bibinfo {author} {\bibfnamefont {S.}~\bibnamefont {Ellis}}, \bibinfo {author} {\bibfnamefont {D.~G.}\ \bibnamefont {Figueroa}}, \bibinfo {author} {\bibfnamefont {G.}~\bibnamefont {Franciolini}}, \bibinfo {author} {\bibfnamefont {C.}~\bibnamefont {García-Cely}}, \bibinfo {author} {\bibfnamefont {A.}~\bibnamefont {Geraci}}, \bibinfo {author} {\bibfnamefont {M.}~\bibnamefont {Goryachev}}, \bibinfo {author} {\bibfnamefont {H.}~\bibnamefont {Grote}}, \bibinfo {author} {\bibfnamefont
  {M.}~\bibnamefont {Hindmarsh}}, \bibinfo {author} {\bibfnamefont {A.}~\bibnamefont {Ito}}, \bibinfo {author} {\bibfnamefont {J.}~\bibnamefont {Kopp}}, \bibinfo {author} {\bibfnamefont {S.~M.}\ \bibnamefont {Lee}}, \bibinfo {author} {\bibfnamefont {K.}~\bibnamefont {Martineau}}, \bibinfo {author} {\bibfnamefont {J.}~\bibnamefont {McDonald}}, \bibinfo {author} {\bibfnamefont {F.}~\bibnamefont {Muia}}, \bibinfo {author} {\bibfnamefont {N.}~\bibnamefont {Mukund}}, \bibinfo {author} {\bibfnamefont {D.}~\bibnamefont {Ottaway}}, \bibinfo {author} {\bibfnamefont {M.}~\bibnamefont {Peloso}}, \bibinfo {author} {\bibfnamefont {K.}~\bibnamefont {Peters}}, \bibinfo {author} {\bibfnamefont {F.}~\bibnamefont {Quevedo}}, \bibinfo {author} {\bibfnamefont {A.}~\bibnamefont {Ricciardone}}, \bibinfo {author} {\bibfnamefont {A.}~\bibnamefont {Ringwald}}, \bibinfo {author} {\bibfnamefont {J.}~\bibnamefont {Steinlechner}}, \bibinfo {author} {\bibfnamefont {S.}~\bibnamefont {Steinlechner}}, \bibinfo {author} {\bibfnamefont
  {S.}~\bibnamefont {Sun}}, \bibinfo {author} {\bibfnamefont {C.}~\bibnamefont {Tamarit}}, \bibinfo {author} {\bibfnamefont {M.~E.}\ \bibnamefont {Tobar}}, \bibinfo {author} {\bibfnamefont {F.}~\bibnamefont {Torrenti}}, \bibinfo {author} {\bibfnamefont {C.}~\bibnamefont {Ünal}},\ and\ \bibinfo {author} {\bibfnamefont {G.}~\bibnamefont {White}},\ }\bibfield  {title} {\bibinfo {title} {Challenges and opportunities of gravitational-wave searches above 10 {kHz}},\ }\href {https://doi.org/10.1007/s41114-025-00060-5} {\bibfield  {journal} {\bibinfo  {journal} {Living Reviews in Relativity}\ }\textbf {\bibinfo {volume} {28}},\ \bibinfo {pages} {10} (\bibinfo {year} {2025})}\BibitemShut {NoStop}%
\bibitem [{\citenamefont {Schrinski}\ \emph {et~al.}(2023)\citenamefont {Schrinski}, \citenamefont {Yang}, \citenamefont {von L\"upke}, \citenamefont {Bild}, \citenamefont {Chu}, \citenamefont {Hornberger}, \citenamefont {Nimmrichter},\ and\ \citenamefont {Fadel}}]{Schrinski2023}%
  \BibitemOpen
  \bibfield  {author} {\bibinfo {author} {\bibfnamefont {B.}~\bibnamefont {Schrinski}}, \bibinfo {author} {\bibfnamefont {Y.}~\bibnamefont {Yang}}, \bibinfo {author} {\bibfnamefont {U.}~\bibnamefont {von L\"upke}}, \bibinfo {author} {\bibfnamefont {M.}~\bibnamefont {Bild}}, \bibinfo {author} {\bibfnamefont {Y.}~\bibnamefont {Chu}}, \bibinfo {author} {\bibfnamefont {K.}~\bibnamefont {Hornberger}}, \bibinfo {author} {\bibfnamefont {S.}~\bibnamefont {Nimmrichter}},\ and\ \bibinfo {author} {\bibfnamefont {M.}~\bibnamefont {Fadel}},\ }\bibfield  {title} {\bibinfo {title} {Macroscopic quantum test with bulk acoustic wave resonators},\ }\href {https://doi.org/10.1103/PhysRevLett.130.133604} {\bibfield  {journal} {\bibinfo  {journal} {Phys. Rev. Lett.}\ }\textbf {\bibinfo {volume} {130}},\ \bibinfo {pages} {133604} (\bibinfo {year} {2023})}\BibitemShut {NoStop}%
\bibitem [{\citenamefont {Chu}\ \emph {et~al.}(2018)\citenamefont {Chu}, \citenamefont {Kharel}, \citenamefont {Yoon}, \citenamefont {Frunzio}, \citenamefont {Rakich},\ and\ \citenamefont {Schoelkopf}}]{Chu2018}%
  \BibitemOpen
  \bibfield  {author} {\bibinfo {author} {\bibfnamefont {Y.}~\bibnamefont {Chu}}, \bibinfo {author} {\bibfnamefont {P.}~\bibnamefont {Kharel}}, \bibinfo {author} {\bibfnamefont {T.}~\bibnamefont {Yoon}}, \bibinfo {author} {\bibfnamefont {L.}~\bibnamefont {Frunzio}}, \bibinfo {author} {\bibfnamefont {P.~T.}\ \bibnamefont {Rakich}},\ and\ \bibinfo {author} {\bibfnamefont {R.~J.}\ \bibnamefont {Schoelkopf}},\ }\bibfield  {title} {\bibinfo {title} {Creation and control of multi-phonon fock states in a bulk acoustic-wave resonator},\ }\href {https://doi.org/10.1038/s41586-018-0717-7} {\bibfield  {journal} {\bibinfo  {journal} {Nature}\ }\textbf {\bibinfo {volume} {563}},\ \bibinfo {pages} {666} (\bibinfo {year} {2018})}\BibitemShut {NoStop}%
\bibitem [{\citenamefont {Bild}\ \emph {et~al.}(2023)\citenamefont {Bild}, \citenamefont {Fadel}, \citenamefont {Yang}, \citenamefont {von L\"upke}, \citenamefont {Martin}, \citenamefont {Bruno},\ and\ \citenamefont {Chu}}]{Bild2023}%
  \BibitemOpen
  \bibfield  {author} {\bibinfo {author} {\bibfnamefont {M.}~\bibnamefont {Bild}}, \bibinfo {author} {\bibfnamefont {M.}~\bibnamefont {Fadel}}, \bibinfo {author} {\bibfnamefont {Y.}~\bibnamefont {Yang}}, \bibinfo {author} {\bibfnamefont {U.}~\bibnamefont {von L\"upke}}, \bibinfo {author} {\bibfnamefont {P.}~\bibnamefont {Martin}}, \bibinfo {author} {\bibfnamefont {A.}~\bibnamefont {Bruno}},\ and\ \bibinfo {author} {\bibfnamefont {Y.}~\bibnamefont {Chu}},\ }\bibfield  {title} {\bibinfo {title} {Schr\"{o}dinger cat states of a 16-microgram mechanical oscillator},\ }\href {https://doi.org/10.1126/science.adf7553} {\bibfield  {journal} {\bibinfo  {journal} {Science}\ }\textbf {\bibinfo {volume} {380}},\ \bibinfo {pages} {274} (\bibinfo {year} {2023})}\BibitemShut {NoStop}%
\bibitem [{\citenamefont {Yadin}\ and\ \citenamefont {Fadel}(2025)}]{Fadel2025}%
  \BibitemOpen
  \bibfield  {author} {\bibinfo {author} {\bibfnamefont {B.}~\bibnamefont {Yadin}}\ and\ \bibinfo {author} {\bibfnamefont {M.}~\bibnamefont {Fadel}},\ }\href {https://arxiv.org/abs/2503.08324} {\bibinfo {title} {Macroscopic quantum coherence and entanglement in mechanical systems}} (\bibinfo {year} {2025}),\ \Eprint {https://arxiv.org/abs/2503.08324} {arXiv:2503.08324 [quant-ph]} \BibitemShut {NoStop}%
\bibitem [{\citenamefont {Luo}\ \emph {et~al.}(2025)\citenamefont {Luo}, \citenamefont {Diamandi}, \citenamefont {Li}, \citenamefont {Bi}, \citenamefont {Mason}, \citenamefont {Yoon}, \citenamefont {Guo}, \citenamefont {Tang}, \citenamefont {Behunin}, \citenamefont {Walker}, \citenamefont {Ahn},\ and\ \citenamefont {Rakich}}]{luo2025}%
  \BibitemOpen
  \bibfield  {author} {\bibinfo {author} {\bibfnamefont {Y.}~\bibnamefont {Luo}}, \bibinfo {author} {\bibfnamefont {H.~H.}\ \bibnamefont {Diamandi}}, \bibinfo {author} {\bibfnamefont {H.}~\bibnamefont {Li}}, \bibinfo {author} {\bibfnamefont {R.}~\bibnamefont {Bi}}, \bibinfo {author} {\bibfnamefont {D.}~\bibnamefont {Mason}}, \bibinfo {author} {\bibfnamefont {T.}~\bibnamefont {Yoon}}, \bibinfo {author} {\bibfnamefont {X.}~\bibnamefont {Guo}}, \bibinfo {author} {\bibfnamefont {H.}~\bibnamefont {Tang}}, \bibinfo {author} {\bibfnamefont {R.~O.}\ \bibnamefont {Behunin}}, \bibinfo {author} {\bibfnamefont {F.~J.}\ \bibnamefont {Walker}}, \bibinfo {author} {\bibfnamefont {C.}~\bibnamefont {Ahn}},\ and\ \bibinfo {author} {\bibfnamefont {P.~T.}\ \bibnamefont {Rakich}},\ }\href {https://doi.org/10.48550/arXiv.2504.07523} {\bibinfo {title} {Lifetime-limited {{Gigahertz-frequency Mechanical Oscillators}} with {{Millisecond Coherence Times}}}} (\bibinfo {year} {2025}),\ \Eprint {https://arxiv.org/abs/2504.07523}
  {arXiv:2504.07523 [quant-ph]} \BibitemShut {NoStop}%
\bibitem [{\citenamefont {Belles}\ \emph {et~al.}(2026)\citenamefont {Belles}, \citenamefont {Anferov}, \citenamefont {Deeg}, \citenamefont {Colicchio}, \citenamefont {Brooks}, \citenamefont {Schatteburg}, \citenamefont {Drimmer}, \citenamefont {Rodrigues}, \citenamefont {Benevides}, \citenamefont {Liffredo}, \citenamefont {Patidar}, \citenamefont {Pshyk}, \citenamefont {Fadel}, \citenamefont {Villanueva}, \citenamefont {Siol}, \citenamefont {Kirchmair},\ and\ \citenamefont {Chu}}]{Garcia-Belles2025}%
  \BibitemOpen
  \bibfield  {author} {\bibinfo {author} {\bibfnamefont {R.~G.}\ \bibnamefont {Belles}}, \bibinfo {author} {\bibfnamefont {A.}~\bibnamefont {Anferov}}, \bibinfo {author} {\bibfnamefont {L.~F.}\ \bibnamefont {Deeg}}, \bibinfo {author} {\bibfnamefont {L.}~\bibnamefont {Colicchio}}, \bibinfo {author} {\bibfnamefont {A.}~\bibnamefont {Brooks}}, \bibinfo {author} {\bibfnamefont {T.}~\bibnamefont {Schatteburg}}, \bibinfo {author} {\bibfnamefont {M.}~\bibnamefont {Drimmer}}, \bibinfo {author} {\bibfnamefont {I.~C.}\ \bibnamefont {Rodrigues}}, \bibinfo {author} {\bibfnamefont {R.}~\bibnamefont {Benevides}}, \bibinfo {author} {\bibfnamefont {M.}~\bibnamefont {Liffredo}}, \bibinfo {author} {\bibfnamefont {J.}~\bibnamefont {Patidar}}, \bibinfo {author} {\bibfnamefont {O.}~\bibnamefont {Pshyk}}, \bibinfo {author} {\bibfnamefont {M.}~\bibnamefont {Fadel}}, \bibinfo {author} {\bibfnamefont {L.~G.}\ \bibnamefont {Villanueva}}, \bibinfo {author} {\bibfnamefont {S.}~\bibnamefont {Siol}}, \bibinfo {author} {\bibfnamefont
  {G.}~\bibnamefont {Kirchmair}},\ and\ \bibinfo {author} {\bibfnamefont {Y.}~\bibnamefont {Chu}},\ }\href {https://arxiv.org/abs/2602.22117} {\bibinfo {title} {Loss mechanisms in high-coherence multimode mechanical resonators coupled to superconducting circuits}} (\bibinfo {year} {2026}),\ \Eprint {https://arxiv.org/abs/2602.22117} {arXiv:2602.22117 [quant-ph]} \BibitemShut {NoStop}%
\bibitem [{\citenamefont {von L{\"u}pke}\ \emph {et~al.}(2024)\citenamefont {von L{\"u}pke}, \citenamefont {Rodrigues}, \citenamefont {Yang}, \citenamefont {Fadel},\ and\ \citenamefont {Chu}}]{vonLuepke2024}%
  \BibitemOpen
  \bibfield  {author} {\bibinfo {author} {\bibfnamefont {U.}~\bibnamefont {von L{\"u}pke}}, \bibinfo {author} {\bibfnamefont {I.~C.}\ \bibnamefont {Rodrigues}}, \bibinfo {author} {\bibfnamefont {Y.}~\bibnamefont {Yang}}, \bibinfo {author} {\bibfnamefont {M.}~\bibnamefont {Fadel}},\ and\ \bibinfo {author} {\bibfnamefont {Y.}~\bibnamefont {Chu}},\ }\bibfield  {title} {\bibinfo {title} {Engineering multimode interactions in circuit quantum acoustodynamics},\ }\href {https://doi.org/10.1038/s41567-023-02377-w} {\bibfield  {journal} {\bibinfo  {journal} {Nature Physics}\ }\textbf {\bibinfo {volume} {20}},\ \bibinfo {pages} {564} (\bibinfo {year} {2024})}\BibitemShut {NoStop}%
\bibitem [{\citenamefont {Marti}\ \emph {et~al.}(2024)\citenamefont {Marti}, \citenamefont {von L{\"u}pke}, \citenamefont {Joshi}, \citenamefont {Yang}, \citenamefont {Bild}, \citenamefont {Omahen}, \citenamefont {Chu},\ and\ \citenamefont {Fadel}}]{Marti2024}%
  \BibitemOpen
  \bibfield  {author} {\bibinfo {author} {\bibfnamefont {S.}~\bibnamefont {Marti}}, \bibinfo {author} {\bibfnamefont {U.}~\bibnamefont {von L{\"u}pke}}, \bibinfo {author} {\bibfnamefont {O.}~\bibnamefont {Joshi}}, \bibinfo {author} {\bibfnamefont {Y.}~\bibnamefont {Yang}}, \bibinfo {author} {\bibfnamefont {M.}~\bibnamefont {Bild}}, \bibinfo {author} {\bibfnamefont {A.}~\bibnamefont {Omahen}}, \bibinfo {author} {\bibfnamefont {Y.}~\bibnamefont {Chu}},\ and\ \bibinfo {author} {\bibfnamefont {M.}~\bibnamefont {Fadel}},\ }\bibfield  {title} {\bibinfo {title} {Quantum squeezing in a nonlinear mechanical oscillator},\ }\href {https://doi.org/10.1038/s41567-024-02545-6} {\bibfield  {journal} {\bibinfo  {journal} {Nature Physics}\ }\textbf {\bibinfo {volume} {20}},\ \bibinfo {pages} {1448} (\bibinfo {year} {2024})}\BibitemShut {NoStop}%
\bibitem [{\citenamefont {Geerlings}\ \emph {et~al.}(2013)\citenamefont {Geerlings}, \citenamefont {Leghtas}, \citenamefont {Pop}, \citenamefont {Shankar}, \citenamefont {Frunzio}, \citenamefont {Schoelkopf}, \citenamefont {Mirrahimi},\ and\ \citenamefont {Devoret}}]{Geerlings2013}%
  \BibitemOpen
  \bibfield  {author} {\bibinfo {author} {\bibfnamefont {K.}~\bibnamefont {Geerlings}}, \bibinfo {author} {\bibfnamefont {Z.}~\bibnamefont {Leghtas}}, \bibinfo {author} {\bibfnamefont {I.~M.}\ \bibnamefont {Pop}}, \bibinfo {author} {\bibfnamefont {S.}~\bibnamefont {Shankar}}, \bibinfo {author} {\bibfnamefont {L.}~\bibnamefont {Frunzio}}, \bibinfo {author} {\bibfnamefont {R.~J.}\ \bibnamefont {Schoelkopf}}, \bibinfo {author} {\bibfnamefont {M.}~\bibnamefont {Mirrahimi}},\ and\ \bibinfo {author} {\bibfnamefont {M.~H.}\ \bibnamefont {Devoret}},\ }\bibfield  {title} {\bibinfo {title} {Demonstrating a driven reset protocol for a superconducting qubit},\ }\href {https://doi.org/10.1103/PhysRevLett.110.120501} {\bibfield  {journal} {\bibinfo  {journal} {Phys. Rev. Lett.}\ }\textbf {\bibinfo {volume} {110}},\ \bibinfo {pages} {120501} (\bibinfo {year} {2013})}\BibitemShut {NoStop}%
\bibitem [{\citenamefont {Jin}\ \emph {et~al.}(2015)\citenamefont {Jin}, \citenamefont {Kamal}, \citenamefont {Sears}, \citenamefont {Gudmundsen}, \citenamefont {Hover}, \citenamefont {Miloshi}, \citenamefont {Slattery}, \citenamefont {Yan}, \citenamefont {Yoder}, \citenamefont {Orlando}, \citenamefont {Gustavsson},\ and\ \citenamefont {Oliver}}]{Jin2015}%
  \BibitemOpen
  \bibfield  {author} {\bibinfo {author} {\bibfnamefont {X.~Y.}\ \bibnamefont {Jin}}, \bibinfo {author} {\bibfnamefont {A.}~\bibnamefont {Kamal}}, \bibinfo {author} {\bibfnamefont {A.~P.}\ \bibnamefont {Sears}}, \bibinfo {author} {\bibfnamefont {T.}~\bibnamefont {Gudmundsen}}, \bibinfo {author} {\bibfnamefont {D.}~\bibnamefont {Hover}}, \bibinfo {author} {\bibfnamefont {J.}~\bibnamefont {Miloshi}}, \bibinfo {author} {\bibfnamefont {R.}~\bibnamefont {Slattery}}, \bibinfo {author} {\bibfnamefont {F.}~\bibnamefont {Yan}}, \bibinfo {author} {\bibfnamefont {J.}~\bibnamefont {Yoder}}, \bibinfo {author} {\bibfnamefont {T.~P.}\ \bibnamefont {Orlando}}, \bibinfo {author} {\bibfnamefont {S.}~\bibnamefont {Gustavsson}},\ and\ \bibinfo {author} {\bibfnamefont {W.~D.}\ \bibnamefont {Oliver}},\ }\bibfield  {title} {\bibinfo {title} {Thermal and residual excited-state population in a 3d transmon qubit},\ }\href {https://doi.org/10.1103/PhysRevLett.114.240501} {\bibfield  {journal} {\bibinfo  {journal} {Phys. Rev. Lett.}\
  }\textbf {\bibinfo {volume} {114}},\ \bibinfo {pages} {240501} (\bibinfo {year} {2015})}\BibitemShut {NoStop}%
\bibitem [{\citenamefont {Yang}\ \emph {et~al.}(2024)\citenamefont {Yang}, \citenamefont {Kladaric}, \citenamefont {Drimmer}, \citenamefont {von L\"upke}, \citenamefont {Lenterman}, \citenamefont {Bus}, \citenamefont {Marti}, \citenamefont {Fadel},\ and\ \citenamefont {Chu}}]{Yang2024}%
  \BibitemOpen
  \bibfield  {author} {\bibinfo {author} {\bibfnamefont {Y.}~\bibnamefont {Yang}}, \bibinfo {author} {\bibfnamefont {I.}~\bibnamefont {Kladaric}}, \bibinfo {author} {\bibfnamefont {M.}~\bibnamefont {Drimmer}}, \bibinfo {author} {\bibfnamefont {U.}~\bibnamefont {von L\"upke}}, \bibinfo {author} {\bibfnamefont {D.}~\bibnamefont {Lenterman}}, \bibinfo {author} {\bibfnamefont {J.}~\bibnamefont {Bus}}, \bibinfo {author} {\bibfnamefont {S.}~\bibnamefont {Marti}}, \bibinfo {author} {\bibfnamefont {M.}~\bibnamefont {Fadel}},\ and\ \bibinfo {author} {\bibfnamefont {Y.}~\bibnamefont {Chu}},\ }\bibfield  {title} {\bibinfo {title} {A mechanical qubit},\ }\href {https://doi.org/10.1126/science.adr2464} {\bibfield  {journal} {\bibinfo  {journal} {Science}\ }\textbf {\bibinfo {volume} {386}},\ \bibinfo {pages} {783} (\bibinfo {year} {2024})}\BibitemShut {NoStop}%
\bibitem [{\citenamefont {Lambert}\ \emph {et~al.}(2024)\citenamefont {Lambert}, \citenamefont {Gigu\`{e}re}, \citenamefont {Menczel}, \citenamefont {Li}, \citenamefont {Hopf}, \citenamefont {Su\'{a}rez}, \citenamefont {Gali}, \citenamefont {Lishman}, \citenamefont {Gadhvi}, \citenamefont {Agarwal}, \citenamefont {Galicia}, \citenamefont {Shammah}, \citenamefont {Nation}, \citenamefont {Johansson}, \citenamefont {Ahmed}, \citenamefont {Cross}, \citenamefont {Pitchford},\ and\ \citenamefont {Nori}}]{QuTip2024}%
  \BibitemOpen
  \bibfield  {author} {\bibinfo {author} {\bibfnamefont {N.}~\bibnamefont {Lambert}}, \bibinfo {author} {\bibfnamefont {E.}~\bibnamefont {Gigu\`{e}re}}, \bibinfo {author} {\bibfnamefont {P.}~\bibnamefont {Menczel}}, \bibinfo {author} {\bibfnamefont {B.}~\bibnamefont {Li}}, \bibinfo {author} {\bibfnamefont {P.}~\bibnamefont {Hopf}}, \bibinfo {author} {\bibfnamefont {G.}~\bibnamefont {Su\'{a}rez}}, \bibinfo {author} {\bibfnamefont {M.}~\bibnamefont {Gali}}, \bibinfo {author} {\bibfnamefont {J.}~\bibnamefont {Lishman}}, \bibinfo {author} {\bibfnamefont {R.}~\bibnamefont {Gadhvi}}, \bibinfo {author} {\bibfnamefont {R.}~\bibnamefont {Agarwal}}, \bibinfo {author} {\bibfnamefont {A.}~\bibnamefont {Galicia}}, \bibinfo {author} {\bibfnamefont {N.}~\bibnamefont {Shammah}}, \bibinfo {author} {\bibfnamefont {P.}~\bibnamefont {Nation}}, \bibinfo {author} {\bibfnamefont {J.~R.}\ \bibnamefont {Johansson}}, \bibinfo {author} {\bibfnamefont {S.}~\bibnamefont {Ahmed}}, \bibinfo {author} {\bibfnamefont {S.}~\bibnamefont
  {Cross}}, \bibinfo {author} {\bibfnamefont {A.}~\bibnamefont {Pitchford}},\ and\ \bibinfo {author} {\bibfnamefont {F.}~\bibnamefont {Nori}},\ }\href {https://arxiv.org/abs/2412.04705} {\bibinfo {title} {Qutip 5: The quantum toolbox in python}} (\bibinfo {year} {2024}),\ \Eprint {https://arxiv.org/abs/2412.04705} {arXiv:2412.04705 [quant-ph]} \BibitemShut {NoStop}%
\bibitem [{\citenamefont {Collaboration}\ and\ \citenamefont {Collaboration}(2016)}]{Ligo2015}%
  \BibitemOpen
  \bibfield  {author} {\bibinfo {author} {\bibfnamefont {L.~S.}\ \bibnamefont {Collaboration}}\ and\ \bibinfo {author} {\bibfnamefont {V.}~\bibnamefont {Collaboration}},\ }\bibfield  {title} {\bibinfo {title} {Observation of gravitational waves from a binary black hole merger},\ }\href {https://doi.org/10.1103/PhysRevLett.116.061102} {\bibfield  {journal} {\bibinfo  {journal} {Phys. Rev. Lett.}\ }\textbf {\bibinfo {volume} {116}},\ \bibinfo {pages} {061102} (\bibinfo {year} {2016})}\BibitemShut {NoStop}%
\bibitem [{\citenamefont {Goryachev}\ \emph {et~al.}(2021)\citenamefont {Goryachev}, \citenamefont {Campbell}, \citenamefont {Heng}, \citenamefont {Galliou}, \citenamefont {Ivanov},\ and\ \citenamefont {Tobar}}]{Goryachev2021}%
  \BibitemOpen
  \bibfield  {author} {\bibinfo {author} {\bibfnamefont {M.}~\bibnamefont {Goryachev}}, \bibinfo {author} {\bibfnamefont {W.~M.}\ \bibnamefont {Campbell}}, \bibinfo {author} {\bibfnamefont {I.~S.}\ \bibnamefont {Heng}}, \bibinfo {author} {\bibfnamefont {S.}~\bibnamefont {Galliou}}, \bibinfo {author} {\bibfnamefont {E.~N.}\ \bibnamefont {Ivanov}},\ and\ \bibinfo {author} {\bibfnamefont {M.~E.}\ \bibnamefont {Tobar}},\ }\bibfield  {title} {\bibinfo {title} {Rare events detected with a bulk acoustic wave high frequency gravitational wave antenna},\ }\href {https://doi.org/10.1103/PhysRevLett.127.071102} {\bibfield  {journal} {\bibinfo  {journal} {Phys. Rev. Lett.}\ }\textbf {\bibinfo {volume} {127}},\ \bibinfo {pages} {071102} (\bibinfo {year} {2021})}\BibitemShut {NoStop}%
\bibitem [{\citenamefont {Aggarwal}\ \emph {et~al.}(2021)\citenamefont {Aggarwal}, \citenamefont {Aguiar}, \citenamefont {Bauswein}, \citenamefont {Cella}, \citenamefont {Clesse}, \citenamefont {Cruise}, \citenamefont {Domcke}, \citenamefont {Figueroa}, \citenamefont {Geraci}, \citenamefont {Goryachev}, \citenamefont {Grote}, \citenamefont {Hindmarsh}, \citenamefont {Muia}, \citenamefont {Mukund}, \citenamefont {Ottaway}, \citenamefont {Peloso}, \citenamefont {Quevedo}, \citenamefont {Ricciardone}, \citenamefont {Steinlechner}, \citenamefont {Steinlechner}, \citenamefont {Sun}, \citenamefont {Tobar}, \citenamefont {Torrenti}, \citenamefont {{\"U}nal},\ and\ \citenamefont {White}}]{Aggarwal2021}%
  \BibitemOpen
  \bibfield  {author} {\bibinfo {author} {\bibfnamefont {N.}~\bibnamefont {Aggarwal}}, \bibinfo {author} {\bibfnamefont {O.~D.}\ \bibnamefont {Aguiar}}, \bibinfo {author} {\bibfnamefont {A.}~\bibnamefont {Bauswein}}, \bibinfo {author} {\bibfnamefont {G.}~\bibnamefont {Cella}}, \bibinfo {author} {\bibfnamefont {S.}~\bibnamefont {Clesse}}, \bibinfo {author} {\bibfnamefont {A.~M.}\ \bibnamefont {Cruise}}, \bibinfo {author} {\bibfnamefont {V.}~\bibnamefont {Domcke}}, \bibinfo {author} {\bibfnamefont {D.~G.}\ \bibnamefont {Figueroa}}, \bibinfo {author} {\bibfnamefont {A.}~\bibnamefont {Geraci}}, \bibinfo {author} {\bibfnamefont {M.}~\bibnamefont {Goryachev}}, \bibinfo {author} {\bibfnamefont {H.}~\bibnamefont {Grote}}, \bibinfo {author} {\bibfnamefont {M.}~\bibnamefont {Hindmarsh}}, \bibinfo {author} {\bibfnamefont {F.}~\bibnamefont {Muia}}, \bibinfo {author} {\bibfnamefont {N.}~\bibnamefont {Mukund}}, \bibinfo {author} {\bibfnamefont {D.}~\bibnamefont {Ottaway}}, \bibinfo {author} {\bibfnamefont {M.}~\bibnamefont
  {Peloso}}, \bibinfo {author} {\bibfnamefont {F.}~\bibnamefont {Quevedo}}, \bibinfo {author} {\bibfnamefont {A.}~\bibnamefont {Ricciardone}}, \bibinfo {author} {\bibfnamefont {J.}~\bibnamefont {Steinlechner}}, \bibinfo {author} {\bibfnamefont {S.}~\bibnamefont {Steinlechner}}, \bibinfo {author} {\bibfnamefont {S.}~\bibnamefont {Sun}}, \bibinfo {author} {\bibfnamefont {M.~E.}\ \bibnamefont {Tobar}}, \bibinfo {author} {\bibfnamefont {F.}~\bibnamefont {Torrenti}}, \bibinfo {author} {\bibfnamefont {C.}~\bibnamefont {{\"U}nal}},\ and\ \bibinfo {author} {\bibfnamefont {G.}~\bibnamefont {White}},\ }\bibfield  {title} {\bibinfo {title} {Challenges and opportunities of gravitational-wave searches at mhz to ghz frequencies},\ }\href {https://doi.org/10.1007/s41114-021-00032-5} {\bibfield  {journal} {\bibinfo  {journal} {Living Reviews in Relativity}\ }\textbf {\bibinfo {volume} {24}},\ \bibinfo {pages} {4} (\bibinfo {year} {2021})}\BibitemShut {NoStop}%
\bibitem [{\citenamefont {Bozorgnia}\ \emph {et~al.}(2024)\citenamefont {Bozorgnia}, \citenamefont {Bramante}, \citenamefont {Cline}, \citenamefont {Curtin}, \citenamefont {McKeen}, \citenamefont {Morrissey}, \citenamefont {Ritz}, \citenamefont {Viel}, \citenamefont {Vincent},\ and\ \citenamefont {Zhang}}]{Bozorgnia2024}%
  \BibitemOpen
  \bibfield  {author} {\bibinfo {author} {\bibfnamefont {N.}~\bibnamefont {Bozorgnia}}, \bibinfo {author} {\bibfnamefont {J.}~\bibnamefont {Bramante}}, \bibinfo {author} {\bibfnamefont {J.~M.}\ \bibnamefont {Cline}}, \bibinfo {author} {\bibfnamefont {D.}~\bibnamefont {Curtin}}, \bibinfo {author} {\bibfnamefont {D.}~\bibnamefont {McKeen}}, \bibinfo {author} {\bibfnamefont {D.~E.}\ \bibnamefont {Morrissey}}, \bibinfo {author} {\bibfnamefont {A.}~\bibnamefont {Ritz}}, \bibinfo {author} {\bibfnamefont {S.}~\bibnamefont {Viel}}, \bibinfo {author} {\bibfnamefont {A.~C.}\ \bibnamefont {Vincent}},\ and\ \bibinfo {author} {\bibfnamefont {Y.}~\bibnamefont {Zhang}},\ }\href {https://arxiv.org/abs/2410.23454} {\bibinfo {title} {Dark matter candidates and searches}} (\bibinfo {year} {2024}),\ \Eprint {https://arxiv.org/abs/2410.23454} {arXiv:2410.23454 [hep-ph]} \BibitemShut {NoStop}%
\bibitem [{\citenamefont {Carney}\ \emph {et~al.}(2021)\citenamefont {Carney}, \citenamefont {Krnjaic}, \citenamefont {Moore}, \citenamefont {Regal}, \citenamefont {Afek}, \citenamefont {Bhave}, \citenamefont {Brubaker}, \citenamefont {Corbitt}, \citenamefont {Cripe}, \citenamefont {Crisosto}, \citenamefont {Geraci}, \citenamefont {Ghosh}, \citenamefont {Harris}, \citenamefont {Hook}, \citenamefont {Kolb}, \citenamefont {Kunjummen}, \citenamefont {Lang}, \citenamefont {Li}, \citenamefont {Lin}, \citenamefont {Liu}, \citenamefont {Lykken}, \citenamefont {Magrini}, \citenamefont {Manley}, \citenamefont {Matsumoto}, \citenamefont {Monte}, \citenamefont {Monteiro}, \citenamefont {Purdy}, \citenamefont {Riedel}, \citenamefont {Singh}, \citenamefont {Singh}, \citenamefont {Sinha}, \citenamefont {Taylor}, \citenamefont {Qin}, \citenamefont {Wilson},\ and\ \citenamefont {Zhao}}]{carney2021}%
  \BibitemOpen
  \bibfield  {author} {\bibinfo {author} {\bibfnamefont {D.}~\bibnamefont {Carney}}, \bibinfo {author} {\bibfnamefont {G.}~\bibnamefont {Krnjaic}}, \bibinfo {author} {\bibfnamefont {D.~C.}\ \bibnamefont {Moore}}, \bibinfo {author} {\bibfnamefont {C.~A.}\ \bibnamefont {Regal}}, \bibinfo {author} {\bibfnamefont {G.}~\bibnamefont {Afek}}, \bibinfo {author} {\bibfnamefont {S.}~\bibnamefont {Bhave}}, \bibinfo {author} {\bibfnamefont {B.}~\bibnamefont {Brubaker}}, \bibinfo {author} {\bibfnamefont {T.}~\bibnamefont {Corbitt}}, \bibinfo {author} {\bibfnamefont {J.}~\bibnamefont {Cripe}}, \bibinfo {author} {\bibfnamefont {N.}~\bibnamefont {Crisosto}}, \bibinfo {author} {\bibfnamefont {A.}~\bibnamefont {Geraci}}, \bibinfo {author} {\bibfnamefont {S.}~\bibnamefont {Ghosh}}, \bibinfo {author} {\bibfnamefont {J.~G.~E.}\ \bibnamefont {Harris}}, \bibinfo {author} {\bibfnamefont {A.}~\bibnamefont {Hook}}, \bibinfo {author} {\bibfnamefont {E.~W.}\ \bibnamefont {Kolb}}, \bibinfo {author} {\bibfnamefont {J.}~\bibnamefont
  {Kunjummen}}, \bibinfo {author} {\bibfnamefont {R.~F.}\ \bibnamefont {Lang}}, \bibinfo {author} {\bibfnamefont {T.}~\bibnamefont {Li}}, \bibinfo {author} {\bibfnamefont {T.}~\bibnamefont {Lin}}, \bibinfo {author} {\bibfnamefont {Z.}~\bibnamefont {Liu}}, \bibinfo {author} {\bibfnamefont {J.}~\bibnamefont {Lykken}}, \bibinfo {author} {\bibfnamefont {L.}~\bibnamefont {Magrini}}, \bibinfo {author} {\bibfnamefont {J.}~\bibnamefont {Manley}}, \bibinfo {author} {\bibfnamefont {N.}~\bibnamefont {Matsumoto}}, \bibinfo {author} {\bibfnamefont {A.}~\bibnamefont {Monte}}, \bibinfo {author} {\bibfnamefont {F.}~\bibnamefont {Monteiro}}, \bibinfo {author} {\bibfnamefont {T.}~\bibnamefont {Purdy}}, \bibinfo {author} {\bibfnamefont {C.~J.}\ \bibnamefont {Riedel}}, \bibinfo {author} {\bibfnamefont {R.}~\bibnamefont {Singh}}, \bibinfo {author} {\bibfnamefont {S.}~\bibnamefont {Singh}}, \bibinfo {author} {\bibfnamefont {K.}~\bibnamefont {Sinha}}, \bibinfo {author} {\bibfnamefont {J.~M.}\ \bibnamefont {Taylor}}, \bibinfo
  {author} {\bibfnamefont {J.}~\bibnamefont {Qin}}, \bibinfo {author} {\bibfnamefont {D.~J.}\ \bibnamefont {Wilson}},\ and\ \bibinfo {author} {\bibfnamefont {Y.}~\bibnamefont {Zhao}},\ }\bibfield  {title} {\bibinfo {title} {Mechanical quantum sensing in the search for dark matter},\ }\href {https://doi.org/10.1088/2058-9565/abcfcd} {\bibfield  {journal} {\bibinfo  {journal} {Quantum Science and Technology}\ }\textbf {\bibinfo {volume} {6}},\ \bibinfo {pages} {024002} (\bibinfo {year} {2021})}\BibitemShut {NoStop}%
\bibitem [{\citenamefont {Martin}\ \emph {et~al.}(2004)\citenamefont {Martin}, \citenamefont {Muralt}, \citenamefont {Dubois},\ and\ \citenamefont {Pezous}}]{martin2004}%
  \BibitemOpen
  \bibfield  {author} {\bibinfo {author} {\bibfnamefont {F.}~\bibnamefont {Martin}}, \bibinfo {author} {\bibfnamefont {P.}~\bibnamefont {Muralt}}, \bibinfo {author} {\bibfnamefont {M.-A.}\ \bibnamefont {Dubois}},\ and\ \bibinfo {author} {\bibfnamefont {A.}~\bibnamefont {Pezous}},\ }\bibfield  {title} {\bibinfo {title} {Thickness dependence of the properties of highly {\emph{c}} -axis textured {{AlN}} thin films},\ }\href {https://doi.org/10.1116/1.1649343} {\bibfield  {journal} {\bibinfo  {journal} {Journal of Vacuum Science \& Technology A: Vacuum, Surfaces, and Films}\ }\textbf {\bibinfo {volume} {22}},\ \bibinfo {pages} {361} (\bibinfo {year} {2004})}\BibitemShut {NoStop}%
\bibitem [{\citenamefont {O'Hare}(2024)}]{AxionLimits}%
  \BibitemOpen
  \bibfield  {author} {\bibinfo {author} {\bibfnamefont {C.}~\bibnamefont {O'Hare}},\ }\href@noop {} {\bibinfo {title} {{AxionLimits}: Code to produce axion limit plots from experimental and astrophysical data}},\ \bibinfo {howpublished} {\url{https://github.com/cajohare/AxionLimits}} (\bibinfo {year} {2024}),\ \bibinfo {note} {accessed: 2025-06-11}\BibitemShut {NoStop}%
\bibitem [{\citenamefont {Lee}\ \emph {et~al.}(2023)\citenamefont {Lee}, \citenamefont {Guo}, \citenamefont {Cleland}, \citenamefont {Wollack}, \citenamefont {Gruenke}, \citenamefont {Makihara}, \citenamefont {Wang}, \citenamefont {Rajabzadeh}, \citenamefont {Jiang}, \citenamefont {Mayor}, \citenamefont {{Arrangoiz-Arriola}}, \citenamefont {Sarabalis},\ and\ \citenamefont {{Safavi-Naeini}}}]{lee2023}%
  \BibitemOpen
  \bibfield  {author} {\bibinfo {author} {\bibfnamefont {N.~R.}\ \bibnamefont {Lee}}, \bibinfo {author} {\bibfnamefont {Y.}~\bibnamefont {Guo}}, \bibinfo {author} {\bibfnamefont {A.~Y.}\ \bibnamefont {Cleland}}, \bibinfo {author} {\bibfnamefont {E.~A.}\ \bibnamefont {Wollack}}, \bibinfo {author} {\bibfnamefont {R.~G.}\ \bibnamefont {Gruenke}}, \bibinfo {author} {\bibfnamefont {T.}~\bibnamefont {Makihara}}, \bibinfo {author} {\bibfnamefont {Z.}~\bibnamefont {Wang}}, \bibinfo {author} {\bibfnamefont {T.}~\bibnamefont {Rajabzadeh}}, \bibinfo {author} {\bibfnamefont {W.}~\bibnamefont {Jiang}}, \bibinfo {author} {\bibfnamefont {F.~M.}\ \bibnamefont {Mayor}}, \bibinfo {author} {\bibfnamefont {P.}~\bibnamefont {{Arrangoiz-Arriola}}}, \bibinfo {author} {\bibfnamefont {C.~J.}\ \bibnamefont {Sarabalis}},\ and\ \bibinfo {author} {\bibfnamefont {A.~H.}\ \bibnamefont {{Safavi-Naeini}}},\ }\bibfield  {title} {\bibinfo {title} {Strong {{Dispersive Coupling Between}} a {{Mechanical Resonator}} and a {{Fluxonium
  Superconducting Qubit}}},\ }\href {https://doi.org/10.1103/PRXQuantum.4.040342} {\bibfield  {journal} {\bibinfo  {journal} {PRX Quantum}\ }\textbf {\bibinfo {volume} {4}},\ \bibinfo {pages} {040342} (\bibinfo {year} {2023})}\BibitemShut {NoStop}%
\bibitem [{\citenamefont {{Najera-Santos}}\ \emph {et~al.}(2024)\citenamefont {{Najera-Santos}}, \citenamefont {Rousseau}, \citenamefont {Gerashchenko}, \citenamefont {Patange}, \citenamefont {Riva}, \citenamefont {Villiers}, \citenamefont {Briant}, \citenamefont {Cohadon}, \citenamefont {Heidmann}, \citenamefont {Palomo}, \citenamefont {Rosticher}, \citenamefont {{le Sueur}}, \citenamefont {Sarlette}, \citenamefont {Smith}, \citenamefont {Leghtas}, \citenamefont {Flurin}, \citenamefont {Jacqmin},\ and\ \citenamefont {Del{\'e}glise}}]{najera2024}%
  \BibitemOpen
  \bibfield  {author} {\bibinfo {author} {\bibfnamefont {B.-L.}\ \bibnamefont {{Najera-Santos}}}, \bibinfo {author} {\bibfnamefont {R.}~\bibnamefont {Rousseau}}, \bibinfo {author} {\bibfnamefont {K.}~\bibnamefont {Gerashchenko}}, \bibinfo {author} {\bibfnamefont {H.}~\bibnamefont {Patange}}, \bibinfo {author} {\bibfnamefont {A.}~\bibnamefont {Riva}}, \bibinfo {author} {\bibfnamefont {M.}~\bibnamefont {Villiers}}, \bibinfo {author} {\bibfnamefont {T.}~\bibnamefont {Briant}}, \bibinfo {author} {\bibfnamefont {P.-F.}\ \bibnamefont {Cohadon}}, \bibinfo {author} {\bibfnamefont {A.}~\bibnamefont {Heidmann}}, \bibinfo {author} {\bibfnamefont {J.}~\bibnamefont {Palomo}}, \bibinfo {author} {\bibfnamefont {M.}~\bibnamefont {Rosticher}}, \bibinfo {author} {\bibfnamefont {H.}~\bibnamefont {{le Sueur}}}, \bibinfo {author} {\bibfnamefont {A.}~\bibnamefont {Sarlette}}, \bibinfo {author} {\bibfnamefont {W.~C.}\ \bibnamefont {Smith}}, \bibinfo {author} {\bibfnamefont {Z.}~\bibnamefont {Leghtas}}, \bibinfo {author}
  {\bibfnamefont {E.}~\bibnamefont {Flurin}}, \bibinfo {author} {\bibfnamefont {T.}~\bibnamefont {Jacqmin}},\ and\ \bibinfo {author} {\bibfnamefont {S.}~\bibnamefont {Del{\'e}glise}},\ }\bibfield  {title} {\bibinfo {title} {High-{{Sensitivity}} ac-{{Charge Detection}} with a {{MHz-Frequency Fluxonium Qubit}}},\ }\href {https://doi.org/10.1103/PhysRevX.14.011007} {\bibfield  {journal} {\bibinfo  {journal} {Physical Review X}\ }\textbf {\bibinfo {volume} {14}},\ \bibinfo {pages} {011007} (\bibinfo {year} {2024})}\BibitemShut {NoStop}%
\bibitem [{\citenamefont {Galliou}\ \emph {et~al.}(2013)\citenamefont {Galliou}, \citenamefont {Goryachev}, \citenamefont {Bourquin}, \citenamefont {Abb{\'e}}, \citenamefont {Aubry},\ and\ \citenamefont {Tobar}}]{galliou2013}%
  \BibitemOpen
  \bibfield  {author} {\bibinfo {author} {\bibfnamefont {S.}~\bibnamefont {Galliou}}, \bibinfo {author} {\bibfnamefont {M.}~\bibnamefont {Goryachev}}, \bibinfo {author} {\bibfnamefont {R.}~\bibnamefont {Bourquin}}, \bibinfo {author} {\bibfnamefont {P.}~\bibnamefont {Abb{\'e}}}, \bibinfo {author} {\bibfnamefont {J.~P.}\ \bibnamefont {Aubry}},\ and\ \bibinfo {author} {\bibfnamefont {M.~E.}\ \bibnamefont {Tobar}},\ }\bibfield  {title} {\bibinfo {title} {Extremely low loss phonon-trapping cryogenic acoustic cavities for future physical experiments.},\ }\href {https://doi.org/10.1038/srep02132} {\bibfield  {journal} {\bibinfo  {journal} {Scientific reports}\ }\textbf {\bibinfo {volume} {3}},\ \bibinfo {pages} {2132} (\bibinfo {year} {2013})}\BibitemShut {NoStop}%
\bibitem [{\citenamefont {Ghirardi}\ \emph {et~al.}(1986)\citenamefont {Ghirardi}, \citenamefont {Rimini},\ and\ \citenamefont {Weber}}]{GhirardiPRD86}%
  \BibitemOpen
  \bibfield  {author} {\bibinfo {author} {\bibfnamefont {G.~C.}\ \bibnamefont {Ghirardi}}, \bibinfo {author} {\bibfnamefont {A.}~\bibnamefont {Rimini}},\ and\ \bibinfo {author} {\bibfnamefont {T.}~\bibnamefont {Weber}},\ }\bibfield  {title} {\bibinfo {title} {Unified dynamics for microscopic and macroscopic systems},\ }\href {https://doi.org/10.1103/PhysRevD.34.470} {\bibfield  {journal} {\bibinfo  {journal} {Phys. Rev. D}\ }\textbf {\bibinfo {volume} {34}},\ \bibinfo {pages} {470} (\bibinfo {year} {1986})}\BibitemShut {NoStop}%
\bibitem [{\citenamefont {Vinante}\ \emph {et~al.}(2020)\citenamefont {Vinante}, \citenamefont {Carlesso}, \citenamefont {Bassi}, \citenamefont {Chiasera}, \citenamefont {Varas}, \citenamefont {Falferi}, \citenamefont {Margesin}, \citenamefont {Mezzena},\ and\ \citenamefont {Ulbricht}}]{VinantePRL20}%
  \BibitemOpen
  \bibfield  {author} {\bibinfo {author} {\bibfnamefont {A.}~\bibnamefont {Vinante}}, \bibinfo {author} {\bibfnamefont {M.}~\bibnamefont {Carlesso}}, \bibinfo {author} {\bibfnamefont {A.}~\bibnamefont {Bassi}}, \bibinfo {author} {\bibfnamefont {A.}~\bibnamefont {Chiasera}}, \bibinfo {author} {\bibfnamefont {S.}~\bibnamefont {Varas}}, \bibinfo {author} {\bibfnamefont {P.}~\bibnamefont {Falferi}}, \bibinfo {author} {\bibfnamefont {B.}~\bibnamefont {Margesin}}, \bibinfo {author} {\bibfnamefont {R.}~\bibnamefont {Mezzena}},\ and\ \bibinfo {author} {\bibfnamefont {H.}~\bibnamefont {Ulbricht}},\ }\bibfield  {title} {\bibinfo {title} {Narrowing the parameter space of collapse models with ultracold layered force sensors},\ }\href {https://doi.org/10.1103/PhysRevLett.125.100404} {\bibfield  {journal} {\bibinfo  {journal} {Phys. Rev. Lett.}\ }\textbf {\bibinfo {volume} {125}},\ \bibinfo {pages} {100404} (\bibinfo {year} {2020})}\BibitemShut {NoStop}%
\bibitem [{\citenamefont {Bosso}\ \emph {et~al.}(2023)\citenamefont {Bosso}, \citenamefont {Gaetano~Luciano}, \citenamefont {Petruzziello},\ and\ \citenamefont {Wagner}}]{Bosso2023}%
  \BibitemOpen
  \bibfield  {author} {\bibinfo {author} {\bibfnamefont {P.}~\bibnamefont {Bosso}}, \bibinfo {author} {\bibfnamefont {G.}~\bibnamefont {Gaetano~Luciano}}, \bibinfo {author} {\bibfnamefont {L.}~\bibnamefont {Petruzziello}},\ and\ \bibinfo {author} {\bibfnamefont {F.}~\bibnamefont {Wagner}},\ }\bibfield  {title} {\bibinfo {title} {30 years in: Quo vadis generalized uncertainty principle?},\ }\href {https://doi.org/10.1088/1361-6382/acf021} {\bibfield  {journal} {\bibinfo  {journal} {Classical and Quantum Gravity}\ }\textbf {\bibinfo {volume} {40}},\ \bibinfo {pages} {195014} (\bibinfo {year} {2023})}\BibitemShut {NoStop}%
\bibitem [{\citenamefont {Percival}(1995)}]{Porcino95}%
  \BibitemOpen
  \bibfield  {author} {\bibinfo {author} {\bibfnamefont {I.~C.}\ \bibnamefont {Percival}},\ }\bibfield  {title} {\bibinfo {title} {Quantum spacetime fluctuations and primary state diffusion},\ }\href {https://doi.org/10.1098/rspa.1995.0139} {\bibfield  {journal} {\bibinfo  {journal} {Proceedings of the Royal Society of London. Series A: Mathematical and Physical Sciences}\ }\textbf {\bibinfo {volume} {451}},\ \bibinfo {pages} {503} (\bibinfo {year} {1995})}\BibitemShut {NoStop}%
\bibitem [{\citenamefont {Donadi}\ and\ \citenamefont {Fadel}(2025)}]{DonadiPRD25}%
  \BibitemOpen
  \bibfield  {author} {\bibinfo {author} {\bibfnamefont {S.}~\bibnamefont {Donadi}}\ and\ \bibinfo {author} {\bibfnamefont {M.}~\bibnamefont {Fadel}},\ }\bibfield  {title} {\bibinfo {title} {Quantum gravitational decoherence of a mechanical oscillator from spacetime fluctuations},\ }\href {https://doi.org/10.1103/PhysRevD.111.026009} {\bibfield  {journal} {\bibinfo  {journal} {Phys. Rev. D}\ }\textbf {\bibinfo {volume} {111}},\ \bibinfo {pages} {026009} (\bibinfo {year} {2025})}\BibitemShut {NoStop}%
\bibitem [{\citenamefont {Roberts}\ \emph {et~al.}(2017)\citenamefont {Roberts}, \citenamefont {Blewitt}, \citenamefont {Dailey}, \citenamefont {Murphy}, \citenamefont {Pospelov}, \citenamefont {Rollings}, \citenamefont {Sherman}, \citenamefont {Williams},\ and\ \citenamefont {Derevianko}}]{Roberts2017}%
  \BibitemOpen
  \bibfield  {author} {\bibinfo {author} {\bibfnamefont {B.~M.}\ \bibnamefont {Roberts}}, \bibinfo {author} {\bibfnamefont {G.}~\bibnamefont {Blewitt}}, \bibinfo {author} {\bibfnamefont {C.}~\bibnamefont {Dailey}}, \bibinfo {author} {\bibfnamefont {M.}~\bibnamefont {Murphy}}, \bibinfo {author} {\bibfnamefont {M.}~\bibnamefont {Pospelov}}, \bibinfo {author} {\bibfnamefont {A.}~\bibnamefont {Rollings}}, \bibinfo {author} {\bibfnamefont {J.}~\bibnamefont {Sherman}}, \bibinfo {author} {\bibfnamefont {W.}~\bibnamefont {Williams}},\ and\ \bibinfo {author} {\bibfnamefont {A.}~\bibnamefont {Derevianko}},\ }\bibfield  {title} {\bibinfo {title} {Search for domain wall dark matter with atomic clocks on board global positioning system satellites},\ }\href {https://doi.org/10.1038/s41467-017-01440-4} {\bibfield  {journal} {\bibinfo  {journal} {Nature Communications}\ }\textbf {\bibinfo {volume} {8}},\ \bibinfo {pages} {1195} (\bibinfo {year} {2017})}\BibitemShut {NoStop}%
\bibitem [{\citenamefont {Hogan}(2010)}]{Hogan2010}%
  \BibitemOpen
  \bibfield  {author} {\bibinfo {author} {\bibfnamefont {C.~J.}\ \bibnamefont {Hogan}},\ }\href {https://arxiv.org/abs/0905.4803} {\bibinfo {title} {Holographic noise in interferometers}} (\bibinfo {year} {2010}),\ \Eprint {https://arxiv.org/abs/0905.4803} {arXiv:0905.4803 [gr-qc]} \BibitemShut {NoStop}%
\bibitem [{\citenamefont {Bose}\ \emph {et~al.}(2025)\citenamefont {Bose}, \citenamefont {Fuentes}, \citenamefont {Geraci}, \citenamefont {Khan}, \citenamefont {Qvarfort}, \citenamefont {Rademacher}, \citenamefont {Rashid}, \citenamefont {Toro\ifmmode~\check{s}\else \v{s}\fi{}}, \citenamefont {Ulbricht},\ and\ \citenamefont {Wanjura}}]{Bose2024}%
  \BibitemOpen
  \bibfield  {author} {\bibinfo {author} {\bibfnamefont {S.}~\bibnamefont {Bose}}, \bibinfo {author} {\bibfnamefont {I.}~\bibnamefont {Fuentes}}, \bibinfo {author} {\bibfnamefont {A.~A.}\ \bibnamefont {Geraci}}, \bibinfo {author} {\bibfnamefont {S.~M.}\ \bibnamefont {Khan}}, \bibinfo {author} {\bibfnamefont {S.}~\bibnamefont {Qvarfort}}, \bibinfo {author} {\bibfnamefont {M.}~\bibnamefont {Rademacher}}, \bibinfo {author} {\bibfnamefont {M.}~\bibnamefont {Rashid}}, \bibinfo {author} {\bibfnamefont {M.}~\bibnamefont {Toro\ifmmode~\check{s}\else \v{s}\fi{}}}, \bibinfo {author} {\bibfnamefont {H.}~\bibnamefont {Ulbricht}},\ and\ \bibinfo {author} {\bibfnamefont {C.~C.}\ \bibnamefont {Wanjura}},\ }\bibfield  {title} {\bibinfo {title} {Massive quantum systems as interfaces of quantum mechanics and gravity},\ }\href {https://doi.org/10.1103/RevModPhys.97.015003} {\bibfield  {journal} {\bibinfo  {journal} {Rev. Mod. Phys.}\ }\textbf {\bibinfo {volume} {97}},\ \bibinfo {pages} {015003} (\bibinfo {year}
  {2025})}\BibitemShut {NoStop}%
\bibitem [{\citenamefont {Clifton}\ \emph {et~al.}(2012)\citenamefont {Clifton}, \citenamefont {Ferreira}, \citenamefont {Padilla},\ and\ \citenamefont {Skordis}}]{Clifton2012}%
  \BibitemOpen
  \bibfield  {author} {\bibinfo {author} {\bibfnamefont {T.}~\bibnamefont {Clifton}}, \bibinfo {author} {\bibfnamefont {P.~G.}\ \bibnamefont {Ferreira}}, \bibinfo {author} {\bibfnamefont {A.}~\bibnamefont {Padilla}},\ and\ \bibinfo {author} {\bibfnamefont {C.}~\bibnamefont {Skordis}},\ }\bibfield  {title} {\bibinfo {title} {Modified gravity and cosmology},\ }\href {https://doi.org/https://doi.org/10.1016/j.physrep.2012.01.001} {\bibfield  {journal} {\bibinfo  {journal} {Physics Reports}\ }\textbf {\bibinfo {volume} {513}},\ \bibinfo {pages} {1} (\bibinfo {year} {2012})},\ \bibinfo {note} {modified Gravity and Cosmology}\BibitemShut {NoStop}%
\bibitem [{\citenamefont {Tobar}\ \emph {et~al.}(2024)\citenamefont {Tobar}, \citenamefont {Manikandan}, \citenamefont {Beitel},\ and\ \citenamefont {Pikovski}}]{Tobar2024}%
  \BibitemOpen
  \bibfield  {author} {\bibinfo {author} {\bibfnamefont {G.}~\bibnamefont {Tobar}}, \bibinfo {author} {\bibfnamefont {S.~K.}\ \bibnamefont {Manikandan}}, \bibinfo {author} {\bibfnamefont {T.}~\bibnamefont {Beitel}},\ and\ \bibinfo {author} {\bibfnamefont {I.}~\bibnamefont {Pikovski}},\ }\bibfield  {title} {\bibinfo {title} {Detecting single gravitons with quantum sensing},\ }\href {https://doi.org/10.1038/s41467-024-51420-8} {\bibfield  {journal} {\bibinfo  {journal} {Nature Communications}\ }\textbf {\bibinfo {volume} {15}},\ \bibinfo {pages} {7229} (\bibinfo {year} {2024})}\BibitemShut {NoStop}%
\bibitem [{\citenamefont {Bose}\ \emph {et~al.}(2017)\citenamefont {Bose}, \citenamefont {Mazumdar}, \citenamefont {Morley}, \citenamefont {Ulbricht}, \citenamefont {Toro\ifmmode~\check{s}\else \v{s}\fi{}}, \citenamefont {Paternostro}, \citenamefont {Geraci}, \citenamefont {Barker}, \citenamefont {Kim},\ and\ \citenamefont {Milburn}}]{Bose2017}%
  \BibitemOpen
  \bibfield  {author} {\bibinfo {author} {\bibfnamefont {S.}~\bibnamefont {Bose}}, \bibinfo {author} {\bibfnamefont {A.}~\bibnamefont {Mazumdar}}, \bibinfo {author} {\bibfnamefont {G.~W.}\ \bibnamefont {Morley}}, \bibinfo {author} {\bibfnamefont {H.}~\bibnamefont {Ulbricht}}, \bibinfo {author} {\bibfnamefont {M.}~\bibnamefont {Toro\ifmmode~\check{s}\else \v{s}\fi{}}}, \bibinfo {author} {\bibfnamefont {M.}~\bibnamefont {Paternostro}}, \bibinfo {author} {\bibfnamefont {A.~A.}\ \bibnamefont {Geraci}}, \bibinfo {author} {\bibfnamefont {P.~F.}\ \bibnamefont {Barker}}, \bibinfo {author} {\bibfnamefont {M.~S.}\ \bibnamefont {Kim}},\ and\ \bibinfo {author} {\bibfnamefont {G.}~\bibnamefont {Milburn}},\ }\bibfield  {title} {\bibinfo {title} {Spin entanglement witness for quantum gravity},\ }\href {https://doi.org/10.1103/PhysRevLett.119.240401} {\bibfield  {journal} {\bibinfo  {journal} {Phys. Rev. Lett.}\ }\textbf {\bibinfo {volume} {119}},\ \bibinfo {pages} {240401} (\bibinfo {year} {2017})}\BibitemShut {NoStop}%
\bibitem [{\citenamefont {Marletto}\ and\ \citenamefont {Vedral}(2017)}]{Marletto2017}%
  \BibitemOpen
  \bibfield  {author} {\bibinfo {author} {\bibfnamefont {C.}~\bibnamefont {Marletto}}\ and\ \bibinfo {author} {\bibfnamefont {V.}~\bibnamefont {Vedral}},\ }\bibfield  {title} {\bibinfo {title} {Gravitationally induced entanglement between two massive particles is sufficient evidence of quantum effects in gravity},\ }\href {https://doi.org/10.1103/PhysRevLett.119.240402} {\bibfield  {journal} {\bibinfo  {journal} {Phys. Rev. Lett.}\ }\textbf {\bibinfo {volume} {119}},\ \bibinfo {pages} {240402} (\bibinfo {year} {2017})}\BibitemShut {NoStop}%
\bibitem [{\citenamefont {Bassi}\ \emph {et~al.}(2017)\citenamefont {Bassi}, \citenamefont {Gro\ss{}ardt},\ and\ \citenamefont {Ulbricht}}]{Bassi2017}%
  \BibitemOpen
  \bibfield  {author} {\bibinfo {author} {\bibfnamefont {A.}~\bibnamefont {Bassi}}, \bibinfo {author} {\bibfnamefont {A.}~\bibnamefont {Gro\ss{}ardt}},\ and\ \bibinfo {author} {\bibfnamefont {H.}~\bibnamefont {Ulbricht}},\ }\bibfield  {title} {\bibinfo {title} {Gravitational decoherence},\ }\href {https://doi.org/10.1088/1361-6382/aa864f} {\bibfield  {journal} {\bibinfo  {journal} {Classical and Quantum Gravity}\ }\textbf {\bibinfo {volume} {34}},\ \bibinfo {pages} {193002} (\bibinfo {year} {2017})}\BibitemShut {NoStop}%
\bibitem [{\citenamefont {(ed.)}\ and\ \citenamefont {(ed.)}(2011)}]{ChapelHill2011}%
  \BibitemOpen
  \bibfield  {author} {\bibinfo {author} {\bibfnamefont {C.~M.~D.}\ \bibnamefont {(ed.)}}\ and\ \bibinfo {author} {\bibfnamefont {D.~R.}\ \bibnamefont {(ed.)}},\ }\href {http://lib.ugent.be/catalog/ebk01:9870000000000549} {\emph {\bibinfo {title} {The Role of Gravitation in Physics: Report from the 1957 Chapel Hill Conference}}}\ (\bibinfo  {publisher} {Edition Open Access 2011},\ \bibinfo {year} {2011})\BibitemShut {NoStop}%
\bibitem [{\citenamefont {Aamir}\ \emph {et~al.}(2025)\citenamefont {Aamir}, \citenamefont {Jamet~Suria}, \citenamefont {Mar{\'i}n~Guzm{\'a}n}, \citenamefont {Castillo-Moreno}, \citenamefont {Epstein}, \citenamefont {Yunger~Halpern},\ and\ \citenamefont {Gasparinetti}}]{Aamir2025}%
  \BibitemOpen
  \bibfield  {author} {\bibinfo {author} {\bibfnamefont {M.~A.}\ \bibnamefont {Aamir}}, \bibinfo {author} {\bibfnamefont {P.}~\bibnamefont {Jamet~Suria}}, \bibinfo {author} {\bibfnamefont {J.~A.}\ \bibnamefont {Mar{\'i}n~Guzm{\'a}n}}, \bibinfo {author} {\bibfnamefont {C.}~\bibnamefont {Castillo-Moreno}}, \bibinfo {author} {\bibfnamefont {J.~M.}\ \bibnamefont {Epstein}}, \bibinfo {author} {\bibfnamefont {N.}~\bibnamefont {Yunger~Halpern}},\ and\ \bibinfo {author} {\bibfnamefont {S.}~\bibnamefont {Gasparinetti}},\ }\bibfield  {title} {\bibinfo {title} {Thermally driven quantum refrigerator autonomously resets a superconducting qubit},\ }\href {https://doi.org/10.1038/s41567-024-02708-5} {\bibfield  {journal} {\bibinfo  {journal} {Nature Physics}\ }\textbf {\bibinfo {volume} {21}},\ \bibinfo {pages} {318} (\bibinfo {year} {2025})}\BibitemShut {NoStop}%
\bibitem [{\citenamefont {Somoroff}\ \emph {et~al.}(2023)\citenamefont {Somoroff}, \citenamefont {Ficheux}, \citenamefont {Mencia}, \citenamefont {Xiong}, \citenamefont {Kuzmin},\ and\ \citenamefont {Manucharyan}}]{Somoroff2023}%
  \BibitemOpen
  \bibfield  {author} {\bibinfo {author} {\bibfnamefont {A.}~\bibnamefont {Somoroff}}, \bibinfo {author} {\bibfnamefont {Q.}~\bibnamefont {Ficheux}}, \bibinfo {author} {\bibfnamefont {R.~A.}\ \bibnamefont {Mencia}}, \bibinfo {author} {\bibfnamefont {H.}~\bibnamefont {Xiong}}, \bibinfo {author} {\bibfnamefont {R.}~\bibnamefont {Kuzmin}},\ and\ \bibinfo {author} {\bibfnamefont {V.~E.}\ \bibnamefont {Manucharyan}},\ }\bibfield  {title} {\bibinfo {title} {Millisecond coherence in a superconducting qubit},\ }\href {https://doi.org/10.1103/PhysRevLett.130.267001} {\bibfield  {journal} {\bibinfo  {journal} {Phys. Rev. Lett.}\ }\textbf {\bibinfo {volume} {130}},\ \bibinfo {pages} {267001} (\bibinfo {year} {2023})}\BibitemShut {NoStop}%
\bibitem [{\citenamefont {Keith}\ \emph {et~al.}(2022)\citenamefont {Keith}, \citenamefont {Chung}, \citenamefont {Kranz}, \citenamefont {Thorgrimsson}, \citenamefont {Gorman},\ and\ \citenamefont {Simmons}}]{Keith2022}%
  \BibitemOpen
  \bibfield  {author} {\bibinfo {author} {\bibfnamefont {D.}~\bibnamefont {Keith}}, \bibinfo {author} {\bibfnamefont {Y.}~\bibnamefont {Chung}}, \bibinfo {author} {\bibfnamefont {L.}~\bibnamefont {Kranz}}, \bibinfo {author} {\bibfnamefont {B.}~\bibnamefont {Thorgrimsson}}, \bibinfo {author} {\bibfnamefont {S.~K.}\ \bibnamefont {Gorman}},\ and\ \bibinfo {author} {\bibfnamefont {M.~Y.}\ \bibnamefont {Simmons}},\ }\bibfield  {title} {\bibinfo {title} {Ramped measurement technique for robust high-fidelity spin qubit readout},\ }\href {https://doi.org/10.1126/sciadv.abq0455} {\bibfield  {journal} {\bibinfo  {journal} {Science Advances}\ }\textbf {\bibinfo {volume} {8}},\ \bibinfo {pages} {eabq0455} (\bibinfo {year} {2022})}\BibitemShut {NoStop}%
\bibitem [{\citenamefont {Chou}\ \emph {et~al.}(2024)\citenamefont {Chou}, \citenamefont {Shemma}, \citenamefont {McCarrick}, \citenamefont {Chien}, \citenamefont {Teoh}, \citenamefont {Winkel}, \citenamefont {Anderson}, \citenamefont {Chen}, \citenamefont {Curtis}, \citenamefont {de~Graaf}, \citenamefont {Garmon}, \citenamefont {Gudlewski}, \citenamefont {Kalfus}, \citenamefont {Keen}, \citenamefont {Khedkar}, \citenamefont {Lei}, \citenamefont {Liu}, \citenamefont {Lu}, \citenamefont {Lu}, \citenamefont {Maiti}, \citenamefont {Mastalli-Kelly}, \citenamefont {Mehta}, \citenamefont {Mundhada}, \citenamefont {Narla}, \citenamefont {Noh}, \citenamefont {Tsunoda}, \citenamefont {Xue}, \citenamefont {Yuan}, \citenamefont {Frunzio}, \citenamefont {Aumentado}, \citenamefont {Puri}, \citenamefont {Girvin}, \citenamefont {Moseley},\ and\ \citenamefont {Schoelkopf}}]{Chou2024}%
  \BibitemOpen
  \bibfield  {author} {\bibinfo {author} {\bibfnamefont {K.~S.}\ \bibnamefont {Chou}}, \bibinfo {author} {\bibfnamefont {T.}~\bibnamefont {Shemma}}, \bibinfo {author} {\bibfnamefont {H.}~\bibnamefont {McCarrick}}, \bibinfo {author} {\bibfnamefont {T.-C.}\ \bibnamefont {Chien}}, \bibinfo {author} {\bibfnamefont {J.~D.}\ \bibnamefont {Teoh}}, \bibinfo {author} {\bibfnamefont {P.}~\bibnamefont {Winkel}}, \bibinfo {author} {\bibfnamefont {A.}~\bibnamefont {Anderson}}, \bibinfo {author} {\bibfnamefont {J.}~\bibnamefont {Chen}}, \bibinfo {author} {\bibfnamefont {J.~C.}\ \bibnamefont {Curtis}}, \bibinfo {author} {\bibfnamefont {S.~J.}\ \bibnamefont {de~Graaf}}, \bibinfo {author} {\bibfnamefont {J.~W.~O.}\ \bibnamefont {Garmon}}, \bibinfo {author} {\bibfnamefont {B.}~\bibnamefont {Gudlewski}}, \bibinfo {author} {\bibfnamefont {W.~D.}\ \bibnamefont {Kalfus}}, \bibinfo {author} {\bibfnamefont {T.}~\bibnamefont {Keen}}, \bibinfo {author} {\bibfnamefont {N.}~\bibnamefont {Khedkar}}, \bibinfo {author} {\bibfnamefont
  {C.~U.}\ \bibnamefont {Lei}}, \bibinfo {author} {\bibfnamefont {G.}~\bibnamefont {Liu}}, \bibinfo {author} {\bibfnamefont {P.}~\bibnamefont {Lu}}, \bibinfo {author} {\bibfnamefont {Y.}~\bibnamefont {Lu}}, \bibinfo {author} {\bibfnamefont {A.}~\bibnamefont {Maiti}}, \bibinfo {author} {\bibfnamefont {L.}~\bibnamefont {Mastalli-Kelly}}, \bibinfo {author} {\bibfnamefont {N.}~\bibnamefont {Mehta}}, \bibinfo {author} {\bibfnamefont {S.~O.}\ \bibnamefont {Mundhada}}, \bibinfo {author} {\bibfnamefont {A.}~\bibnamefont {Narla}}, \bibinfo {author} {\bibfnamefont {T.}~\bibnamefont {Noh}}, \bibinfo {author} {\bibfnamefont {T.}~\bibnamefont {Tsunoda}}, \bibinfo {author} {\bibfnamefont {S.~H.}\ \bibnamefont {Xue}}, \bibinfo {author} {\bibfnamefont {J.~O.}\ \bibnamefont {Yuan}}, \bibinfo {author} {\bibfnamefont {L.}~\bibnamefont {Frunzio}}, \bibinfo {author} {\bibfnamefont {J.}~\bibnamefont {Aumentado}}, \bibinfo {author} {\bibfnamefont {S.}~\bibnamefont {Puri}}, \bibinfo {author} {\bibfnamefont {S.~M.}\ \bibnamefont
  {Girvin}}, \bibinfo {author} {\bibfnamefont {S.~H.}\ \bibnamefont {Moseley}},\ and\ \bibinfo {author} {\bibfnamefont {R.~J.}\ \bibnamefont {Schoelkopf}},\ }\bibfield  {title} {\bibinfo {title} {A superconducting dual-rail cavity qubit with erasure-detected logical measurements},\ }\href {https://doi.org/10.1038/s41567-024-02539-4} {\bibfield  {journal} {\bibinfo  {journal} {Nature Physics}\ }\textbf {\bibinfo {volume} {20}},\ \bibinfo {pages} {1454} (\bibinfo {year} {2024})}\BibitemShut {NoStop}%
\bibitem [{\citenamefont {Salath\'e}\ \emph {et~al.}(2018)\citenamefont {Salath\'e}, \citenamefont {Kurpiers}, \citenamefont {Karg}, \citenamefont {Lang}, \citenamefont {Andersen}, \citenamefont {Akin}, \citenamefont {Krinner}, \citenamefont {Eichler},\ and\ \citenamefont {Wallraff}}]{Salathe2018}%
  \BibitemOpen
  \bibfield  {author} {\bibinfo {author} {\bibfnamefont {Y.}~\bibnamefont {Salath\'e}}, \bibinfo {author} {\bibfnamefont {P.}~\bibnamefont {Kurpiers}}, \bibinfo {author} {\bibfnamefont {T.}~\bibnamefont {Karg}}, \bibinfo {author} {\bibfnamefont {C.}~\bibnamefont {Lang}}, \bibinfo {author} {\bibfnamefont {C.~K.}\ \bibnamefont {Andersen}}, \bibinfo {author} {\bibfnamefont {A.}~\bibnamefont {Akin}}, \bibinfo {author} {\bibfnamefont {S.}~\bibnamefont {Krinner}}, \bibinfo {author} {\bibfnamefont {C.}~\bibnamefont {Eichler}},\ and\ \bibinfo {author} {\bibfnamefont {A.}~\bibnamefont {Wallraff}},\ }\bibfield  {title} {\bibinfo {title} {Low-latency digital signal processing for feedback and feedforward in quantum computing and communication},\ }\href {https://doi.org/10.1103/PhysRevApplied.9.034011} {\bibfield  {journal} {\bibinfo  {journal} {Phys. Rev. Appl.}\ }\textbf {\bibinfo {volume} {9}},\ \bibinfo {pages} {034011} (\bibinfo {year} {2018})}\BibitemShut {NoStop}%
\bibitem [{\citenamefont {Rist\`e}\ \emph {et~al.}(2012)\citenamefont {Rist\`e}, \citenamefont {Bultink}, \citenamefont {Lehnert},\ and\ \citenamefont {DiCarlo}}]{Riste2012}%
  \BibitemOpen
  \bibfield  {author} {\bibinfo {author} {\bibfnamefont {D.}~\bibnamefont {Rist\`e}}, \bibinfo {author} {\bibfnamefont {C.~C.}\ \bibnamefont {Bultink}}, \bibinfo {author} {\bibfnamefont {K.~W.}\ \bibnamefont {Lehnert}},\ and\ \bibinfo {author} {\bibfnamefont {L.}~\bibnamefont {DiCarlo}},\ }\bibfield  {title} {\bibinfo {title} {Feedback control of a solid-state qubit using high-fidelity projective measurement},\ }\href {https://doi.org/10.1103/PhysRevLett.109.240502} {\bibfield  {journal} {\bibinfo  {journal} {Phys. Rev. Lett.}\ }\textbf {\bibinfo {volume} {109}},\ \bibinfo {pages} {240502} (\bibinfo {year} {2012})}\BibitemShut {NoStop}%
\bibitem [{\citenamefont {Reed}\ \emph {et~al.}(2010)\citenamefont {Reed}, \citenamefont {Johnson}, \citenamefont {Houck}, \citenamefont {DiCarlo}, \citenamefont {Chow}, \citenamefont {Schuster}, \citenamefont {Frunzio},\ and\ \citenamefont {Schoelkopf}}]{Reed2010}%
  \BibitemOpen
  \bibfield  {author} {\bibinfo {author} {\bibfnamefont {M.~D.}\ \bibnamefont {Reed}}, \bibinfo {author} {\bibfnamefont {B.~R.}\ \bibnamefont {Johnson}}, \bibinfo {author} {\bibfnamefont {A.~A.}\ \bibnamefont {Houck}}, \bibinfo {author} {\bibfnamefont {L.}~\bibnamefont {DiCarlo}}, \bibinfo {author} {\bibfnamefont {J.~M.}\ \bibnamefont {Chow}}, \bibinfo {author} {\bibfnamefont {D.~I.}\ \bibnamefont {Schuster}}, \bibinfo {author} {\bibfnamefont {L.}~\bibnamefont {Frunzio}},\ and\ \bibinfo {author} {\bibfnamefont {R.~J.}\ \bibnamefont {Schoelkopf}},\ }\bibfield  {title} {\bibinfo {title} {Fast reset and suppressing spontaneous emission of a superconducting qubit},\ }\href {https://doi.org/10.1063/1.3435463} {\bibfield  {journal} {\bibinfo  {journal} {Applied Physics Letters}\ }\textbf {\bibinfo {volume} {96}},\ \bibinfo {pages} {203110} (\bibinfo {year} {2010})}\BibitemShut {NoStop}%
\bibitem [{\citenamefont {Magnard}\ \emph {et~al.}(2018)\citenamefont {Magnard}, \citenamefont {Kurpiers}, \citenamefont {Royer}, \citenamefont {Walter}, \citenamefont {Besse}, \citenamefont {Gasparinetti}, \citenamefont {Pechal}, \citenamefont {Heinsoo}, \citenamefont {Storz}, \citenamefont {Blais},\ and\ \citenamefont {Wallraff}}]{Magnard2018}%
  \BibitemOpen
  \bibfield  {author} {\bibinfo {author} {\bibfnamefont {P.}~\bibnamefont {Magnard}}, \bibinfo {author} {\bibfnamefont {P.}~\bibnamefont {Kurpiers}}, \bibinfo {author} {\bibfnamefont {B.}~\bibnamefont {Royer}}, \bibinfo {author} {\bibfnamefont {T.}~\bibnamefont {Walter}}, \bibinfo {author} {\bibfnamefont {J.-C.}\ \bibnamefont {Besse}}, \bibinfo {author} {\bibfnamefont {S.}~\bibnamefont {Gasparinetti}}, \bibinfo {author} {\bibfnamefont {M.}~\bibnamefont {Pechal}}, \bibinfo {author} {\bibfnamefont {J.}~\bibnamefont {Heinsoo}}, \bibinfo {author} {\bibfnamefont {S.}~\bibnamefont {Storz}}, \bibinfo {author} {\bibfnamefont {A.}~\bibnamefont {Blais}},\ and\ \bibinfo {author} {\bibfnamefont {A.}~\bibnamefont {Wallraff}},\ }\bibfield  {title} {\bibinfo {title} {Fast and unconditional all-microwave reset of a superconducting qubit},\ }\href {https://doi.org/10.1103/PhysRevLett.121.060502} {\bibfield  {journal} {\bibinfo  {journal} {Phys. Rev. Lett.}\ }\textbf {\bibinfo {volume} {121}},\ \bibinfo {pages} {060502}
  (\bibinfo {year} {2018})}\BibitemShut {NoStop}%
\bibitem [{\citenamefont {Egger}\ \emph {et~al.}(2018)\citenamefont {Egger}, \citenamefont {Werninghaus}, \citenamefont {Ganzhorn}, \citenamefont {Salis}, \citenamefont {Fuhrer}, \citenamefont {M\"uller},\ and\ \citenamefont {Filipp}}]{Egger2018}%
  \BibitemOpen
  \bibfield  {author} {\bibinfo {author} {\bibfnamefont {D.}~\bibnamefont {Egger}}, \bibinfo {author} {\bibfnamefont {M.}~\bibnamefont {Werninghaus}}, \bibinfo {author} {\bibfnamefont {M.}~\bibnamefont {Ganzhorn}}, \bibinfo {author} {\bibfnamefont {G.}~\bibnamefont {Salis}}, \bibinfo {author} {\bibfnamefont {A.}~\bibnamefont {Fuhrer}}, \bibinfo {author} {\bibfnamefont {P.}~\bibnamefont {M\"uller}},\ and\ \bibinfo {author} {\bibfnamefont {S.}~\bibnamefont {Filipp}},\ }\bibfield  {title} {\bibinfo {title} {Pulsed reset protocol for fixed-frequency superconducting qubits},\ }\href {https://doi.org/10.1103/PhysRevApplied.10.044030} {\bibfield  {journal} {\bibinfo  {journal} {Phys. Rev. Appl.}\ }\textbf {\bibinfo {volume} {10}},\ \bibinfo {pages} {044030} (\bibinfo {year} {2018})}\BibitemShut {NoStop}%
\bibitem [{\citenamefont {Zhou}\ \emph {et~al.}(2021)\citenamefont {Zhou}, \citenamefont {Zhang}, \citenamefont {Yin}, \citenamefont {Huai}, \citenamefont {Gu}, \citenamefont {Xu}, \citenamefont {Allcock}, \citenamefont {Liu}, \citenamefont {Xi}, \citenamefont {Yu}, \citenamefont {Zhang}, \citenamefont {Zhang}, \citenamefont {Li}, \citenamefont {Song}, \citenamefont {Wang}, \citenamefont {Zheng}, \citenamefont {An}, \citenamefont {Zheng},\ and\ \citenamefont {Zhang}}]{Zhou2021}%
  \BibitemOpen
  \bibfield  {author} {\bibinfo {author} {\bibfnamefont {Y.}~\bibnamefont {Zhou}}, \bibinfo {author} {\bibfnamefont {Z.}~\bibnamefont {Zhang}}, \bibinfo {author} {\bibfnamefont {Z.}~\bibnamefont {Yin}}, \bibinfo {author} {\bibfnamefont {S.}~\bibnamefont {Huai}}, \bibinfo {author} {\bibfnamefont {X.}~\bibnamefont {Gu}}, \bibinfo {author} {\bibfnamefont {X.}~\bibnamefont {Xu}}, \bibinfo {author} {\bibfnamefont {J.}~\bibnamefont {Allcock}}, \bibinfo {author} {\bibfnamefont {F.}~\bibnamefont {Liu}}, \bibinfo {author} {\bibfnamefont {G.}~\bibnamefont {Xi}}, \bibinfo {author} {\bibfnamefont {Q.}~\bibnamefont {Yu}}, \bibinfo {author} {\bibfnamefont {H.}~\bibnamefont {Zhang}}, \bibinfo {author} {\bibfnamefont {M.}~\bibnamefont {Zhang}}, \bibinfo {author} {\bibfnamefont {H.}~\bibnamefont {Li}}, \bibinfo {author} {\bibfnamefont {X.}~\bibnamefont {Song}}, \bibinfo {author} {\bibfnamefont {Z.}~\bibnamefont {Wang}}, \bibinfo {author} {\bibfnamefont {D.}~\bibnamefont {Zheng}}, \bibinfo {author} {\bibfnamefont
  {S.}~\bibnamefont {An}}, \bibinfo {author} {\bibfnamefont {Y.}~\bibnamefont {Zheng}},\ and\ \bibinfo {author} {\bibfnamefont {S.}~\bibnamefont {Zhang}},\ }\bibfield  {title} {\bibinfo {title} {Rapid and unconditional parametric reset protocol for tunable superconducting qubits},\ }\href {https://doi.org/10.1038/s41467-021-26205-y} {\bibfield  {journal} {\bibinfo  {journal} {Nature Communications}\ }\textbf {\bibinfo {volume} {12}},\ \bibinfo {pages} {5924} (\bibinfo {year} {2021})}\BibitemShut {NoStop}%
\bibitem [{\citenamefont {Ding}\ \emph {et~al.}(2025)\citenamefont {Ding}, \citenamefont {Li}, \citenamefont {Wang}, \citenamefont {Xue}, \citenamefont {Su}, \citenamefont {Wang}, \citenamefont {Sun}, \citenamefont {Li}, \citenamefont {Zhang}, \citenamefont {Gao}, \citenamefont {Peng}, \citenamefont {Jiang}, \citenamefont {Yu}, \citenamefont {Yu},\ and\ \citenamefont {Yan}}]{Ding2025}%
  \BibitemOpen
  \bibfield  {author} {\bibinfo {author} {\bibfnamefont {J.}~\bibnamefont {Ding}}, \bibinfo {author} {\bibfnamefont {Y.}~\bibnamefont {Li}}, \bibinfo {author} {\bibfnamefont {H.}~\bibnamefont {Wang}}, \bibinfo {author} {\bibfnamefont {G.}~\bibnamefont {Xue}}, \bibinfo {author} {\bibfnamefont {T.}~\bibnamefont {Su}}, \bibinfo {author} {\bibfnamefont {C.}~\bibnamefont {Wang}}, \bibinfo {author} {\bibfnamefont {W.}~\bibnamefont {Sun}}, \bibinfo {author} {\bibfnamefont {F.}~\bibnamefont {Li}}, \bibinfo {author} {\bibfnamefont {Y.}~\bibnamefont {Zhang}}, \bibinfo {author} {\bibfnamefont {Y.}~\bibnamefont {Gao}}, \bibinfo {author} {\bibfnamefont {J.}~\bibnamefont {Peng}}, \bibinfo {author} {\bibfnamefont {Z.~H.}\ \bibnamefont {Jiang}}, \bibinfo {author} {\bibfnamefont {Y.}~\bibnamefont {Yu}}, \bibinfo {author} {\bibfnamefont {H.}~\bibnamefont {Yu}},\ and\ \bibinfo {author} {\bibfnamefont {F.}~\bibnamefont {Yan}},\ }\bibfield  {title} {\bibinfo {title} {Multipurpose architecture for fast reset and protective readout
  of superconducting qubits},\ }\href {https://doi.org/10.1103/PhysRevApplied.23.014012} {\bibfield  {journal} {\bibinfo  {journal} {Phys. Rev. Appl.}\ }\textbf {\bibinfo {volume} {23}},\ \bibinfo {pages} {014012} (\bibinfo {year} {2025})}\BibitemShut {NoStop}%
\bibitem [{\citenamefont {le~Floch}\ \emph {et~al.}(2008)\citenamefont {le~Floch}, \citenamefont {Tobar}, \citenamefont {Cros},\ and\ \citenamefont {Krupka}}]{Floch2008}%
  \BibitemOpen
  \bibfield  {author} {\bibinfo {author} {\bibfnamefont {J.-M.}\ \bibnamefont {le~Floch}}, \bibinfo {author} {\bibfnamefont {M.~E.}\ \bibnamefont {Tobar}}, \bibinfo {author} {\bibfnamefont {D.}~\bibnamefont {Cros}},\ and\ \bibinfo {author} {\bibfnamefont {J.}~\bibnamefont {Krupka}},\ }\bibfield  {title} {\bibinfo {title} {Low-loss materials for high q-factor bragg reflector resonators},\ }\href {https://doi.org/10.1063/1.2828025} {\bibfield  {journal} {\bibinfo  {journal} {Applied Physics Letters}\ }\textbf {\bibinfo {volume} {92}},\ \bibinfo {pages} {032901} (\bibinfo {year} {2008})}\BibitemShut {NoStop}%
\bibitem [{\citenamefont {Hovis}\ and\ \citenamefont {Reddy}(2006)}]{Hovis2006}%
  \BibitemOpen
  \bibfield  {author} {\bibinfo {author} {\bibfnamefont {D.}~\bibnamefont {Hovis}}\ and\ \bibinfo {author} {\bibfnamefont {A.}~\bibnamefont {Reddy}},\ }\bibfield  {title} {\bibinfo {title} {X-ray elastic constants for $\alpha$-al2o3},\ }\href {https://doi.org/10.1063/1.2189071} {\bibfield  {journal} {\bibinfo  {journal} {Applied Physics Letters}\ }\textbf {\bibinfo {volume} {88}} (\bibinfo {year} {2006})}\BibitemShut {NoStop}%
\bibitem [{\citenamefont {Vodenitcharova}\ \emph {et~al.}(2007)\citenamefont {Vodenitcharova}, \citenamefont {Zhang}, \citenamefont {Zarudi}, \citenamefont {Yin}, \citenamefont {Domyo}, \citenamefont {Ho},\ and\ \citenamefont {Sato}}]{Vodenitcharova2007}%
  \BibitemOpen
  \bibfield  {author} {\bibinfo {author} {\bibfnamefont {T.}~\bibnamefont {Vodenitcharova}}, \bibinfo {author} {\bibfnamefont {L.}~\bibnamefont {Zhang}}, \bibinfo {author} {\bibfnamefont {I.}~\bibnamefont {Zarudi}}, \bibinfo {author} {\bibfnamefont {Y.}~\bibnamefont {Yin}}, \bibinfo {author} {\bibfnamefont {H.}~\bibnamefont {Domyo}}, \bibinfo {author} {\bibfnamefont {T.}~\bibnamefont {Ho}},\ and\ \bibinfo {author} {\bibfnamefont {M.}~\bibnamefont {Sato}},\ }\bibfield  {title} {\bibinfo {title} {The effect of anisotropy on the deformation and fracture of sapphire wafers subjected to thermal shocks},\ }\href {https://doi.org/https://doi.org/10.1016/j.jmatprotec.2007.03.125} {\bibfield  {journal} {\bibinfo  {journal} {Journal of Materials Processing Technology}\ }\textbf {\bibinfo {volume} {194}},\ \bibinfo {pages} {52} (\bibinfo {year} {2007})}\BibitemShut {NoStop}%
\bibitem [{\citenamefont {{Roditi International Corporation}}(2025)}]{roditi_2025}%
  \BibitemOpen
  \bibfield  {author} {\bibinfo {author} {\bibnamefont {{Roditi International Corporation}}},\ }\href@noop {} {\bibinfo {title} {{Sapphire Properties}}},\ \bibinfo {howpublished} {\url{https://www.roditi.com/SingleCrystal/Sapphire/Properties.html}} (\bibinfo {year} {2025}),\ \bibinfo {note} {accessed: 2025-06-19}\BibitemShut {NoStop}%
\bibitem [{\citenamefont {Maggiore}(2007)}]{Maggiore2007}%
  \BibitemOpen
  \bibfield  {author} {\bibinfo {author} {\bibfnamefont {M.}~\bibnamefont {Maggiore}},\ }\href {https://doi.org/10.1093/acprof:oso/9780198570745.001.0001} {\emph {\bibinfo {title} {Gravitational Waves: Volume 1: Theory and Experiments}}}\ (\bibinfo  {publisher} {Oxford University Press},\ \bibinfo {year} {2007})\BibitemShut {NoStop}%
\bibitem [{\citenamefont {Goryachev}\ and\ \citenamefont {Tobar}(2014)}]{goryachev_gravitational_2014}%
  \BibitemOpen
  \bibfield  {author} {\bibinfo {author} {\bibfnamefont {M.}~\bibnamefont {Goryachev}}\ and\ \bibinfo {author} {\bibfnamefont {M.~E.}\ \bibnamefont {Tobar}},\ }\bibfield  {title} {\bibinfo {title} {Gravitational wave detection with high frequency phonon trapping acoustic cavities},\ }\href {https://doi.org/10.1103/PhysRevD.90.102005} {\bibfield  {journal} {\bibinfo  {journal} {Physical Review D}\ }\textbf {\bibinfo {volume} {90}},\ \bibinfo {pages} {102005} (\bibinfo {year} {2014})},\ \bibinfo {note} {publisher: American Physical Society}\BibitemShut {NoStop}%
\bibitem [{\citenamefont {Muia}\ \emph {et~al.}(2025)\citenamefont {Muia}, \citenamefont {Ringwald},\ and\ \citenamefont {Tamarit}}]{HFGWPlotter_Omega}%
  \BibitemOpen
  \bibfield  {author} {\bibinfo {author} {\bibfnamefont {F.}~\bibnamefont {Muia}}, \bibinfo {author} {\bibfnamefont {A.}~\bibnamefont {Ringwald}},\ and\ \bibinfo {author} {\bibfnamefont {C.}~\bibnamefont {Tamarit}},\ }\href@noop {} {\bibinfo {title} {{HFGWPlotter}: A python tool to visualize stochastic gravitational wave backgrounds}},\ \bibinfo {howpublished} {\url{https://github.com/ctamaritd/HFGWPlotter_Sh}} (\bibinfo {year} {2025}),\ \bibinfo {note} {accessed: 2025-06-12}\BibitemShut {NoStop}%
\bibitem [{\citenamefont {Zhang}\ \emph {et~al.}(2021)\citenamefont {Zhang}, \citenamefont {Chakram}, \citenamefont {Roy}, \citenamefont {Earnest}, \citenamefont {Lu}, \citenamefont {Huang}, \citenamefont {Weiss}, \citenamefont {Koch},\ and\ \citenamefont {Schuster}}]{Zhang_2021}%
  \BibitemOpen
  \bibfield  {author} {\bibinfo {author} {\bibfnamefont {H.}~\bibnamefont {Zhang}}, \bibinfo {author} {\bibfnamefont {S.}~\bibnamefont {Chakram}}, \bibinfo {author} {\bibfnamefont {T.}~\bibnamefont {Roy}}, \bibinfo {author} {\bibfnamefont {N.}~\bibnamefont {Earnest}}, \bibinfo {author} {\bibfnamefont {Y.}~\bibnamefont {Lu}}, \bibinfo {author} {\bibfnamefont {Z.}~\bibnamefont {Huang}}, \bibinfo {author} {\bibfnamefont {D.~K.}\ \bibnamefont {Weiss}}, \bibinfo {author} {\bibfnamefont {J.}~\bibnamefont {Koch}},\ and\ \bibinfo {author} {\bibfnamefont {D.~I.}\ \bibnamefont {Schuster}},\ }\bibfield  {title} {\bibinfo {title} {Universal fast-flux control of a coherent, low-frequency qubit},\ }\href {https://doi.org/10.1103/PhysRevX.11.011010} {\bibfield  {journal} {\bibinfo  {journal} {Phys. Rev. X}\ }\textbf {\bibinfo {volume} {11}},\ \bibinfo {pages} {011010} (\bibinfo {year} {2021})}\BibitemShut {NoStop}%
\bibitem [{\citenamefont {Arias}\ \emph {et~al.}(2012)\citenamefont {Arias}, \citenamefont {Cadamuro}, \citenamefont {Goodsell}, \citenamefont {Jaeckel}, \citenamefont {Redondo},\ and\ \citenamefont {Ringwald}}]{PaolaArias_2012}%
  \BibitemOpen
  \bibfield  {author} {\bibinfo {author} {\bibfnamefont {P.}~\bibnamefont {Arias}}, \bibinfo {author} {\bibfnamefont {D.}~\bibnamefont {Cadamuro}}, \bibinfo {author} {\bibfnamefont {M.}~\bibnamefont {Goodsell}}, \bibinfo {author} {\bibfnamefont {J.}~\bibnamefont {Jaeckel}}, \bibinfo {author} {\bibfnamefont {J.}~\bibnamefont {Redondo}},\ and\ \bibinfo {author} {\bibfnamefont {A.}~\bibnamefont {Ringwald}},\ }\bibfield  {title} {\bibinfo {title} {Wispy cold dark matter},\ }\href {https://doi.org/10.1088/1475-7516/2012/06/013} {\bibfield  {journal} {\bibinfo  {journal} {Journal of Cosmology and Astroparticle Physics}\ }\textbf {\bibinfo {volume} {2012}}\bibinfo  {number} { (06)},\ \bibinfo {pages} {013}}\BibitemShut {NoStop}%
\bibitem [{\citenamefont {An}\ \emph {et~al.}(2025)\citenamefont {An}, \citenamefont {Ge}, \citenamefont {Liu},\ and\ \citenamefont {Liu}}]{An2025}%
  \BibitemOpen
\bibfield  {number} {  }\bibfield  {author} {\bibinfo {author} {\bibfnamefont {H.}~\bibnamefont {An}}, \bibinfo {author} {\bibfnamefont {S.}~\bibnamefont {Ge}}, \bibinfo {author} {\bibfnamefont {J.}~\bibnamefont {Liu}},\ and\ \bibinfo {author} {\bibfnamefont {M.}~\bibnamefont {Liu}},\ }\bibfield  {title} {\bibinfo {title} {In situ measurements of dark photon dark matter using parker solar probe: Going beyond the radio window},\ }\href {https://doi.org/10.1103/PhysRevLett.134.171001} {\bibfield  {journal} {\bibinfo  {journal} {Phys. Rev. Lett.}\ }\textbf {\bibinfo {volume} {134}},\ \bibinfo {pages} {171001} (\bibinfo {year} {2025})}\BibitemShut {NoStop}%
\bibitem [{\citenamefont {Semertzidis}\ and\ \citenamefont {Youn}(2022)}]{Semertzidis2022}%
  \BibitemOpen
  \bibfield  {author} {\bibinfo {author} {\bibfnamefont {Y.~K.}\ \bibnamefont {Semertzidis}}\ and\ \bibinfo {author} {\bibfnamefont {S.}~\bibnamefont {Youn}},\ }\bibfield  {title} {\bibinfo {title} {Axion dark matter: How to see it?},\ }\href {https://doi.org/10.1126/sciadv.abm9928} {\bibfield  {journal} {\bibinfo  {journal} {Science Advances}\ }\textbf {\bibinfo {volume} {8}},\ \bibinfo {pages} {eabm9928} (\bibinfo {year} {2022})}\BibitemShut {NoStop}%
\bibitem [{\citenamefont {{Irwin Lab, Stanford University}}(2025)}]{Irwin_DMRadio}%
  \BibitemOpen
  \bibfield  {author} {\bibinfo {author} {\bibnamefont {{Irwin Lab, Stanford University}}},\ }\href@noop {} {\bibinfo {title} {{Dark Matter Radio (DM Radio)}}},\ \bibinfo {howpublished} {\url{https://irwinlab.stanford.edu/dark-matter-radio-dm-radio}} (\bibinfo {year} {2025}),\ \bibinfo {note} {accessed: 2025-06-11}\BibitemShut {NoStop}%
\bibitem [{\citenamefont {Godfrey}\ \emph {et~al.}(2021)\citenamefont {Godfrey}, \citenamefont {Tyson}, \citenamefont {Hillbrand}, \citenamefont {Balajthy}, \citenamefont {Polin}, \citenamefont {Tripathi}, \citenamefont {Klomp}, \citenamefont {Levine}, \citenamefont {MacFadden}, \citenamefont {Kolner}, \citenamefont {Smith}, \citenamefont {Stucky}, \citenamefont {Phipps}, \citenamefont {Graham},\ and\ \citenamefont {Irwin}}]{Godfrey2021}%
  \BibitemOpen
  \bibfield  {author} {\bibinfo {author} {\bibfnamefont {B.}~\bibnamefont {Godfrey}}, \bibinfo {author} {\bibfnamefont {J.~A.}\ \bibnamefont {Tyson}}, \bibinfo {author} {\bibfnamefont {S.}~\bibnamefont {Hillbrand}}, \bibinfo {author} {\bibfnamefont {J.}~\bibnamefont {Balajthy}}, \bibinfo {author} {\bibfnamefont {D.}~\bibnamefont {Polin}}, \bibinfo {author} {\bibfnamefont {S.~M.}\ \bibnamefont {Tripathi}}, \bibinfo {author} {\bibfnamefont {S.}~\bibnamefont {Klomp}}, \bibinfo {author} {\bibfnamefont {J.}~\bibnamefont {Levine}}, \bibinfo {author} {\bibfnamefont {N.}~\bibnamefont {MacFadden}}, \bibinfo {author} {\bibfnamefont {B.~H.}\ \bibnamefont {Kolner}}, \bibinfo {author} {\bibfnamefont {M.~R.}\ \bibnamefont {Smith}}, \bibinfo {author} {\bibfnamefont {P.}~\bibnamefont {Stucky}}, \bibinfo {author} {\bibfnamefont {A.}~\bibnamefont {Phipps}}, \bibinfo {author} {\bibfnamefont {P.}~\bibnamefont {Graham}},\ and\ \bibinfo {author} {\bibfnamefont {K.}~\bibnamefont {Irwin}},\ }\bibfield  {title} {\bibinfo {title}
  {Search for dark photon dark matter: Dark $e$ field radio pilot experiment},\ }\href {https://doi.org/10.1103/PhysRevD.104.012013} {\bibfield  {journal} {\bibinfo  {journal} {Phys. Rev. D}\ }\textbf {\bibinfo {volume} {104}},\ \bibinfo {pages} {012013} (\bibinfo {year} {2021})}\BibitemShut {NoStop}%
\bibitem [{\citenamefont {Millar}\ \emph {et~al.}(2023)\citenamefont {Millar}, \citenamefont {Anlage}, \citenamefont {Balafendiev}, \citenamefont {Belov}, \citenamefont {van Bibber}, \citenamefont {Conrad}, \citenamefont {Demarteau}, \citenamefont {Droster}, \citenamefont {Dunne}, \citenamefont {Rosso}, \citenamefont {Gudmundsson}, \citenamefont {Jackson}, \citenamefont {Kaur}, \citenamefont {Klaesson}, \citenamefont {Kowitt}, \citenamefont {Lawson}, \citenamefont {Leder}, \citenamefont {Miyazaki}, \citenamefont {Morampudi}, \citenamefont {Peiris}, \citenamefont {R\o{}ising}, \citenamefont {Singh}, \citenamefont {Sun}, \citenamefont {Thomas}, \citenamefont {Wilczek}, \citenamefont {Withington}, \citenamefont {Wooten}, \citenamefont {Dilling}, \citenamefont {Febbraro}, \citenamefont {Knirck},\ and\ \citenamefont {Marvinney}}]{Millar2023}%
  \BibitemOpen
  \bibfield  {author} {\bibinfo {author} {\bibfnamefont {A.~J.}\ \bibnamefont {Millar}}, \bibinfo {author} {\bibfnamefont {S.~M.}\ \bibnamefont {Anlage}}, \bibinfo {author} {\bibfnamefont {R.}~\bibnamefont {Balafendiev}}, \bibinfo {author} {\bibfnamefont {P.}~\bibnamefont {Belov}}, \bibinfo {author} {\bibfnamefont {K.}~\bibnamefont {van Bibber}}, \bibinfo {author} {\bibfnamefont {J.}~\bibnamefont {Conrad}}, \bibinfo {author} {\bibfnamefont {M.}~\bibnamefont {Demarteau}}, \bibinfo {author} {\bibfnamefont {A.}~\bibnamefont {Droster}}, \bibinfo {author} {\bibfnamefont {K.}~\bibnamefont {Dunne}}, \bibinfo {author} {\bibfnamefont {A.~G.}\ \bibnamefont {Rosso}}, \bibinfo {author} {\bibfnamefont {J.~E.}\ \bibnamefont {Gudmundsson}}, \bibinfo {author} {\bibfnamefont {H.}~\bibnamefont {Jackson}}, \bibinfo {author} {\bibfnamefont {G.}~\bibnamefont {Kaur}}, \bibinfo {author} {\bibfnamefont {T.}~\bibnamefont {Klaesson}}, \bibinfo {author} {\bibfnamefont {N.}~\bibnamefont {Kowitt}}, \bibinfo {author} {\bibfnamefont
  {M.}~\bibnamefont {Lawson}}, \bibinfo {author} {\bibfnamefont {A.}~\bibnamefont {Leder}}, \bibinfo {author} {\bibfnamefont {A.}~\bibnamefont {Miyazaki}}, \bibinfo {author} {\bibfnamefont {S.}~\bibnamefont {Morampudi}}, \bibinfo {author} {\bibfnamefont {H.~V.}\ \bibnamefont {Peiris}}, \bibinfo {author} {\bibfnamefont {H.~S.}\ \bibnamefont {R\o{}ising}}, \bibinfo {author} {\bibfnamefont {G.}~\bibnamefont {Singh}}, \bibinfo {author} {\bibfnamefont {D.}~\bibnamefont {Sun}}, \bibinfo {author} {\bibfnamefont {J.~H.}\ \bibnamefont {Thomas}}, \bibinfo {author} {\bibfnamefont {F.}~\bibnamefont {Wilczek}}, \bibinfo {author} {\bibfnamefont {S.}~\bibnamefont {Withington}}, \bibinfo {author} {\bibfnamefont {M.}~\bibnamefont {Wooten}}, \bibinfo {author} {\bibfnamefont {J.}~\bibnamefont {Dilling}}, \bibinfo {author} {\bibfnamefont {M.}~\bibnamefont {Febbraro}}, \bibinfo {author} {\bibfnamefont {S.}~\bibnamefont {Knirck}},\ and\ \bibinfo {author} {\bibfnamefont {C.}~\bibnamefont {Marvinney}} (\bibinfo {collaboration}
  {Endorsers}),\ }\bibfield  {title} {\bibinfo {title} {Searching for dark matter with plasma haloscopes},\ }\href {https://doi.org/10.1103/PhysRevD.107.055013} {\bibfield  {journal} {\bibinfo  {journal} {Phys. Rev. D}\ }\textbf {\bibinfo {volume} {107}},\ \bibinfo {pages} {055013} (\bibinfo {year} {2023})}\BibitemShut {NoStop}%
\bibitem [{\citenamefont {Brun}\ \emph {et~al.}(2019)\citenamefont {Brun}, \citenamefont {Caldwell}, \citenamefont {Chevalier}, \citenamefont {Dvali}, \citenamefont {Freire}, \citenamefont {Garutti}, \citenamefont {Heyminck}, \citenamefont {Jochum}, \citenamefont {Knirck}, \citenamefont {Kramer}, \citenamefont {Krieger}, \citenamefont {Lasserre}, \citenamefont {Lee}, \citenamefont {Li}, \citenamefont {Lindner}, \citenamefont {Majorovits}, \citenamefont {Martens}, \citenamefont {Matysek}, \citenamefont {Millar}, \citenamefont {Raffelt}, \citenamefont {Redondo}, \citenamefont {Reimann}, \citenamefont {Ringwald}, \citenamefont {Saikawa}, \citenamefont {Schaffran}, \citenamefont {Schmidt}, \citenamefont {Sch{\"u}tte-Engel}, \citenamefont {Steffen}, \citenamefont {Strandhagen}, \citenamefont {Wieching},\ and\ \citenamefont {Collaboration}}]{Brun2019}%
  \BibitemOpen
  \bibfield  {author} {\bibinfo {author} {\bibfnamefont {P.}~\bibnamefont {Brun}}, \bibinfo {author} {\bibfnamefont {A.}~\bibnamefont {Caldwell}}, \bibinfo {author} {\bibfnamefont {L.}~\bibnamefont {Chevalier}}, \bibinfo {author} {\bibfnamefont {G.}~\bibnamefont {Dvali}}, \bibinfo {author} {\bibfnamefont {P.}~\bibnamefont {Freire}}, \bibinfo {author} {\bibfnamefont {E.}~\bibnamefont {Garutti}}, \bibinfo {author} {\bibfnamefont {S.}~\bibnamefont {Heyminck}}, \bibinfo {author} {\bibfnamefont {J.}~\bibnamefont {Jochum}}, \bibinfo {author} {\bibfnamefont {S.}~\bibnamefont {Knirck}}, \bibinfo {author} {\bibfnamefont {M.}~\bibnamefont {Kramer}}, \bibinfo {author} {\bibfnamefont {C.}~\bibnamefont {Krieger}}, \bibinfo {author} {\bibfnamefont {T.}~\bibnamefont {Lasserre}}, \bibinfo {author} {\bibfnamefont {C.}~\bibnamefont {Lee}}, \bibinfo {author} {\bibfnamefont {X.}~\bibnamefont {Li}}, \bibinfo {author} {\bibfnamefont {A.}~\bibnamefont {Lindner}}, \bibinfo {author} {\bibfnamefont {B.}~\bibnamefont {Majorovits}},
  \bibinfo {author} {\bibfnamefont {S.}~\bibnamefont {Martens}}, \bibinfo {author} {\bibfnamefont {M.}~\bibnamefont {Matysek}}, \bibinfo {author} {\bibfnamefont {A.}~\bibnamefont {Millar}}, \bibinfo {author} {\bibfnamefont {G.}~\bibnamefont {Raffelt}}, \bibinfo {author} {\bibfnamefont {J.}~\bibnamefont {Redondo}}, \bibinfo {author} {\bibfnamefont {O.}~\bibnamefont {Reimann}}, \bibinfo {author} {\bibfnamefont {A.}~\bibnamefont {Ringwald}}, \bibinfo {author} {\bibfnamefont {K.}~\bibnamefont {Saikawa}}, \bibinfo {author} {\bibfnamefont {J.}~\bibnamefont {Schaffran}}, \bibinfo {author} {\bibfnamefont {A.}~\bibnamefont {Schmidt}}, \bibinfo {author} {\bibfnamefont {J.}~\bibnamefont {Sch{\"u}tte-Engel}}, \bibinfo {author} {\bibfnamefont {F.}~\bibnamefont {Steffen}}, \bibinfo {author} {\bibfnamefont {C.}~\bibnamefont {Strandhagen}}, \bibinfo {author} {\bibfnamefont {G.}~\bibnamefont {Wieching}},\ and\ \bibinfo {author} {\bibfnamefont {M.~A. D. M. A.~X.}\ \bibnamefont {Collaboration}},\ }\bibfield  {title} {\bibinfo
  {title} {A new experimental approach to probe qcd axion dark matter in the mass range above 40$\mu$ev},\ }\href {https://doi.org/10.1140/epjc/s10052-019-6683-x} {\bibfield  {journal} {\bibinfo  {journal} {The European Physical Journal C}\ }\textbf {\bibinfo {volume} {79}},\ \bibinfo {pages} {186} (\bibinfo {year} {2019})}\BibitemShut {NoStop}%
\bibitem [{\citenamefont {Donadi}\ \emph {et~al.}(2021)\citenamefont {Donadi}, \citenamefont {Piscicchia}, \citenamefont {Del~Grande}, \citenamefont {Curceanu}, \citenamefont {Laubenstein},\ and\ \citenamefont {Bassi}}]{Donadi_Xray_21}%
  \BibitemOpen
  \bibfield  {author} {\bibinfo {author} {\bibfnamefont {S.}~\bibnamefont {Donadi}}, \bibinfo {author} {\bibfnamefont {K.}~\bibnamefont {Piscicchia}}, \bibinfo {author} {\bibfnamefont {R.}~\bibnamefont {Del~Grande}}, \bibinfo {author} {\bibfnamefont {C.}~\bibnamefont {Curceanu}}, \bibinfo {author} {\bibfnamefont {M.}~\bibnamefont {Laubenstein}},\ and\ \bibinfo {author} {\bibfnamefont {A.}~\bibnamefont {Bassi}},\ }\bibfield  {title} {\bibinfo {title} {Novel csl bounds from the noise-induced radiation emission from atoms},\ }\href {https://doi.org/10.1140/epjc/s10052-021-09556-0} {\bibfield  {journal} {\bibinfo  {journal} {The European Physical Journal C}\ }\textbf {\bibinfo {volume} {81}},\ \bibinfo {pages} {773} (\bibinfo {year} {2021})}\BibitemShut {NoStop}%
\bibitem [{\citenamefont {Adler}(2007)}]{Adler_2007}%
  \BibitemOpen
  \bibfield  {author} {\bibinfo {author} {\bibfnamefont {S.~L.}\ \bibnamefont {Adler}},\ }\bibfield  {title} {\bibinfo {title} {Lower and upper bounds on csl parameters from latent image formation and igm heating},\ }\href {https://doi.org/10.1088/1751-8113/40/12/S03} {\bibfield  {journal} {\bibinfo  {journal} {Journal of Physics A: Mathematical and Theoretical}\ }\textbf {\bibinfo {volume} {40}},\ \bibinfo {pages} {2935} (\bibinfo {year} {2007})}\BibitemShut {NoStop}%
\end{thebibliography}%
\end{document}